# Reliability in Source Coding with Side Information

Benjamin G. Kelly and Aaron B. Wagner

School of Electrical and Computer Engineering

Cornell University

Ithaca, NY 14853

bgk6@cornell.edu, wagner@ece.cornell.edu

### Abstract

We study error exponents for source coding with side information. Both achievable exponents and converse bounds are obtained for the following two cases: lossless source coding with coded information (SCCSI) and lossy source coding with full side information (Wyner-Ziv). These results recover and extend several existing results on source-coding error exponents and are tight in some circumstances. Our bounds have a natural interpretation as a two-player game between nature and the code designer, with nature seeking to minimize the exponent and the code designer seeking to maximize it. In the Wyner-Ziv problem our analysis exposes a tension in the choice of test channel with the optimal test channel balancing two competing error events. The Gaussian and binary-erasure cases are examined in detail.

## I. Introduction

In a typical lossy data compression problem a source is to be compressed by an encoder at a prescribed rate so that a decoder may reproduce the source to within some desired fidelity (distortion). Sometimes present, in addition to the data to be compressed, is some correlated information that can be utilized by a second encoder, that is able to send a separate message to the decoder. We refer to this kind of problem as source coding with side information (SCSI). The set-up is depicted in Fig. 1, where a source $X$ is compressed by encoder one to a rate $R_1$ with the decoder having access to encoded side information $Y$, compressed at rate $R_2$ by encoder two, as well as the compressed version of $X$ from the first encoder.

The SCSI scenario arises in a variety of applications. For example, in video applications [1] $X$ can represent a current frame, and $Y$ a separate correlated frame sent from a second encoder. $Y$ can even represent the frame(s) preceding the current frame $X$ in the stream: while the previous frames are certainly available to the encoder, the encoder's coding scheme can be simplified by not making use of this information and leaving the decoder to exploit the interframe dependence. A second example can be found in communication in networks with relays [2]. A source sends a message $X$ to a sink in a network containing a relay. One mode of operation for the relay is "compress and forward", i.e. for the relay to send a compressed version of its observation, $Y$, of the source-sink message to the sink. This compressed message can be used by the sink to further aid its decoding. SCSI appears in applications even beyond communication, for example (with minor changes) it has been proposed as a model for rate-constrained pattern recognition [3].

For the lossless problem with coded side information (SCCSI)[1], and the lossy problem with full side information (Wyner-Ziv), the "rate region" problem, i.e. determining the rates required to meet a given average distortion constraint, is solved. In this paper, we study these two problems from an error-exponent standpoint. Our motivation for doing so is three-fold:

- In the applications mentioned above the average distortion of a compression scheme is not the only important metric. Indeed, a video compression system with good average performance but that frequently yields poor images, or a communication system that suffers from frequent outages is usually deemed unacceptable. In addition to minimizing the average distortion, one would like to minimize the fraction of time in which the images are poor or the relay is unable to help.

---

[1]Also known as the "One Helper" problem, Wyner's problem [4] or the Ahlswede-Körner problem [5].

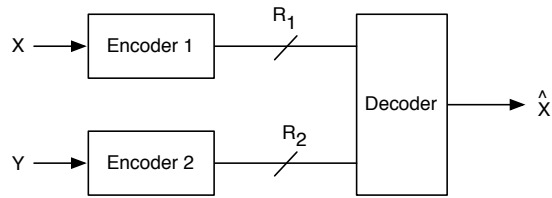

Fig. 1. The Source Coding with Side Information Problem

- In some important cases, there is no *rate loss*, meaning that there is no difference in the rate-distortion performance between the SCSI problem and the problem in which the side information $Y$ is available to the encoder as well as the decoder. In particular, it is well known that this is true of both the binary erasure and quadratic Gaussian forms of the Wyner-Ziv problem [6]. This raises the question of whether these two systems are equivalent when performance is measured via error exponents instead of the average distortion.

- Recently a connection has been established between error exponents in channel coding and the stabilization of linear systems over noisy channels [7], and there is a known interdependence between source- and channel-coding error exponents. Thus new techniques in source-coding error exponents could aid our understanding of problems at the intersection of communication and control [8].

### A. Contributions and Overview

Our key contributions are achievable exponents and converse bounds for the SCCSI and Wyner-Ziv problems. The conventional approach to proving coding theorems for the these problems [9] relies on typicality-based arguments and yields error exponents that are essentially zero. By using more sophisticated coding techniques, we obtain lower bounds that are strictly positive for all achievable rates and distortions. Both achievable exponents have a natural interpretation as a dynamic, two-player, zero-sum game between nature and the code designer, with nature seeking to minimize the exponent and the code designer seeking to maximize it. Play alternates between the two players, and the available actions for each stage correspond to marginal or conditional probability distributions. At the end of the game, the actions selected by the players together determine the joint distribution of all of the relevant random variables, which in turn determines the achievable exponent. See Sections III-A and IV-A for more detail.

For the SCCSI problem, our upper bound uses a change-of-measure argument that is more refined than the conventional approach [10, p.g. 268] and yields a formally better bound. This bound more accurately captures the structure of the problem and might be applicable to other network information setups. The proof also uses the Karush-Kuhn-Tucker (KKT) conditions in a novel way to obtain cardinality bounds on the auxiliary random variable.

For the Wyner-Ziv problem, we supply results for both the discrete-memoryless and Gaussian versions of the problem. Our analysis indicates that the optimization of the coding scheme is a richer problem than it is when the goal is to minimize the average distortion. In both cases, the encoder performs vector quantization, with an associated test channel, followed by binning. When the goal is to minimize the average distortion, the test channel should be chosen to be "clean" so that the binning error probability vanishes but with a negligible exponent [9, Thm. 15.9.1]. When optimizing the error exponent, on the other hand, this choice is poor because the overall exponent is dominated by binning errors. Choosing a "noisy" test channel, leads to a large binning error exponent but results in little information transmission from the encoder to the decoder. This also leads to a poor overall exponent, because small atypicalities in $(X^n, Y^n)$ lead to a distortion error. The optimum choice of the test channel balances these two competing error events. This is illustrated numerically in section V-A for the binary erasure version of the problem. A similar tension arises in the problem of compression for distributed hypothesis testing [11].

Our results present evidence that, for both the binary erasure and Gaussian cases, there is likely a difference in the error exponents between conventional Wyner-Ziv and the version of the problem in which the side information is available at both encoder and decoder (an "exponent loss"). This is in contrast to the rate-distortion version of the problem, for which the two scenarios have identical performance. Determining whether the reliability functions are indeed different is an interesting topic for future work.

An application of our results on discrete-memoryless Wyner-Ziv allows us to determine the reliability function exactly (for a range of rates) for the lossless functional source coding problem, in which the goal is to reproduce a function $g(X)$ at the decoder (see section IV-A).

In our coding scheme the optimum test channel depends crucially upon the source statistics (see Fig. 4), which for the applications mentioned at the outset may not be known exactly. Thus, another implication of our results is that video coding or relaying systems based on a Wyner-Ziv scheme are likely to require detailed knowledge of the source distribution. This provides a theoretical justification for the observation that good estimates of the correlation between source and side information are "critical to the performance" of practical Wyner-Ziv coding schemes [12].

## B. Other Prior Work

Error exponents for both SCCSI and Wyner-Ziv were studied by Arutyunyan and Marutyan [13]. However, their results were not proven rigorously and appear to be unduly strong; they have recently been retracted [14]. Kochman and Wornell [15] have recently studied achievable exponents for the Gaussian Wyner-Ziv problem using lattices, and have conjecture an exponent loss in certain settings. Eswaran and Gastpar [16] have established an achievable exponent for the Berger-Yeung problem [17], which subsumes both of the problems studied here. Their approach is based on determining the rate of convergence in the Markov lemma and is fundamentally different from the approach used here. It is not difficult to find cases for which the achieveable exponent presented here exceeds theirs[2]. Morever, we shall see that approach used here reveals greater insight into both the design of coding schemes for these problems and theoretical questions such as the exponent loss for the Binary Erasure and Gaussian Wyner-Ziv problems.

For the SCCSI problem, Csiszár and Körner [10, p.g. 268] provide an upper bound on the reliability function. This bound is formally improved in the present paper by using a more refined change-of-measure argument. For the Wyner-Ziv problem, Jayaraman and Berger [18] studied the exponent associated with the binning error probability. One of the goals of this paper is to show that a binning error is only one of two competing error events. In this sense, at the error exponent level the Wyner-Ziv problem resembles the problem of distributed hypothesis testing [19].

The Wyner-Ziv problem is in a sense "dual" to the problem of channel coding with side information (CCSI) (see [20], [21] for a precise statement). Comparing the results in this paper to error exponent studies of the CCSI problem [22], [23], however, show that this duality breaks down at the level of error exponents. In particular, in the CCSI problem, the encoder can force the realization of the auxiliary random variable to have a specified joint distribution with the side information. In the Wyner-Ziv problem, however, the encoder must rely on the law of large numbers to ensure this. At the rate level, atypical realizations can be ignored and this difference is immaterial. At the level of error exponents, on the other hand, the two are quite different, and the Wyner-Ziv setup is more challenging. There is a substantial literature on error exponents for simpler source coding problems such as lossless compression with full side information [24], [25], [26], the Slepian-Wolf problem [27], [28], [29], and lossy compression without side information [30], [31]. None of the these problems involve optimization over an auxiliary random variable, however, and we shall see that the presence of auxiliary random variables makes the error exponent problem significantly richer.

---

[2] In fact, it is not difficult to find examples for which our exponents are infinite, while theirs is always finite.

## C. Outline

The rest of the paper proceeds as follows. Section II gives definitions. Section III formally states the SCCSI problem and contains our results and discussion. Similarly, section IV formally states the Wyner-Ziv problem and contains our results and discussion. Section V applies the Wyner-Ziv results to the binary erasure and Gaussian problems. The proofs of the theorems are somewhat involved and can be found in the appendices.

## II. PRELIMINARIES

### A. Definitions and Notations

We use $\mathcal{P}(\mathcal{X})$ to denote the set of discrete probability distributions on $\mathcal{X}$ and $\mathcal{C}(\mathcal{X} \to \mathcal{Y})$ to denote all channels from $\mathcal{X}$ to $\mathcal{Y}$. For $P \in \mathcal{P}(\mathcal{X})$ and $V \in \mathcal{C}(\mathcal{X} \to \mathcal{Y})$, we write $P \times V$ to denote the distribution of the pair $(X, Y) \in \mathcal{X} \times \mathcal{Y}$ in which $X$ is generated according to $P(\cdot)$ and $Y$ is taken as the output of the channel $V$ whose input is $X$. For $P \in \mathcal{P}(\mathcal{X})$ and $P_{Y|X} \in \mathcal{C}(\mathcal{X} \to \mathcal{Y})$ we use $P_{XY}$ as shorthand for $P_X \times P_{Y|X}$.

We use $\mathbf{x}$ to denote vectors in $\mathcal{X}^n$; usually the length of the vector is clear from the context. For any $\mathbf{x} \in \mathcal{X}^n$ we write $Q_{\mathbf{x}}(\cdot)$ as the empirical distribution or *type* of $\mathbf{x}$. The set of all sequences of length $n$ with type $Q$ is denoted $T_Q^n$. The set of all type variables $Q \in \mathcal{P}(\mathcal{X})$, i.e. those for which $T_Q^n \neq \emptyset$, is denoted $\mathcal{P}^n(\mathcal{X})$. For $Q \in \mathcal{P}^n(\mathcal{X})$, we let $\mathcal{C}^n(Q, \mathcal{Y})$ denote the set of all $W \in \mathcal{C}(\mathcal{X} \to \mathcal{Y})$ for which (1) $T_{Q \times W}^n$ is non-empty; and (2) in the case that $Q(x) = 0$, $W(\cdot|x)$ takes the form $W(y|x) = |\mathcal{Y}|^{-1}$. For $\mathbf{x} \in \mathcal{X}^n$ and $V \in \mathcal{C}(\mathcal{X} \to \mathcal{Y})$ we denote by $T_V^n(\mathbf{x})$, the set of sequences in $\mathcal{Y}^n$ having conditional type $V$ given $\mathbf{x}$. For a type $Q_Y \in \mathcal{P}^n(\mathcal{Y})$, we use the function $k(Q_Y)$ to refer to a unique index in $\{1, \ldots, |\mathcal{P}^n(\mathcal{Y})|\}$ for that type.

When dealing with discrete random variables, all logarithms and exponents have base 2. We take $0 \log 0 = 0$ and $\log 0 = -\infty$ based on continuity arguments. For a distribution or type $P$ we let $H(P)$ denote entropy. For strings $\mathbf{x}, \mathbf{y}$, we write $H(\mathbf{x}|\mathbf{y})$ as the conditional empirical entropy. For a distribution $P_X$ and a channel $P_{Y|X}$ we write $I(P_X; P_{Y|X})$ for the mutual information between $X$ and $Y$ supposing that $P_X \times P_{Y|X}$ governs the pair. $D(P||Q)$ denotes the Kullback-Leibler (KL) divergence between distributions $P$ and $Q$. We also use the standard definitions of conditional entropy, conditional mutual information, and conditional KL divergence.

Whenever the range of a summation, maximization or minimization is clear we will use shorthand, e.g. $\sum_{Q_X \in \mathcal{P}^n(\mathcal{X})} = \sum_{Q_X}$. We define $[x]^+ \triangleq \max(0, x)$.

For the Gaussian Wyner-Ziv problem logarithms and exponents have base $e$. For $K$ a variance or covariance matrix, we write $f_K$ as a shorthand for a $\mathcal{N}(0, K)$ Gaussian random density. For $(X, Y) \sim f_K$, we write $f_{K_{Y|X}}$ for the conditional distribution of $Y$ given $X$ and write $K_{Y|X}$ for the conditional covariance (matrix). $h(K)$ denotes the differential entropy of a Gaussian random variable with distribution $f_K$. A subscript $K$ denotes that expectation or mutual information should be computed using $f_K$, e.g $I_K(X; Y)$ is the mutual information between jointly Gaussian random variables $X$ and $Y$ whose joint density is $f_K$. $D(K||\bar{K})$ denotes the KL divergence between two Gaussian random variables/vectors with densities $f_K$ and $f_{\bar{K}}$.

## III. SCCSI RESULTS AND DISCUSSION

Let $(X_i, Y_i)$ be the output of a memoryless source with distribution $P_{XY}(x, y)$ on a finite alphabet $\mathcal{X} \times \mathcal{Y}$. The encoders are deterministic functions $f_1^n : \mathcal{X}^n \to \mathcal{M}_1$ and $f_2^n : \mathcal{Y}^n \to \mathcal{M}_2$. The first encoder observes only the i.i.d sequence $X^n$, the second encoder observes only the i.i.d sequence $Y^n$. The decoder, $g^n : \mathcal{M}_1 \times \mathcal{M}_2 \to \mathcal{X}^n$ must reproduce $X^n$ using the messages from the encoders.

For this problem the rate region was determined by Ahlswede and Körner [5] and by Wyner [4] who showed that $R_1, R_2$ are achievable if

$$\exists S - Y - X \text{ s.t. } R_1 \geq H(X|S), R_2 \geq I(Y; S).$$

The closure of the union of the pairs over all such $S$ gives the entire rate region.

Let the decoder output be denoted $\hat{X}^n = g^n(f_1^n(X^n), f_2^n(Y^n))$. Then error probability is

$$P_e(f_1^n, f_2^n, g^n) = P_{XY}^n(\hat{X}^n \neq X^n),$$

and we define the source coding with coded side information error exponent as

$$\eta(P_{XY}, R_1, R_2) = \lim_{\epsilon \downarrow 0} \limsup_{n \to \infty} -\frac{1}{n} \log \left[ \min_{f_1^n, f_2^n, g^n} P_e(f_1^n, f_2^n, g^n) \right], \tag{1}$$

where the minimization ranges over all encoders and decoders $f_1^n, f_2^n, g^n$, such that

$$\log \mathcal{M}_i \leq n(R_i + \epsilon). \tag{2}$$

Our main results for SCCSI are as follows.

**Theorem 1.** *Let $R_1, R_2, P_{XY} \in \mathcal{P}(\mathcal{X} \times \mathcal{Y})$ be given. Then*

$$\eta(P_{XY}, R_1, R_2) \geq \eta_L(P_{XY}, R_1, R_2) \triangleq \inf_{Q_Y} \sup_{Q_{S|Y}} \inf_{\substack{Q_{X|YS}: \\ H(Q_X) \geq R_1}} D(Q_{XYS} || P_{XY} Q_{S|Y})$$

$$+ \begin{cases} [R_1 + R_2 - H(Q_{X|S}|Q_S) \\ \quad - I(Q_Y; Q_{S|Y})]^+ & \text{if } I(Q_Y; Q_{S|Y}) \geq R_2 \\ [R_1 - H(Q_{X|S}|Q_S)]^+ & \text{if } I(Q_Y; Q_{S|Y}) < R_2 \end{cases} \tag{3}$$

*where the joint distribution of $X, Y, S$ is $Q_Y Q_{S|Y} Q_{X|YS}$ and $S$ takes finitely many[3] values.*

The scheme to achieve this exponent is explained in detail in Appendix A. In brief, operating on a type by type basis the scheme is as follows. The second encoder quantizes its observation using the test channel $Q_{S|Y}$ and if necessary bins the quantizations into $2^{nR_2}$ bins. The primary encoder, assigns an index from $\{1, \ldots, 2^{nR_1}\}$ to each string in the typeclass, using binning if necessary. In the case that the primary encoder was able to communicate without binning the decoder will make no error. Otherwise, the the decoder finds the pair of sequences with the smallest joint empirical entropy in the received bins and outputs the $X$ string.

**Theorem 2.** *Let $R_1, R_2, P_{XY} \in \mathcal{P}(\mathcal{X} \times \mathcal{Y})$ be given, and suppose that $P_{XY}(x,y) > 0$ for all $x$ and $y$. Then*

$$\eta(P_{XY}, R_1, R_2) \leq \eta_U(P_{XY}, R_1, R_2) \triangleq \inf_{Q_Y} \sup_{\substack{Q_{S|Y}: \\ I(Q_Y; Q_{S|Y}) \leq R_2}} \inf_{\substack{Q_{X|Y}: \\ H(Q_{X|S}|Q_S) > R_1}} D(Q_{XY} || P_{XY}) \tag{4}$$

*where the joint distribution of $X, Y, S$ is $Q_Y Q_{X|Y} Q_{S|Y}$, i.e. $X, Y$ and $S$ form a Markov chain in that order, and $S$ satisfies*

$$|\mathcal{S}| \leq |\mathcal{X}| \cdot |\mathcal{Y}| + |\mathcal{Y}| + 2. \tag{5}$$

## A. Discussion

Both theorems can be viewed as a competitive game between two players, nature and the code designer. Nature's goal is to minimize the exponent and the code designer's goal is to maximize it. The structure of the problem determines the parameters and order of the plays. For example in Theorem 1, nature plays first, choosing a "worst-case" side information distribution. Then knowing nature's choice, the code designer picks the best codebook (via its choice of test channel). Nature plays last, choosing the worst possible consistent joint distribution. Notice that the choices at each step match the "information" available to the players.

---

[3]Note that any choice of cardinality for $S$ yields a valid achievable exponent.

A standard application of the change-of-measure argument [10, p.g. 268] provides the following upper-bound on the SCCSI exponent

$$\eta(P_{XY}, R_1, R_2) \leq \eta_{SP}(P_{XY}, R_1, R_2) \triangleq \inf_{Q_{XY}} \sup_{\substack{Q_{S|Y}: \\ I(Q_Y; Q_{S|Y}) \leq R_2}} \begin{cases} D(Q_{XY} || P_{XY}) & H(Q_{X|S}|Q_S) > R_1 \\ \infty & H(Q_{X|S}|Q_S) \leq R_1, \end{cases} \quad (6)$$

where the sup is actually a maximum since the objective is either $\infty$ or $D(Q_{XY} || P_{XY})$. It is straightforward to verify that $\eta_U \leq \eta_{SP}$, and so formally $\eta_U$ provides an improvement upon the standard sphere-packing upper bound. In the game theoretic interpretation the $\eta_{SP}$ exponent is obtained by letting nature's play reveal the *joint distribution* of the source and side information, and then the code designer plays, choosing the best codebook. But in the SCCSI problem, the helper's test channel can only depend on the marginal type of the side information. Thus our improved upper bound better captures the inherent structure of the problem.

We remark that in this and the next section, the solutions to the optimization problems in the theorems can be approximated arbitrarily well by searching over a fine grid. We have not studied conditions under which the optimization problems may be solved more efficiently (e.g. using convexity), nor for conditions under which a min-max theorem may simplify the problems. This may be interesting future work.

The optimizations in Theorems 1 and 2 differ in several respects. Foremost, in Theorem 2 the inner-most optimization is over $Q_{X|Y}$, so that $X, Y, S$ adhere to the Markov structure, yet in the achievable exponent this Markov constraint is not present. This differing Markov structure is also present in the partial Wyner-Ziv exponent results of Jayaraman and Berger [18], [32] who attribute the gap between the sphere packing and random exponents (present even at low rates) in the binning exponent problem they studied to this type of difference in the Markov structure. The other differences between $\eta_L$ and $\eta_U$ are the range of the inner most optimization, the presence of the binning term in the achievable exponent and the fact that the choice of test channel is restricted in the upper bound. (This latter difference can be eliminated by adding the restriction $I(Q_Y; Q_{S|Y}) \leq R_2$ to the choice of test channel in the lower bound, which only weakens the result.)

Despite these differences, the bounds provided by the theorems do allow us to determine the error exponent exactly in some special cases. When $R_2 = 0$, there is no possibility of encoding the side information. Taking $S$ to be constant in both exponents, one recovers the standard point to point exponent

$$\inf_{\substack{Q_X: \\ H(Q_X) \geq R_1}} D(Q_X || P_X).$$

More generally, if $R_2$ is sufficiently large and $R_1$ is sufficiently close to $H(X|Y)$, then one can show that the achievable exponent (3) coincides with the upper bound in (6) and hence also (4). The proof of this fact parallels the proof that the random-coding and sphere-packing bounds for coding coding coincide above the critical rate.

## IV. WYNER-ZIV RESULTS AND DISCUSSION

Let $(X_i, Y_i)$ be the output of a memoryless source with distribution $P_{XY}(x, y)$ on a finite alphabet $\mathcal{X} \times \mathcal{Y}$. Let $\hat{\mathcal{X}}$ be the reproduction alphabet and $d : \mathcal{X} \times \hat{\mathcal{X}} \to \mathbb{R}$ a single letter distortion measure. Define the distortion between two strings as $d(\mathbf{x}, \hat{\mathbf{x}}) = \frac{1}{n} \sum_{i=1}^{n} d(x_i, \hat{x}_i)$.

An encoder observes the i.i.d. source sequence, $X^n$ and communicates a message using $nR$ bits (or nats) to the decoder. The decoder combines the message with the side information $Y^n$ to give its reproduction $\hat{X}^n$. The encoder/decoder pair are functions $\psi : \mathcal{X}^n \to \mathcal{M}$ and $\varphi : \mathcal{M} \times \mathcal{Y}^n \to \hat{\mathcal{X}}^n$, where $\mathcal{M}$ is a fixed set.

The rate region was determined by Wyner and Ziv [33], who showed that if the allowable distortion is $\Delta$, then the required rate is given by

$$R_{WZ}(P_{XY}, \Delta) = \inf I(X; Z) - I(Y; Z),$$

where the infimum is over all auxiliary random variables $Z$ such that (1) $Z$, $X$, and $Y$ form a Markov chain in this order and (2) there exists a function $\lambda$ such that

$$\mathbb{E}[d(X, \lambda(Y, Z))] \leq \Delta.$$

Let $\hat{X}^n = \varphi(\psi(X^n), Y^n)$ be the decoder's output and define the error probability

$$P_e(\psi, \varphi, \Delta, d) = \Pr\left(d(X^n, \hat{X}^n) > \Delta\right). \tag{7}$$

We define the Wyner-Ziv error exponent to be

$$\theta(R, \Delta, P_{XY}, d) = \lim_{\epsilon \downarrow 0} \limsup_{n \to \infty} -\frac{1}{n} \log \left[\min_{(\psi,\varphi)} P_e(\psi, \varphi, \Delta, d)\right] \tag{8}$$

where the minimization ranges over all encoder/decoder pairs satisfying

$$\log |\mathcal{M}| \leq n(R + \epsilon). \tag{9}$$

Our main results for the Wyner-Ziv problem are as follows.

*a) Discrete Memoryless Case:*

**Theorem 3.** *Let $P_{XY} \in \mathcal{P}(\mathcal{X} \times \mathcal{Y})$ and $R > 0$, $\Delta > 0$, $d(\cdot, \cdot)$ be given. Then*

$$\theta(R, \Delta, P_{XY}, d) \geq \inf_{Q_X} \sup_{Q_{Z|X}} \inf_{Q_Y} \sup_{f \in \mathcal{F}} \inf_{Q_{XYZ}} G_D\left[Q_{XYZ}, P_{XY}, f, d, \Delta, R\right] \tag{10}$$

*where*

$$G_D\left[Q_{XYZ}, P_{XY}, f, d, \Delta, R\right] = \begin{cases} D(Q_{XYZ} \| P_{XY} Q_{Z|X}) & \mathbb{E}_Q[d(X, f(Y, Z))] \geq \Delta \\ D(Q_{XYZ} \| P_{XY} Q_{Z|X}) & \\ \quad + \left[R - I(Q_X; Q_{Z|X})\right. & \\ \quad \left. + I(Q_Y; Q_{Z|Y})\right]^+ & \mathbb{E}_Q[d(X, f(Y, Z))] < \Delta \\ & I(Q_X; Q_{Z|X}) \geq R \\ \infty & \text{otherwise}, \end{cases}$$

$\mathcal{F} = \{f | f : \mathcal{Y} \times \mathcal{Z} \to \hat{\mathcal{X}}\}$, *and $Z$ takes finitely many[4] values. Note in the final minimization over $Q_{XYZ}$, $Q_{XZ}$ and $Q_Y$ are fixed to be those specified earlier in the optimization.*

For completeness, we state the upper bound, which can be proved easily following Marton's [30] sphere-packing/change-of-measure proof for the point-to-point case.

**Theorem 4.** *Let $P_{XY} \in \mathcal{P}(\mathcal{X} \times \mathcal{Y})$ and $R > 0$, $\Delta > 0$, $d(\cdot, \cdot)$ be given. Then*

$$\theta(R, \Delta, P_{XY}, d) \leq \inf_{Q_{XY}: R_{WZ}(Q_{XY}, \Delta) > R} D(Q_{XY} \| P_{XY}).$$

This result is analogous to the upper bound in (6) and is therefore not as strong as its SCCSI counterpart (cf. (4)). We expect that this bound can be improved, although the technique used to obtain Theorem 2 does not seem to be applicable here. If this bound can be strictly improved in the binary erasure case, it would imply an exponent loss (see Section V-A).

---

[4] As we are providing an achievable exponent, any choice of cardinality for $Z$ yields a valid achievable exponent

*b) Gaussian Case:*

**Theorem 5.** *Let $(X_i, Y_i)$ be jointly Gaussian with zero means and covariance matrix*

$$\Sigma = \begin{bmatrix} 1 & \zeta_{XY} \\ \zeta_{XY} & 1 \end{bmatrix}, \tag{11}$$

*and let $d(x, \hat{x}) = (x - \hat{x})^2$. Then for any $R > 0$, $\Delta > 0$, and $\Sigma$ as in (11),*

$$\theta(R, \Delta, f_\Sigma, d) \geq \inf_{\sigma_X^2} \sup_{\rho_{xz}} \inf_{\sigma_Y^2} \sup_{\lambda \in \Lambda} \inf_{\rho_{yz}, \rho_{xy}} G_G[K, \Sigma, \lambda, \Delta, R] \tag{12}$$

*where*

$$G_G[K, \Sigma, \lambda, \Delta, R] = \begin{cases} D(K || \bar{K}) & \mathbb{E}_K[(X - \lambda(Y, Z))^2] \geq \Delta \\ D(K || \bar{K}) & \\ \quad + \left[ R - I_K(X; Z) \right. & \mathbb{E}_K[(X - \lambda(Y, Z))^2] < \Delta \\ \quad \left. + I_K(Y; Z) \right]^+ & I_K(X; Z) \geq R \\ \infty & \text{otherwise,} \end{cases} \tag{13}$$

$\Lambda = \{\lambda : \mathbb{R} \times \mathbb{R} \to \mathbb{R} : \lambda(y, z) = \alpha y + \beta z, \alpha, \beta \in [-M_\lambda, M_\lambda]\}$, *the covariance matrix of $(X, Y, Z)$ is*

$$K = \begin{bmatrix} \sigma_X^2 & \sigma_X \sigma_Y \rho_{xy} & \sigma_X \rho_{xz} \\ \sigma_X \sigma_Y \rho_{xy} & \sigma_Y^2 & \sigma_Y \rho_{yz} \\ \sigma_X \rho_{xz} & \sigma_Y \rho_{yz} & 1 \end{bmatrix}$$

*and*

$$\bar{K} = \begin{bmatrix} 1 & \zeta_{XY} & \frac{\rho_{xz}}{\sigma_X} \\ \zeta_{XY} & 1 & \zeta_{XY} \frac{\rho_{xz}}{\sigma_X} \\ \frac{\rho_{xz}}{\sigma_X} & \zeta_{XY} \frac{\rho_{xz}}{\sigma_X} & \frac{\rho_{xz}^2}{\sigma_X^2} + 1 - \rho_{xz}^2 \end{bmatrix}. \tag{14}$$

$M_\lambda > 0$ *is an arbitrary real number. The covariance matrix $\bar{K}$ corresponds to a source $(X, Y, Z)$, where $X, Y \sim \mathcal{N}(0, \Sigma)$, $Z, X$ and $Y$ form a Markov chain in that order, and the distribution of $Z$ conditional on $X$ is taken from $K$.*

The theorem can be proven along the same lines as the discrete memoryless case, using a modified notion of Gaussian types [34]. The full proof can be found in Appendix E.

**Theorem 6.** *Let $(X_i, Y_i)$ be jointly Gaussian with zero means and covariance $\Sigma$ as in (11). Let $R_{X|Y}(f_\Sigma, \Delta)$ denote the conditional rate distortion function. Let $\tilde{\theta}$ denote the error exponent for a modified Gaussian Wyner-Ziv problem in which the side information is also available at the encoder. Then for any, $\Delta > 0$, $R > R_{X|Y}(f_\Sigma, \Delta)$*

$$\tilde{\theta}(R, \Delta, f_\Sigma, d) \leq \inf_{\Pi : R_{X|Y}(f_\Pi, \Delta) \geq R} D(\Pi || \Sigma) \tag{15}$$

*where $\Pi$ is a $2 \times 2$ positive definite covariance matrix and*

$$R_{X|Y}(f_\Pi, \Delta) = R_{WZ}(f_\Pi, \Delta) = \frac{1}{2} \log^+ \left( \frac{\mathrm{Var}_\Pi(X|Y)}{\Delta} \right).$$

*Proof:* See Appendix F. ∎

**Corollary 1.** *Under the assumptions of Theorem 6, we have that*

$$\theta(R, \Delta, f_\Sigma, d) \leq \tilde{\theta}(R, \Delta, f_\Sigma, d).$$

*Proof:* Any code that works for the Wyner-Ziv problem will work when the encoder also sees the side information. This implies that the error exponent for the Wyner-Ziv problem is upper bounded by the error exponent for the problem in which the side information is available at both the encoder and decoder. ∎

The upper bound in the Corollary is identical to the change-of-measure upper bound obtained via Theorem 4. As with that bound, we believe that this upper bound can be improved, and showing a strict improvement would establish an exponent loss.

### A. Discussion

As in the SCCSI case, in the Wyner-Ziv case the same game-theoretic interpretation holds, but there are more parameters and the game becomes more elaborate. Nature plays first, choosing the most "difficult" marginal distribution for $X$. The code designer plays next, selecting the "best" test channel for that difficult source. Nature plays again choosing the worst marginal distribution for the side information. Then, knowing everything chosen so far, the code designer chooses the estimation function. Nature has the final play, choosing the worst consistent joint distribution for triple the $X, Y, Z$. Once again the choices and order of plays match the problem.

The nature of the optimizations in Theorems 3 and 5 give us some insight into the design of practical coding schemes by revealing a tension, which we examine in detail in the next section for the binary erasure and Gaussian problems. Briefly we see that the objective functions $G_D$ (resp. $G_G$) contain three cases which correspond to

- a violation of the distortion constraint even when the codeword is decoded correctly;
- the use of binning, leading to the potential for decoding the wrong codeword;
- no possibility for error.

A large codebook allows for a cleaner quantization and hence lower chance of the first kind of event. But this large codebook comes with the requirement of binning, leading to the potential for the second kind of event. Thus these two kinds of errors are in tension.

Theorem 3 allows us to determine a portion of the reliability function for a certain functional source coding problem. If we wish to reproduce a function $g(X)$ of the source $X$ losslessly at the decoder, who already has $Y$, then the rate required is $H_P(g(X)|Y)$, which follows from the results of Orlitsky and Roche [35]. Setting the distortion measure to be

$$d(X, f(Y, Z)) = d_H(g(X), f(Y, Z))$$

($d_H$ is the hamming measure) and evaluating Theorem 3 in the limit as $\Delta \to 0$ provides an achievable exponent for this problem. This can be seen by always choosing $Q_{Z|X}$ so that $Z = g(X)$ and letting the reproduction function be $f(Y, Z) = Z$. Using the fact that $Z \leftrightarrow X \leftrightarrow Y$, one can show that the limit as $\Delta \to 0$ of the righthand-side of equation (10) is

$$\xi_L(R, P_{XY}) = \inf_{Q_{XY}: H_Q(g(X)) \geq R} D(Q_{XY}||P_{XY}) + [R - H_Q(g(X)|Y)]^+. \tag{16}$$

An upper bound on the error exponent for this problem is given by

$$\xi_U(R, P_{XY}) = \inf_{Q_{XY}: H_Q(g(X)|Y) \geq R} D(Q_{XY}||P_{XY}). \tag{17}$$

On account of the fact that both (16) and (17) are optimizations of a continuous function over a compact sets, the inf is attained. The relationship between these two functions is analogous to the relationship between the sphere-packing and random coding exponents in channel coding [10, Lemma 2.5.4]. Thus for $R \geq 0$ until some critical rate $R_c$ the reliability function for the functional source coding problem is given exactly by

$$\min_{Q_{XY}: H_Q(g(X)|Y) \geq R} D(Q_{XY}||P_{XY}).$$

## V. Examples

### A. Binary Erasure Case

As an application of Theorem 3, we turn to the binary erasure version of the Wyner-Ziv problem. In this case, $X$ is uniformly distributed over the set $\{-1, +1\}$, and $Y$ equals $X$ passed through a binary erasure channel with erasure probability $p$

$$P(Y = 0|X = 1) = p = 1 - P(Y = 1|X = 1)$$
$$P(Y = 0|X = -1) = p = 1 - P(Y = -1|X = -1).$$

We would like to permit the reconstruction string to have erasures but not errors. The reconstruction alphabet is thus

$$\hat{\mathcal{X}} = \{-1, 0, 1\}.$$

One way to avoid errors in the reconstruction string is to use the "erasure" distortion measure

$$d(x, \hat{x}) = \begin{cases} 0 & \text{if } \hat{x} = x \\ 1 & \text{if } \hat{x} = 0 \\ \infty & \text{otherwise.} \end{cases}$$

This distortion measure is overly harsh, however, in that it prohibits all errors. For the Wyner-Ziv problem, higher rates can be achieved if one tolerates a vanishing probability of error. We will therefore consider a finite approximation of this distortion measure,

$$d(x, \hat{x}) = \begin{cases} 0 & \text{if } \hat{x} = x \\ 1 & \text{if } \hat{x} = 0 \\ K & \text{otherwise,} \end{cases}$$

where $K$ is a large but fixed constant. We will examine the rate-distortion and reliability functions in the limit as $K$ tends to infinity.

To determine the rate-distortion function in this case, let $Z$ be the output of a binary erasure channel with input $X$ and erasure probability $\delta$. If $Z$, $X$, and $Y$ form a Markov chain in this order, then it follows that

$$I(X; Z) - I(Y; Z) = p(1 - \delta).$$

There is a natural choice of $f$ for this case

$$f(y, z) = \begin{cases} 1 & \text{if } z = 1 \text{ or } y = 1 \\ 0 & \text{if } z = 0 \text{ and } y = 0 \\ -1 & \text{otherwise.} \end{cases} \tag{18}$$

Then $\mathbb{E}[d(X, f(Y, Z))] = p\delta$, and so any rate

$$R \geq [p - \Delta]^+$$

is achievable. To see that this is in fact the best possible, consider the problem in which the side information $Y^n$ is available to both the encoder and the decoder. The rate-distortion function for this problem is given by

$$\min_{p(\hat{x}|x,y)} I(X; \hat{X}|Y).$$

such that

$$\mathbb{E}[d(X, \hat{X})] \leq \Delta.$$

This minimization can be computed using classical techniques and shown in the limit as $K$ tends to infinity to equal $[p - \Delta]^+$. It follows that $[p - \Delta]^+$ is the rate-distortion function for both problems. In particular, there is no "rate loss" in the sense that the rate-distortion function is the same whether the side information is available at both the encoder and decoder or at the decoder only.

We note that for the problem with side information at both the encoder and decoder, there is a simple scheme that achieves the rate-distortion function $[p - \Delta]^+$. Since the encoder knows the locations of the erasures in $Y^n$, it can simply communicate the value of $X^n$ in the first $nR$ erased locations.

We now turn to the application of Theorem 3 to this set-up. For simplicity of exposition, we will consider the optimization problem in (10) with two restrictions: (1) $Q_X$ is fixed to be the uniform distribution over $\{-1, +1\}$; and (2) we optimize $Q_{Z|X}$ over the class of binary erasure channels, instead of optimizing over the class of all test channels from $\mathcal{X}$ to $\mathcal{Z}$. The optimization problem in (10) then reduces to

$$\sup_{Q_{Z|X}} \min_{Q_{Y|XZ}} G[Q_{XYZ}, P_{XY}, f, \Delta, R].$$

This optimization problem can be written in the following alternative form

$$\sup_{Q_{Z|X}} \min(G_1(Q_{Z|X}), G_2(Q_{Z|X})), \tag{19}$$

where

$$G_1(Q_{Z|X}) = \min_{Q_{Y|XZ}} D(Q_{XYZ}||P_{XY}Q_{Z|X})$$

with the minimization being over all $Q_{Y|XZ}$ such that

$$\mathbb{E}_Q[d(X, f(Y, Z))] \geq \Delta,$$

and

$$G_2(Q_{Z|X}) = \min_{Q_{Y|XZ}} D(Q_{XYZ}||P_{XY}Q_{Z|X}) + [R - I_Q(X; Z) + I_Q(Y; Z)]^+,$$

with the optimization being over all $Q_{Y|XZ}$ such that

$$\mathbb{E}_Q[d(X, f(Y, Z))] < \Delta,$$

and

$$I_Q(X; Z) \geq R.$$

This last condition, of course, either holds for all choices of $Q_{Y|XZ}$ or for none of them.

The alternative form of the optimization problem given in (19) is useful because it shows that maximizing over the binary erasure test channel amounts to maximizing the minimum of the exponents of two error events: the first, $G_1(Q_{Z|X})$, is the exponent on the event that $Y^n$ and $Z^n$ together provide insufficient information about $X^n$ to enable the decoder to meet the distortion constraint. Thus an error will occur even if the codeword $Z^n$ is decoded correctly. The second, $G_2(Q_{Z|X})$, is the exponent on the probability of a binning error.

These two error exponents are in tension in the following sense. Choosing $Q_{Z|X}$ to have a low probability of erasure communicates many of the bits in $X^n$ to the decoder via $Z^n$. This makes it unlikely that $Y^n$ and $Z^n$ will reveal too few bits about $X^n$ for the decoder to meet the distortion constraint, meaning that $G_1(Q_{Z|X})$ will be large. At the same time, choosing $Q_{Z|X}$ to have a low probability of erasure requires the use of large codebook, which makes the binning error probability high, leading to a small $G_2(Q_{Z|X})$. On the other hand, choosing $Q_{Z|X}$ to have a high probability of erasure leads to exactly the opposite behavior: the binning error probability is small since little information is being communicated through $Z^n$, but it is much more likely that the realization of $Y^n$ and $Z^n$ do not collectively reveal enough of the bits in $X^n$ to meet the distortion constraint.

This tension is illustrated in Fig. 2. The optimum choice of $Q_{Z|X}$ is given by a moderate erasure probability that balances the exponents of the two error probabilities. With this choice, both are dominant error events.

The exponent itself is shown for various $R$ in Fig. 3. Since we have not optimized over $Q_X$, this is properly interpreted as an upper bound on the error exponent of the scheme. Fig. 3 also shows the error exponent of the simple scheme mentioned above for achieving the rate-distortion function when the side information is available at both the encoder and the decoder[5]. The error probability of this scheme is simply the probability that $Y^n$ contains more than $n(R+\Delta)$ erasures. Assuming $R > p - \Delta$, the exponent of this event is equal to

$$D(R+\Delta||p),$$

i.e., the relative entropy between two Bernoulli distributions, one with success probability $R+\Delta$ and one with success probability $p$. Fig. 3 shows that when the side information is available at both the encoder and decoder the exponent is higher than for our one-sided scheme. This suggests that there may be exponent loss, although considering non-erasure test channels may close this "gap".

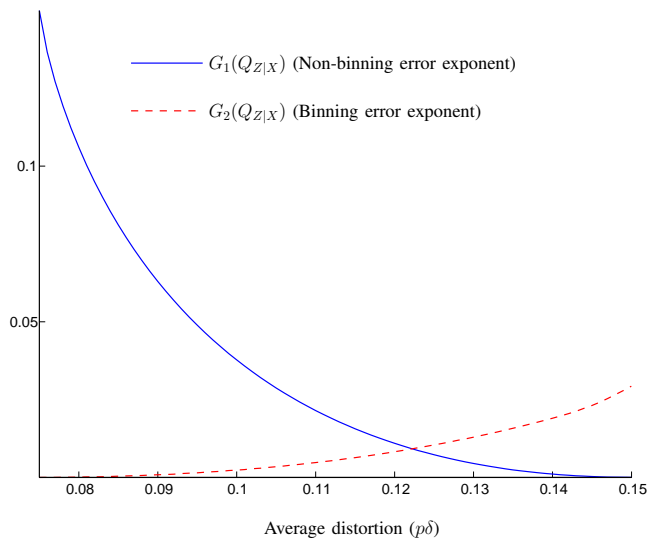

Fig. 2. Tension in choice of the test channel erasure probability $\delta$, revealed by Theorem 3. Note that $p\delta$ is the average distortion of the system. Here $\Delta = 0.15$, $p = 0.5$, and $R = 0.425$.

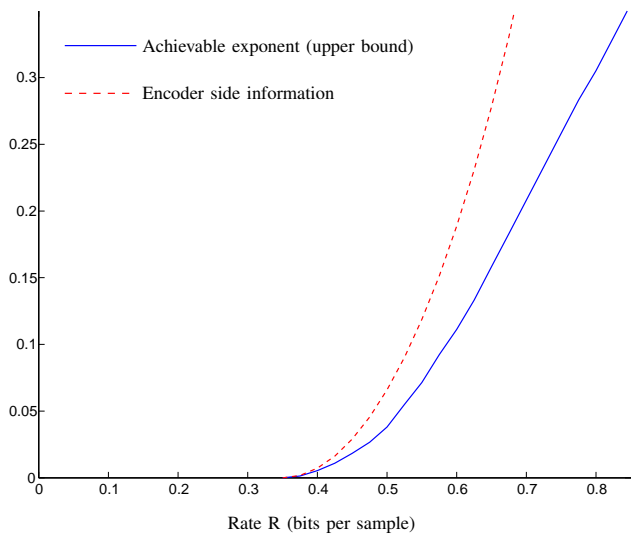

Fig. 3. Upper bound on error exponent of Theorem 3, and the error exponent of the scheme that makes use of side information at the encoder. The parameters $\Delta$, $p$ are the same as those used in Fig. 2.

---

[5]This is also the upper bound in Theorem 4.

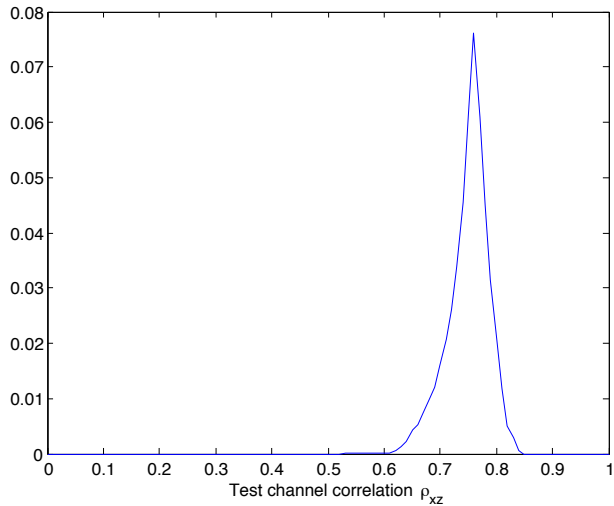

Fig. 4. Test channel optimization for Theorem 5. The plot shows the exponent against $\rho_{xz}$, holding $\sigma_X^2 = 1$ fixed for $R = 0.4, \zeta_{xy} = 0.7$ and $\Delta = 0.4$.

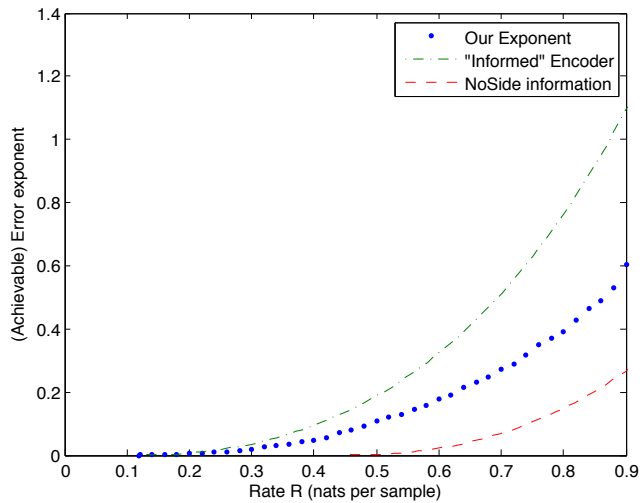

Fig. 5. A plot of the achievable exponent of Theorem 5. Here $\zeta_{xy} = 0.7$ (the correlation coefficient between the source and side information) and $\Delta = 0.4$. $R(\Delta) = 0.121$ nats for these parameters.

### B. Gaussian Case

A similar test channel tension arises in the Gaussian case. This can be seen most clearly by considering the optimization problem over $\rho_{xz}$ for fixed $\sigma_X^2$. In Fig. 4 we plot

$$G_3(\rho_{xz}) = \inf_{\sigma_Y^2} \sup_{\lambda \in \Lambda} \inf_{\rho_{xy}, \rho_{yz}} G\left[K, \Sigma, \lambda, \Delta, R\right]$$

where we hold $\sigma_X^2 = 1$, and $K = K(1, \sigma_Y, 1, \rho_{xy}, \rho_{yz}, \rho_{xz})$ is the covariance matrix of $(X, Y, Z)$.

Intuitively, $\rho_{xz}$ controls the number of different codewords we use to cover the source sequences. At rate $R$ the scheme allows us to identify at most $\exp(nR)$ codewords uniquely, and binning is required to go beyond this. A large codebook has the advantage that each source can be mapped to a better (i.e. closer) codeword. As we increase the size of the codebook beyond this point, the gains from having a "cleaner" codebook are outweighed by the penalty we pay for binning. From the plot we can see there is an optimum choice that occurs around $\rho_{xz} = 0.76$ for the parameters of the plot.

Figure 5 shows the exponent plotted (by numerically solving the optimization problem) against the rate. For comparison the upper bound of Theorem 6 is included, as is the exponent for the no side information case, corresponding to the continuous version of Marton's point-to-point exponent [30]. This result was proved by Ihara and Kubo [31], who showed the exponent is

$$\inf_{\sigma_X : \frac{1}{2} \log(\frac{\sigma_X^2}{\Delta}) > R} D(f_{\sigma_X} || f_1) = \frac{1}{2} \left( \Delta \exp(2R) - \log(\Delta \exp(2R)) - 1 \right). \tag{20}$$

We can show our achievable exponent recovers (20) by taking the side information to be statistically independent i.e. $\zeta = 0$. In this case, one can show that $\rho_{xy} = \rho_{yz} = 0$ solve the inner optimization problem of (10). Further, since $X \perp\!\!\!\perp Y$, $Y$ cannot help achieve the distortion constraint, choosing $\sigma_Y = 1$ is nature's best play. With these choices we see that $D(K || \bar{K}) = D(f_{\sigma_X^2} || f_1)$ and we are left with the following equivalent optimization (where we have written $\hat{X} = \alpha Z$)

$$\inf_{\sigma_X} \sup_{\rho_{x\hat{x}}, \sigma_{\hat{X}}} \begin{cases} D(f_{\sigma_X^2} || f_1) & \mathbb{E}[(X - \hat{X})^2] \geq \Delta \text{ or} \\ & I(X; \hat{X}) \geq R \\ \infty & \text{otherwise.} \end{cases}$$

As nature will always pick $\sigma_X$ such that the supremum is finite, we are left with

$$\inf_{\sigma_X : R(\sigma_X^2, \Delta) \geq R} D(f_{\sigma_X^2} || f_1).$$

Expanding the divergence and appealing to the monotonicity of $x - \log x$ gives (20)[6].

Using equation (20) and Theorem 6 we can determine the error exponent exactly when the side information is available at both the encoder and decoder. In this case, Wyner [36, section 3] provides a simple scheme to achieve the rate distortion function. The encoder simply subtracts the conditional mean $\mathbb{E}[X|Y = y]$ from the source. An achievable exponent then follows by computing the point-to-point exponent for the random variable $X|Y = y$, which is again Gaussian, with mean $-\zeta y$ and variance $1 - \zeta^2$. Our achievable exponent in this case is

$$\inf_{\sigma_X : R(\sigma_X, \Delta) > R} D(f_{\sigma_X^2} || f_{1-\zeta^2}) = \frac{1}{2} \left( \frac{\Delta \exp(2R)}{1 - \zeta^2} - \log \left( \frac{\Delta \exp(2R)}{1 - \zeta^2} \right) - 1 \right) \tag{21}$$

We now show that this is in fact the best we can do, by showing that (21) coincides with the upper bound of Theorem 6. The optimization problem of Theorem 6 can be solved as follows. We first note that if $X, Y$ are zero mean with covariance matrix $K$, then $\text{Var}(X|Y) = \frac{\det(K)}{\text{Var } Y}$. Hence we may write the problem as

$$\inf_{K \succeq 0 : \ g(K, \Delta, R) \leq 0} D(K || \Sigma)$$

where $K \succeq 0$ means the matrix $K$ is positive semi-definite and $g(K, \Delta, R) = -\log \det(K) + \log(\Delta) + \log e_2^T K e_2 + 2R$. The KKT conditions tell us the optimum $K^*$ must satisfy

1) $-\frac{1}{2}(K^*)^{-1} + \frac{1}{2}\Sigma^{-1} + \lambda \left( -(K^*)^{-1} + \begin{bmatrix} 0 & 0 \\ 0 & e_2^T K^* e_2 \end{bmatrix} \right) = 0$

2) $\lambda g(K^*) = 0$.

One can solve to this system to find

$$K^* = \begin{bmatrix} \zeta^2 + \Delta \exp(2R) & \zeta \\ \zeta & 1 \end{bmatrix}.$$

Evaluating $D(K^* || \Sigma)$ yields (21). Therefore, when the side information is available in both places we have determined the exponent exactly is (21).

---

[6]Using a virtually identical argument one can show that exponent of Theorem 3 reduces to Marton's exponent for the discrete-memoryless case when the side information is independent of the source.

# Appendix A
## Proof of Theorem 1

If $R_1 \geq \log |\mathcal{X}_1|$, then clearly $\eta(P_{XY}, R_1, R_2) = \infty$ and the result trivially holds, so we suppose that $R_1 < \log |\mathcal{X}_1|$.

### A. Scheme

We start by describing a scheme and then show the scheme has the performance specified in the theorem. Let $\epsilon > 0$ be given. For a given blocklength $n$, we operate on a type-by-type basis. The encoding and decoding functions are defined as follows.

*Encoder 1:* For each type-class $T^n_{Q_X}$ with $\log |T^n(Q_X)| > nR$ the encoder and decoder agree on a random binning scheme: for every sequence in $T^n_{Q_X}$, a bin index is assigned uniformly at random from $\{1, 2, \ldots, \exp(nR_1)\}$. For the case that $\log |T^n(Q_X)| \leq nR$ each sequence is assigned a unique index. To encode a sequence $\mathbf{x}$, the encoder sends the type $Q_{\mathbf{x}}$ and its index, $U_1(\cdot)$. Mathematically $f^n_1 : \mathcal{X}^n \to \mathcal{M}_1$ is

$$f^n_1(\mathbf{x}) = (U_1(\mathbf{x}), k(Q_{\mathbf{x}})),$$

where

$$\mathcal{M}_1 = \mathcal{M}'_1 \times \mathcal{M}''_1,$$
$$\mathcal{M}'_1 = \{1, 2, \ldots, M_1 \triangleq \exp(nR_1)\},$$
$$\mathcal{M}''_1 = \{1, 2, \ldots, (n+1)^{|\mathcal{X}|}\}.$$

*Encoder 2:* For each type $Q_Y$, fix a conditional type $Q^*_{S|Y}(Q_Y) \in \mathcal{C}^n(Q_Y, \mathcal{S})$ and randomly choose a set of codewords $B^n(Q_Y)$ in the following way. The size of $B^n(Q_Y)$ is an integer satisfying

$$\begin{aligned}
&\exp(nI(Q_Y; Q^*_{S|Y}(Q_Y)) + (|\mathcal{Y}||\mathcal{S}| + 2)\log(n+1)) \\
&\leq |B^n(Q_Y)| \\
&\leq \exp(nI(Q_Y; Q^*_{S|Y}(Q_Y)) + (|\mathcal{Y}||\mathcal{S}| + 4)\log(n+1))
\end{aligned} \tag{22}$$

and the codewords are drawn uniformly, with replacement, from the marginal type class $T^n_{Q^*_S}$ induced by $Q_Y$ and $Q^*_{S|Y}(Q_Y)$. Define $S : T^n_{Q_{\mathbf{y}}} \to B^n(Q_{\mathbf{y}})$ as follows. Let $\mathcal{G}(\mathbf{y}) = B^n(Q_{\mathbf{y}}) \cap T^n_{Q^*_{S|Y}(Q_{\mathbf{y}})}(\mathbf{y})$. If $\mathcal{G}(\mathbf{y})$ is non-empty, then the output of $S(\mathbf{y})$ is drawn uniformly at random from $\mathcal{G}(\mathbf{y})$[7]. If $\mathcal{G}(\mathbf{y})$ is empty the output of $S(\mathbf{y})$ is drawn uniformly at random from $B^n(Q_{\mathbf{y}})$. The function $S(\cdot)$ determines a quantization of $\mathbf{y}$, the observation of the second encoder. We define $S^n = S(Y^n)$.

If $|B^n(Q_{\mathbf{y}})| > \exp(nR_2)$ then the helper encoder assigns an index from the set $\{1, \ldots, \exp(nR_2)\}$ to each unique codeword in $B^n(Q_{\mathbf{y}})$ uniformly at random; in the opposite case each element of $B^n(Q_{\mathbf{y}})$ is assigned a unique index. In either case we let $U_2(\mathbf{s})$ denote the index assigned to $\mathbf{s} \in B^n(Q_{\mathbf{y}})$. To encode a sequence $\mathbf{y} \in T^n_{Q_X}$, the encoder sends the type of $\mathbf{y}$ and the index, $U_2(S(\mathbf{y}))$, of the quantization $S(\mathbf{y})$. Mathematically the second encoder, $f^n_2 : \mathcal{Y}^n \to \mathcal{M}_2$, is specified as follows

$$f^n_2(\mathbf{y}) = (U_2(S(\mathbf{y})), k(Q_{\mathbf{y}}))$$

where

$$\mathcal{M}_2 = \mathcal{M}'_2 \times \mathcal{M}''_2,$$
$$\mathcal{M}'_2 = \{1, 2, \ldots, M_2 \triangleq \exp(nR_2)\},$$
$$\mathcal{M}''_2 = \{1, 2, \ldots, (n+1)^{|\mathcal{Y}|}\}.$$

*Decoder:*

---

[7] Codewords that appear multiple times are proportionally more likely to be selected.

The decoder receives two (bin) indices, $i$ from encoder one and $j$ from encoder two. It then attempts to jointly decode the pair $(\mathbf{x}, \mathbf{s})$ using a minimum empirical entropy rule. That is, the decoder tries to find the pair $(\hat{\mathbf{x}}, \hat{\mathbf{s}})$ among all sequences in corresponding bins satisfying $H(\tilde{\mathbf{x}}, \tilde{\mathbf{s}}) > H(\hat{\mathbf{x}}, \hat{\mathbf{s}})$. If there is no such pair it chooses $(\mathbf{x}, \mathbf{s})$ uniformly at random from the bins consistent with received bin indexes. Mathematically, this is:

$$g'^n(k(Q_X), i, k(Q_Y), j) = \begin{cases} (\hat{\mathbf{x}}, \hat{\mathbf{s}}) & \text{if } U_1(\hat{\mathbf{x}}) = i, U_2(\hat{\mathbf{s}}) = j \text{ and} \\ & \forall (\tilde{\mathbf{x}}, \tilde{\mathbf{s}}) \neq (\hat{\mathbf{x}}, \hat{\mathbf{s}}), U_1(\tilde{\mathbf{x}}) = i, U_2(\tilde{\mathbf{s}}) = j : \\ & \quad H(\tilde{\mathbf{x}}, \tilde{\mathbf{s}}) > H(\hat{\mathbf{x}}, \hat{\mathbf{s}}) \\ \text{any } (\tilde{\mathbf{x}}, \tilde{\mathbf{s}}) \text{ with } U_1(\tilde{\mathbf{x}}) = i, U_2(\tilde{\mathbf{s}}) = j & \text{if no such } (\hat{\mathbf{x}}, \hat{\mathbf{s}}). \end{cases}$$

The decoder's final output is just the first element of the pair $g'^n(k(Q_X), i, k(Q_Y), j)$.

### B. Error Probability Calculation

To begin we define the following sets

$$\mathcal{E}_{r,1} = \{(\mathbf{x}, \mathbf{y}, \mathbf{s}) : H(Q_{\mathbf{x}}) > R_1, |B^n(Q_{\mathbf{y}})| < \exp(nR_2)\},$$
$$\mathcal{D}_{r,1} = \{Q_{XYS} : H(Q_X) > R_1, |B^n(Q_Y)| < \exp(nR_2)\},$$
$$\mathcal{E}_{r,2} = \{(\mathbf{x}, \mathbf{y}, \mathbf{s}) : H(Q_{\mathbf{x}}) > R_1, |B^n(Q_{\mathbf{y}})| \geq \exp(nR_2)\},$$
$$\mathcal{D}_{r,2} = \{Q_{XYS} : H(Q_X) > R_1, |B^n(Q_Y)| \geq \exp(nR_2)\},$$
$$\mathcal{E}_c = \{(\mathbf{x}, \mathbf{y}, \mathbf{s}) : \mathbf{s} \notin T^n_{Q^*_{S|Y}(Q_{\mathbf{y}})}(\mathbf{y})\},$$
$$\mathcal{D}_c = \{Q_{XYS} : Q_{S|Y} \neq Q^*_{S|Y}(Q_Y)\},$$

and the following event $F = \{\exists \; \tilde{\mathbf{s}} \in B^n(Q_{Y^n}) : \tilde{\mathbf{s}} \in T^n_{Q^*_{S|Y}(Q_{Y^n})}(Y^n)\}$.

The following lemmas will be required.

**Lemma 1.** *Let $X^n, Y^n, S^n$ be generated according to our scheme and suppose that $(\mathbf{x}, \mathbf{y}, \mathbf{s})$ is in $(\mathcal{E}_c)^c$, i.e., that $\mathbf{s} \in T^n_{Q^*_{S|Y}(Q_{\mathbf{y}})}(\mathbf{y})$. Then*

$$\Pr(X^n = \mathbf{x}, \, Y^n = \mathbf{y}, S^n = \mathbf{s}) \tag{23}$$
$$\leq P^n_{XY}(\mathbf{x}, \mathbf{y}) \frac{1}{|T^n_{Q^*_{S|Y}(Q_{\mathbf{y}})}(\mathbf{y})|}. \tag{24}$$

*Proof:* For the $\mathbf{x}, \mathbf{y}, \mathbf{s}$ in this lemma, $\{X^n = \mathbf{x}, Y^n = \mathbf{y}, S^n = \mathbf{s}\}$ implies that the event $F$ has occurred. Thus

$$\Pr(X^n = \mathbf{x}, \, Y^n = \mathbf{y}, S^n = \mathbf{s})$$
$$= \Pr(X^n = \mathbf{x}, Y^n = \mathbf{y}, S^n = \mathbf{s}, F)$$
$$= P^n_{XY}(\mathbf{x}, \mathbf{y}) \Pr(F | X^n = \mathbf{x}, Y^n = \mathbf{y})$$
$$\times \Pr(S^n = \mathbf{s} | X^n = \mathbf{x}, Y^n = \mathbf{y}, F)$$
$$\leq P^n_{XY}(\mathbf{x}, \mathbf{y}) \Pr(S^n = \mathbf{s} | X^n = \mathbf{x}, Y^n = \mathbf{y}, F)$$
$$= P^n_{XY}(\mathbf{x}, \mathbf{y}) \frac{1}{|T^n_{Q^*_{S|Y}(Q_{\mathbf{y}})}(\mathbf{y})|}$$

where in the final line we used that conditional on $F$ and $\{Y^n = \mathbf{y}\}$, $S^n$ is uniformly distributed over $T^n_{Q^*_{S|Y}(Q_{\mathbf{y}})}(\mathbf{y})$. ∎

**Lemma 2.** *Let $X^n, Y^n, S^n$ be generated according to our scheme and suppose that $(\mathbf{x}, \mathbf{y}, \mathbf{s}) \in \mathcal{E}_c$. Then*

$$\Pr(X^n = \mathbf{x}, Y^n = \mathbf{y}, S^n = \mathbf{s})$$
$$\leq \exp(-(n+1)^2). \tag{25}$$

*Proof:* For the $\mathbf{x}, \mathbf{y}, \mathbf{s}$ in this lemma, $\{X^n = \mathbf{x}, Y^n = \mathbf{y}, S^n = \mathbf{s}\}$ implies that event $F^c$ has occurred. Thus

$$\Pr(X^n = \mathbf{x}, \ Y^n = \mathbf{y}, S^n = \mathbf{s})$$
$$= \Pr(X^n = \mathbf{x}, Y^n = \mathbf{y}, S^n = \mathbf{s}, F^c)$$
$$= P_Y^n(\mathbf{y}) \Pr(F^c | Y^n = \mathbf{y}) \Pr(X^n = \mathbf{x} | Y^n = \mathbf{y}, F^c)$$
$$\times \Pr(S^n = \mathbf{s} | X^n = \mathbf{x}, Y^n = \mathbf{y}, F^c)$$
$$\leq \Pr(F^c | Y^n = \mathbf{y}).$$

$\Pr(F^c | Y^n = \mathbf{y})$ is the probability that there is no $\tilde{\mathbf{s}} \in B^n(Q_{\mathbf{y}})$ so that $\tilde{\mathbf{s}} \in T^n_{Q^*_{S|Y}(Q_{\mathbf{y}})}(\mathbf{y})$. We will now give an upper bound on this probability using the properties of the codeword set. Let $m = |B^n(Q_{\mathbf{y}})|$ and $B^n(Q_{\mathbf{y}})[i]$ be the $i$th codeword in the set $B^n(Q_{\mathbf{y}})$. Then

$$\Pr(F^c | Y^n = \mathbf{y})$$
$$= \prod_{i=1}^m \Pr(B^n(Q_{\mathbf{y}})[i] \notin T^n_{Q^*_{S|Y}(Q_{\mathbf{y}})}(\mathbf{y}))$$
$$= \prod_{i=1}^m [1 - \Pr(B^n(Q_{\mathbf{y}})[i] \in T^n_{Q^*_{S|Y}(Q_{\mathbf{y}})}(\mathbf{y}))]$$
$$= \left( 1 - \frac{|T^n_{Q^*_{S|Y}(Q_{\mathbf{y}})}(\mathbf{y})|}{|T_{Q^*_S}|} \right)^m$$
$$\leq \exp\left( -\frac{|T^n_{Q^*_{S|Y}(Q_{\mathbf{y}})}(\mathbf{y})|}{|T_{Q^*_S}|} m \right)$$

where the last line followed by applying the inequality $(1-t)^m \leq \exp(-tm)$. Next, using the following bounds on the cardinality of type classes [10, lemmas 2.3 and 2.6],

$$|T^n_{Q_S}| \leq \exp(nH(Q_S))$$
$$|T^n_{Q_{S|Y}}(\mathbf{y})| \geq (n+1)^{-|\mathcal{Y}||\mathcal{S}|} \exp(nH(Q_{S|Y}|Q_Y))$$

and that $I(Q^*_{S|Y}(Q_{\mathbf{y}}); Q_{\mathbf{y}}) = H(Q^*_S) - H(Q^*_{S|Y}(Q_{\mathbf{y}})|Q_{\mathbf{y}})$ we have

$$-\frac{|T^n_{Q^*_{S|Y}(Q_{\mathbf{y}})}(\mathbf{y})|}{|T_{Q^*_S}|} \leq -(n+1)^{-|\mathcal{Y}||\mathcal{S}|} \exp(-nI(Q_{\mathbf{y}}; Q^*_{S|Y}(Q_{\mathbf{y}}))).$$

Thus,

$$\Pr(F^c | Y^n = \mathbf{y})$$
$$\leq \exp\left( -(n+1)^{-|\mathcal{Y}||\mathcal{S}|} \exp(-nI(Q_{\mathbf{y}}; Q^*_{S|Y}(Q_{\mathbf{y}}))) m \right)$$
$$\leq \exp(-(n+1)^2)$$

where the final line followed by substitution our choice of $m$ from (22). $\blacksquare$

**Lemma 3.** *For any pair of strings $\mathbf{x}, \mathbf{y}$, let*

$$S(\mathbf{x}, \mathbf{y}) = \{\tilde{\mathbf{x}}, \tilde{\mathbf{y}} | H(\tilde{\mathbf{x}}, \tilde{\mathbf{y}}) \leq H(\mathbf{x}, \mathbf{y})\}.$$

*Then*

$$|S(\mathbf{x}, \mathbf{y})| \leq (n+1)^{|\mathcal{X}||\mathcal{Y}|} \exp(nH(\mathbf{x}, \mathbf{y})).$$

*Proof:*

$$
\begin{aligned}
|S(\mathbf{x}, \mathbf{y})| &= \sum_{Q_{XY}: H(Q_{XY}) \leq H(\mathbf{x}, \mathbf{y})} |T_{Q_{XY}}^n| \\
&\leq \sum_{Q_{XY}: H(Q_{XY}) \leq H(\mathbf{x}, \mathbf{y})} \exp(nH(Q_{XY})) \\
&\leq (n+1)^{|\mathcal{X}||\mathcal{Y}|} \exp(nH(\mathbf{x}, \mathbf{y})).
\end{aligned}
$$

■

**Lemma 4.** *For any pair of strings* $\mathbf{x}, \mathbf{y}$, *let*

$$S(\mathbf{x}|\mathbf{y}) = \{\tilde{\mathbf{x}} | H(\tilde{\mathbf{x}}|\mathbf{y}) \leq H(\mathbf{x}|\mathbf{y})\}.$$

*Then*

$$|S(\mathbf{x}|\mathbf{y})| \leq (n+1)^{|\mathcal{X}||\mathcal{Y}|} \exp(nH(\mathbf{x}|\mathbf{y})).$$

*Proof:* The proof mirrors that of Lemma 3 and is omitted. ■

**Lemma 5.** *Let* $(\mathbf{x}, \mathbf{y}, \mathbf{s}) \in \mathcal{E}_{r,1} \cap (\mathcal{E}_c)^c$. *Then*

$$
\begin{aligned}
&\Pr(X^n \neq \hat{X}^n | X^n = \mathbf{x}, Y^n = \mathbf{y}, S^n = \mathbf{s}) \\
&\leq \exp\left(-n\left[R_1 - H(Q_{\mathbf{x}|\mathbf{s}}|Q_\mathbf{s}) - \delta_n\right]^+\right)
\end{aligned}
\tag{26}
$$

*where*

$$\delta_n = \frac{1}{n}\log(n+1)^{|\mathcal{S}||\mathcal{X}|}.$$

*Proof:* In the setting of this lemma the decoder knows $S^n$ since the quantization can be decoded unambiguously from the index $U_2(S^n)$. Thus, the decoding rule amounts to finding an $\mathbf{x}$ string with lower conditional empirical entropy in the received bin. The set $S(\mathbf{x}|\mathbf{s})$ (cf. Lemma 4) contains all the sequences with lower conditional empirical entropy (conditioned on $\mathbf{s}$), but having the same type as $\mathbf{x}$. Therefore we can bound the decoding error probability as

$$
\begin{aligned}
&\Pr(X^n \neq \hat{X}^n | X^n = \mathbf{x}, Y^n = \mathbf{y}, S^n = \mathbf{s}) \\
&\leq \sum_{\substack{\tilde{\mathbf{x}} \in S(\mathbf{x}|\mathbf{s}) \\ \tilde{\mathbf{x}} \neq \mathbf{x}}} \Pr(U_1(\tilde{\mathbf{x}}) = U_1(\mathbf{x})) \\
&\leq |S(\mathbf{x}|\mathbf{s})| \exp(-nR_1) \\
&\leq \exp(-n(R_1 - H(Q_{\mathbf{x}|\mathbf{s}}|Q_\mathbf{s}) - \delta_n))
\end{aligned}
$$

where the final line used the result from Lemma 4. Further bounding the probability by one gives the result. ■

**Lemma 6.** *Let* $(\mathbf{x}, \mathbf{y}, \mathbf{s}) \in \mathcal{E}_{r,2} \cap (\mathcal{E}_c)^c$. *Then*

$$
\begin{aligned}
&\Pr(X^n \neq \hat{X}^n | X^n = \mathbf{x}, Y^n = \mathbf{y}, S^n = \mathbf{s}) \\
&\leq 2\exp\left(-n\left[R_1 + R_2 - H(Q_{\mathbf{x}|\mathbf{s}}|Q_\mathbf{s}) - I(Q_{S|Y}^*(Q_\mathbf{y}); Q_\mathbf{y}) - \tilde{\tilde{\delta}}_n\right]^+\right)
\end{aligned}
\tag{27}
$$

*where*

$$\tilde{\tilde{\delta}}_n = \frac{1}{n} \log(n+1)^{|\mathcal{S}||\mathcal{X}|+|\mathcal{Y}||\mathcal{S}|+|\mathcal{S}|+4}.$$

*Proof:* Let us introduce the random variable $\hat{S}^n$ to denote the decoder's guess of the codeword $S^n$. Observe that the probability of the event of interest may be decomposed as follows

$$\Pr(X^n \neq \hat{X}^n | X^n = \mathbf{x}, Y^n = \mathbf{y}, S^n = \mathbf{s}) =$$
$$\Pr(X^n \neq \hat{X}^n, S^n = \hat{S}^n | X^n = \mathbf{x}, Y^n = \mathbf{y}, S^n = \mathbf{s})$$
$$+ \Pr(X^n \neq \hat{X}^n, S^n \neq \hat{S}^n | X^n = \mathbf{x}, Y^n = \mathbf{y}, S^n = \mathbf{s}). \tag{28}$$

We will show that both of the probabilities on the right are exponentially small and that the second summand dominates (in an error exponent sense) the first. To treat the first summand on the righthand side of (28) we begin by upper bounding the probability of $\{S^n = \hat{S}^n\}$ conditional on $\{X^n = \mathbf{x}, Y^n = \mathbf{y}, S^n = \mathbf{s}\}$ by 1; we are then interested in bounding

$$\Pr(X^n \neq \hat{X}^n | X^n = \mathbf{x}, Y^n = \mathbf{y}, S^n = \mathbf{s}, \hat{S}^n = S^n).$$

The analysis in the proof of Lemma 5 shows that this probability is bounded by (26).

For the second summand of (28) the event can occur only if there is a pair $(\tilde{\mathbf{x}}, \tilde{\mathbf{s}})$ with $\tilde{\mathbf{s}} \in B^n(Q_{\mathbf{y}})$ such that the pair have lower joint empirical entropy than the true pair $(\mathbf{x}, \mathbf{s})$ and are the same bins $U_1(\mathbf{x})$ and $U_2(\mathbf{s})$. Using the set $S(\mathbf{x}, \mathbf{s})$ from Lemma 3 we can bound this probability as follows

$$\Pr(X^n \neq \hat{X}^n, S^n \neq \hat{S}^n | X^n = \mathbf{x}, Y^n = \mathbf{y}, S^n = \mathbf{s})$$
$$\leq \sum_{\substack{(\tilde{\mathbf{x}}, \tilde{\mathbf{s}}) \in S(\mathbf{x}, \mathbf{s}) \\ \tilde{\mathbf{x}} \neq \mathbf{x} \text{ and } \tilde{\mathbf{s}} \neq \mathbf{s}}} \Pr(U_1(\tilde{\mathbf{x}}) = U_1(\mathbf{x}), U_2(\tilde{\mathbf{s}}) = U_2(\mathbf{s}), \tilde{\mathbf{s}} \in B^n(Q_{\mathbf{y}}) | X^n = \mathbf{x}, Y^n = \mathbf{y}, S^n = \mathbf{s})$$
$$= \sum_{\substack{(\tilde{\mathbf{x}}, \tilde{\mathbf{s}}) \in S(\mathbf{x}, \mathbf{s}) \\ \tilde{\mathbf{x}} \neq \mathbf{x} \text{ and } \tilde{\mathbf{s}} \neq \mathbf{s}}} \Pr(U_1(\tilde{\mathbf{x}}) = U_1(\mathbf{x})) \Pr(\tilde{\mathbf{s}} \in B^n(Q_{\mathbf{y}}) | X^n = \mathbf{x}, Y^n = \mathbf{y}, S^n = \mathbf{s})$$
$$\times \Pr(U_2(\tilde{\mathbf{s}}) = U_2(\mathbf{s}) | \tilde{\mathbf{s}} \in B^n(Q_{\mathbf{y}}), X^n = \mathbf{x}, Y^n = \mathbf{y}, S^n = \mathbf{s}).$$

We now show that

$$\Pr(\tilde{\mathbf{s}} \in B^n(Q_{\mathbf{y}}) | X^n = \mathbf{x}, Y^n = \mathbf{y}, S^n = \mathbf{s}) \leq \Pr(\tilde{\mathbf{s}} \in B^n(Q_{\mathbf{y}})). \tag{29}$$

To establish this we will show that

$$\Pr(S^n = \mathbf{s} | X^n = \mathbf{x}, Y^n = \mathbf{y}, \tilde{\mathbf{s}} \in B^n(Q_{\mathbf{y}})) \leq \Pr(S^n = \mathbf{s} | X^n = \mathbf{x}, Y^n = \mathbf{y}), \tag{30}$$

which implies the result by reversing the conditioning. Suppose first that $\tilde{\mathbf{s}} \notin T_{Q^*_{S|Y}(Q_{\mathbf{y}})}(\mathbf{y})$, then

$$\Pr(S^n = \mathbf{s} | X^n = \mathbf{x}, Y^n = \mathbf{y}, \tilde{\mathbf{s}} \in B^n(Q_{\mathbf{y}}))$$
$$= \Pr(S^n = \mathbf{s}, F | X^n = \mathbf{x}, Y^n = \mathbf{y}, \tilde{\mathbf{s}} \in B^n(Q_{\mathbf{y}}))$$
$$= \Pr(F | X^n = \mathbf{x}, Y^n = \mathbf{y}, \tilde{\mathbf{s}} \in B^n(Q_{\mathbf{y}})) \Pr(S^n = \mathbf{s} | X^n = \mathbf{x}, Y^n = \mathbf{y}, \tilde{\mathbf{s}} \in B^n(Q_{\mathbf{y}}), F)$$
$$\leq \Pr(F | X^n = \mathbf{x}, Y^n = \mathbf{y}) \Pr(S^n = \mathbf{s} | X^n = \mathbf{x}, Y^n = \mathbf{y}, \tilde{\mathbf{s}} \in B^n(Q_{\mathbf{y}}), F),$$

where the inequality follows because dropping the conditioning event that $\{\tilde{\mathbf{s}} \in B^n(Q_{\mathbf{y}})\}$ frees up a position in the codebook, which increases the probability of $F$. Continuing we obtain

$$\Pr(S^n = \mathbf{s} | X^n = \mathbf{x}, Y^n = \mathbf{y}, \tilde{\mathbf{s}} \in B^n(Q_{\mathbf{y}}))$$
$$\leq \Pr(F | X^n = \mathbf{x}, Y^n = \mathbf{y}) \Pr(S^n = \mathbf{s} | X^n = \mathbf{x}, Y^n = \mathbf{y}, F)$$
$$= \Pr(S^n, F | X^n = \mathbf{x}, Y^n = \mathbf{y})$$
$$= \Pr(S^n | X^n = \mathbf{x}, Y^n = \mathbf{y}),$$

where the inequality used the conditional independence of $\{S^n = \mathbf{s}\}$ and $\{\tilde{\mathbf{s}} \in B^n(Q_{\mathbf{y}})\}$ given $F$. The case that $\tilde{\mathbf{s}} \in T_{Q^*_{S|Y}(Q_{\mathbf{y}})}(\mathbf{y})$ can be handled by a straightforward coupling argument. Equation (30) is established[8].

Applying a standard bound for the cardinality of a typeclass and (22) we see that

$$\Pr(\tilde{\mathbf{s}} \in B^n(Q_{\mathbf{y}})) \le \frac{|B^n(Q_{\mathbf{y}})|}{|T_{Q^*_S}|}$$

$$\le \frac{\exp(nI(Q_{\mathbf{y}}; Q^*_{S|Y}(Q_{\mathbf{y}})) + (|\mathcal{Y}||\mathcal{S}| + 4)\log(n+1))(n+1)^{|\mathcal{S}|}}{\exp(nH(Q^*_S))},$$

where $Q^*_S$ denotes the type induced by $Q_{\mathbf{y}}$ and $Q^*_{S|Y}(Q_{\mathbf{y}})$. Additionally by the code construction, for $\tilde{\mathbf{x}} \ne \mathbf{x}$ and $\tilde{\mathbf{s}} \ne \mathbf{s}$ we have

$$\Pr(U_1(\tilde{\mathbf{x}}) = U_1(\mathbf{x}))\Pr(U_2(\tilde{\mathbf{s}}) = U_2(\mathbf{s})|\tilde{\mathbf{s}} \in B^n(Q_{\mathbf{y}}), X^n = \mathbf{x}, Y^n = \mathbf{y}) = \exp(-n(R_1 + R_2)).$$

These calculations, together with Lemma 3 imply that

$$\Pr(X^n \ne \hat{X}^n, S^n \ne \hat{S}^n | X^n = \mathbf{x}, Y^n = \mathbf{y}, S^n = \mathbf{s})$$
$$\le \exp(-n(R_1 + R_2 - H(Q_{\mathbf{x},\mathbf{s}}) - I(Q_{\mathbf{y}}; Q^*_{S|Y}(Q_{\mathbf{y}})) + H(Q^*_S)$$
$$- n^{-1}(|\mathcal{S}||\mathcal{X}| + |\mathcal{S}| + 4 + |\mathcal{S}||\mathcal{Y}|)\log(n+1))).$$

We now note that $(\mathbf{x}, \mathbf{y}, \mathbf{s}) \in \mathcal{E}^c_c$ implies that $\mathbf{s}$ is a valid codeword and therefore $Q_{\mathbf{s}} = Q^*_S$. By expanding $H(Q_{\mathbf{x},\mathbf{s}})$ using the chain rule and canceling the $H(Q_{\mathbf{s}})$ terms in the previous display we obtain

$$\Pr(X^n \ne \hat{X}^n, S^n \ne \hat{S}^n | X^n = \mathbf{x}, Y^n = \mathbf{y}, S^n = \mathbf{s})$$
$$\le \exp(-n(R_1 + R_2 - H(Q_{\mathbf{x}|\mathbf{s}}|Q_{\mathbf{s}}) - I(Q_{\mathbf{y}}; Q^*_{S|Y}(Q_{\mathbf{y}}))$$
$$- n^{-1}(|\mathcal{S}||\mathcal{X}| + |\mathcal{S}| + 4 + |\mathcal{S}||\mathcal{Y}|)\log(n+1))). \tag{31}$$

We then observe that $(\mathbf{x}, \mathbf{y}, \mathbf{s}) \in \mathcal{E}_{r,2}$ implies

$$\frac{(|\mathcal{Y}||\mathcal{S}| + 4)\log(n+1)}{n} \ge R_2 - I(Q_{\mathbf{y}}; Q^*_{S|Y}(Q_{\mathbf{y}})),$$

and therefore that

$$R_1 + R_2 - H(Q_{\mathbf{x}|\mathbf{s}}|Q_{\mathbf{s}}) - I(Q_{\mathbf{y}}; Q^*_{S|Y}(Q_{\mathbf{y}})) - \tilde{\tilde{\delta}}_n$$
$$\le R_1 - H(Q_{\mathbf{x}|\mathbf{s}}|Q_{\mathbf{s}}) - n^{-1}(|\mathcal{S}||\mathcal{X}| + |\mathcal{S}|)\log(n+1)$$
$$\le R_1 - H(Q_{\mathbf{x}|\mathbf{s}}|Q_{\mathbf{s}}) - n^{-1}(|\mathcal{S}||\mathcal{X}|)\log(n+1).$$

This calculation shows that the righthand side of (31) is larger than the righthand side of (26). To complete the proof we use the fact that $a + b \le 2\max(a, b)$ and keep the summand of (28) with the smaller exponent. ∎

**Lemma 7.** *Let $\delta_n, \tilde{\delta}_n, \tilde{\tilde{\delta}}_n$ be three sequences converging to zero. Let*

$$F_1^n(Q_{XYS}, R_1, R_2) = \begin{cases} [R_1 + R_2 - H(Q_{X|S}|Q_S) & \text{if } H(Q_X) \ge R_1 \\ \quad - I(Q_Y; Q_{S|Y}) - \tilde{\tilde{\delta}}_n]^+ & \text{and } I(Q_{S|Y}; Q_Y) \ge R_2 - \tilde{\delta}_n \\ [R_1 - H(Q_{X|S}|Q_S) - \delta_n]^+ & \text{if } H(Q_X) \ge R_1 \\ & \text{and } I(Q_{S|Y}; Q_Y) < R_2 - \tilde{\delta}_n \\ \infty & \text{otherwise} \end{cases}$$

---

[8]A similar reasoning can be used to verify the final inequality in the proof of Lemma 15 in [26].

$$F_1(Q_{XYS}, R_1, R_2) = \begin{cases} [R_1 + R_2 - H(Q_{X|S}|Q_S) & \text{if } H(Q_X) \geq R_1 \\ \quad - I(Q_Y; Q_{S|Y})]^+ & \text{and } I(Q_{S|Y}; Q_Y) \geq R_2 \\ [R_1 - H(Q_{X|S}|Q_S)]^+ & \text{if } H(Q_X) \geq R_1 \\ & \text{and } I(Q_{S|Y}; Q_Y) < R_2 \\ \infty & \text{otherwise} \end{cases}$$

$$F^n(P_{XY}, R_1, R_2) = \min_{Q_Y} \max_{Q_{S|Y} \in \mathcal{C}^n(Q_Y, \mathcal{S})} \min_{Q_{XYS}} D(Q_{XYS} || P_{XY} Q_{S|Y}) + F_1^n(Q_{XYS}, R_1, R_2) \tag{32}$$

*and*

$$F^\infty(P_{XY}, R_1, R_2) = \inf_{Q_Y} \sup_{Q_{S|Y} \in \mathcal{C}(\mathcal{Y} \to \mathcal{S})} \inf_{Q_{XYS}} D(Q_{XYS} || P_{XY} Q_{S|Y}) + F_1(Q_{XYS}, R_1, R_2). \tag{33}$$

*where in* (32) *the optimizations are over types/conditional types and in* (33) *the optimizations are over all distributions, and in both cases the inner optimizations are compatible with the outer ones, i.e. assume* $Q_{XYS} = Q_Y \times Q_{S|Y} \times Q_{X|SY}$*. Then*

$$\liminf_{n \to \infty} F^n(P_{XY}, R_1, R_2) \geq F^\infty(P_{XY}, R_1, R_2).$$

*Proof:* Let $Q_{XYS}^{(n)} \in \mathcal{P}^n(\mathcal{X} \times \mathcal{Y} \times \mathcal{S})$ solve the optimization problem in (32), i.e.

$$F^n(P_{XY}, R_1, R_2) = D(Q_{XYS}^{(n)} || P_{XY} Q_{S|Y}^{(n)}) + F_1^n(Q_{XYS}^{(n)}, R_1, R_2).$$

Along a subsequence that attains the $\liminf$ in the statement of the Lemma there is a further subsequence $Q_{XYS}^{(n)}$ that converges, and so by relabeling this subsequence we can arrange it so that $Q_{XYS}^{(n)} \to Q_{XYS}^\infty$. Let $\delta > 0$. Then there exists a $\tilde{Q}_{S|Y}^\infty$ so that

$$\inf_{\substack{Q_{XYS}: Q_Y = Q_Y^\infty \\ Q_{S|Y} = \tilde{Q}_{S|Y}^\infty}} D(Q_{XYS} || P_{XY} \tilde{Q}_{S|Y}^\infty) + F_1(Q_{XYS}, R_1, R_2)$$

$$\geq \sup_{\tilde{Q}_{S|Y}} \inf_{\substack{Q_{XYS}: Q_Y = Q_Y^\infty \\ Q_{S|Y} = \tilde{Q}_{S|Y}}} D(Q_{XYS} || P_{XY} Q_{S|Y}) + F_1(Q_{XYS}, R_1, R_2) - \delta.$$

Furthermore, we may find a sequence $\tilde{Q}_{S|Y}^{(n)}$ converging to $\tilde{Q}_{S|Y}^\infty$. We now choose

$$\tilde{Q}_{XYS}^{(n)} = \arg \min_{\substack{Q_{XYS} \in \mathcal{P}^n(\mathcal{X} \times \mathcal{Y} \times \mathcal{S}): \\ Q_Y = Q_Y^{(n)} \\ Q_{S|Y} = \tilde{Q}_{S|Y}^{(n)}}} D(Q_{XYS} || P_{XY} \tilde{Q}_{S|Y}^{(n)}) + F_1^n(Q_{XYS}, R_1, R_2).$$

Again by compactness and relabeling we may arrange it so that $\tilde{Q}_{XYS}^{(n)} \to Q_{XYS}^\infty$. Now we observe that

$$\min_{Q_Y} \max_{Q_{S|Y}} \min_{Q_{XYS}} D(Q_{XYS} || P_{XY} Q_{S|Y}) + F_1^n(Q_{XYS}, R_1, R_2)$$

$$= \max_{\bar{Q}_{S|Y}} \min_{\substack{Q_{XYS}: \\ Q_Y = Q_Y^{(n)} \\ Q_{S|Y} = \bar{Q}_{S|Y}}} D(Q_{XYS} || P_{XY} \bar{Q}_{S|Y}) + F_1^n(Q_{XYS}, R_1, R_2)$$

$$\geq \min_{\substack{Q_{XYS}: \\ Q_Y = Q_Y^{(n)} \\ Q_{S|Y} = \tilde{Q}_{S|Y}^{(n)}}} D(Q_{XYS} || P_{XY} \tilde{Q}_{S|Y}^{(n)}) + F_1^n(Q_{XYS}, R_1, R_2)$$

$$= D(\tilde{Q}_{XYS}^{(n)} || P_{XY} \tilde{Q}_{S|Y}^{(n)}) + F_1^n(\tilde{Q}_{XYS}^{(n)}, R_1, R_2). \tag{34}$$

We now prove that

$$
\liminf_{n \to \infty} D(\tilde{Q}_{XYS}^{(n)} || P_{XY}\tilde{Q}_{S|Y}^{(n)}) + F_1^n(\tilde{Q}_{XYS}^{(n)}, R_1, R_2)
$$
$$
\geq D(\tilde{Q}_{XYS}^{\infty} || P_{XY}\tilde{Q}_{S|Y}^{\infty}) + F_1(\tilde{Q}_{XYS}^{\infty}, R_1, R_2). \tag{35}
$$

The case that $\tilde{Q}_{XYS}^{\infty}$ is such that $H(\tilde{Q}_X^{\infty}) < R_1$, follows by continuity of entropy (since $H(\tilde{Q}_X^{(n)}) < R_1$ for all $n$ sufficiently large). In the opposite case, i.e. $H(\tilde{Q}_X^{\infty}) \geq R_1$, if $H(\tilde{Q}_X^{(n)}) < R_1$ for all $n$ sufficiently large the left side is infinity and so the inequality must hold. For the remaining case that $H(Q_X^{(n)}) \geq R_1$ for infinitely many $n$ we split into sub-cases. Sub-case one: $I_{\tilde{Q}_{SY}^{\infty}}(S;Y) < R_2$, then $I_{\tilde{Q}_{SY}^{(n)}}(S;Y) < R_2 - \tilde{\delta}_n$ for all $n$ sufficiently large so the result is true by lower semi-continuity of the information measures. Sub-case two: $I_{\tilde{Q}_{SY}^{\infty}}(S;Y) \geq R_2$, but then

$$
\liminf_{n \to \infty}[R_1 - H(Q_{X|S}^{(n)}|Q_S^{(n)}) - \delta_n]^+ = [R_1 - H(Q_{X|S}^{\infty}|Q_S^{\infty})]^+
$$
$$
\geq [R_1 - H(Q_{X|S}^{\infty}|Q_S^{\infty}) + R_2 - I(Q_Y^{\infty};Q_{S|Y}^{\infty})]^+.
$$

Therefore (35) is established. Taking the $\liminf$ in (34) and applying (35) yields

$$
\liminf_{n \to \infty} F^n(P_{XY}, R_1, R_2)
$$
$$
\geq D(\tilde{Q}_{XYS}^{\infty} || P_{XY}\tilde{Q}_{S|Y}^{\infty}) + F_1(\tilde{Q}_{XYS}^{\infty}, R_1, R_2)
$$
$$
\geq \inf_{\substack{Q_{XYS}:Q_Y=\tilde{Q}_Y^{\infty} \\ Q_{S|Y}=\tilde{Q}_{S|Y}^{\infty}}} D(Q_{XYS} || P_{XY}\tilde{Q}_{S|Y}^{\infty}) + F_1(Q_{XYS}, R_1, R_2)
$$
$$
\geq \sup_{\tilde{Q}_{S|Y}} \inf_{\substack{Q_{XYS}:Q_Y=Q_Y^{\infty} \\ Q_{S|Y}=\tilde{Q}_{S|Y}}} D(Q_{XYS} || P_{XY}Q_{S|Y}) + F_1(Q_{XYS}, R_1, R_2) - \delta
$$
$$
\geq \inf_{\tilde{Q}_Y} \sup_{\tilde{Q}_{S|Y}} \inf_{\substack{Q_{XYS}:Q_Y=\tilde{Q}_Y \\ Q_{S|Y}=\tilde{Q}_{S|Y}}} D(Q_{XYS} || P_{XY}Q_{S|Y}) + F_1(Q_{XYS}, R_1, R_2) - \delta.
$$

Letting $\delta \to 0$ gives the result. ∎

*Proof of Theorem 1:* To prove the theorem we will upper bound $P_e = \Pr(X^n \neq \hat{X}^n)$, the probability of error for our scheme. For any $\epsilon > 0$, we note that for $n$ sufficiently large the constraints in (2) are met. Define $\mathcal{E}_r = \mathcal{E}_{r,1} \cup \mathcal{E}_{r,2}$. Observe that on $(\mathcal{E}_r)^c$ the scheme makes no error, thus

$$
P_e = \sum_{\mathcal{E}_r} \Pr(X^n \neq \hat{X}^n | X^n = \mathbf{x}, Y^n = \mathbf{y}, S^n = \mathbf{s}) \Pr(X^n = \mathbf{x}, Y^n = \mathbf{y}, S^n = \mathbf{s})
$$
$$
= \sum_{\mathcal{E}_r \cap (\mathcal{E}_c)^c} \Pr(X^n \neq \hat{X}^n | X^n = \mathbf{x}, Y^n = \mathbf{y}, S^n = \mathbf{s}) \Pr(X^n = \mathbf{x}, Y^n = \mathbf{y}, S^n = \mathbf{s})
$$
$$
+ \sum_{\mathcal{E}_r \cap \mathcal{E}_c} \Pr(X^n \neq \hat{X}^n | X^n = \mathbf{x}, Y^n = \mathbf{y}, S^n = \mathbf{s}) \Pr(X^n = \mathbf{x}, Y^n = \mathbf{y}, S^n = \mathbf{s})
$$
$$
\leq \sum_{\mathcal{E}_{r,1} \cap (\mathcal{E}_c)^c} \Pr(X^n \neq \hat{X}^n | X^n = \mathbf{x}, Y^n = \mathbf{y}, S^n = \mathbf{s}) \Pr(X^n = \mathbf{x}, Y^n = \mathbf{y}, S^n = \mathbf{s})
$$
$$
+ \sum_{\mathcal{E}_{r,2} \cap (\mathcal{E}_c)^c} \Pr(X^n \neq \hat{X}^n | X^n = \mathbf{x}, Y^n = \mathbf{y}, S^n = \mathbf{s}) \Pr(X^n = \mathbf{x}, Y^n = \mathbf{y}, S^n = \mathbf{s})
$$
$$
+ \sum_{\mathcal{E}_c} \Pr(X^n = \mathbf{x}, Y^n = \mathbf{y}, S^n = \mathbf{s})
$$

where the final inequality follows by bounding the conditional error probability by 1 on $\mathcal{E}_c$. Applying Lemmas 1 and 5 to the summation over $\mathcal{E}_{r,1} \cap (\mathcal{E}_c)^c$, Lemmas 1 and 6 to the summation over $\mathcal{E}_{r,2}$ and Lemma 2 to summation over $\mathcal{E}_c$ we obtain

$$
\begin{aligned}
P_e \leq & \sum_{\mathcal{E}_{r,1} \cap (\mathcal{E}_c)^c} \exp(-n[R_1 - H(Q_{\mathbf{x}|\mathbf{s}}|Q_{\mathbf{s}}) - \delta_n]^+) \frac{P_{XY}^n(\mathbf{x},\mathbf{y})}{|T_{Q_{S|Y}^*(Q_{\mathbf{y}})}^n(\mathbf{y})|} \\
& + \sum_{\mathcal{E}_{r,2} \cap (\mathcal{E}_c)^c} 2 \exp(-n[R_1 + R_2 - H(Q_{\mathbf{x}|\mathbf{s}}|Q_{\mathbf{s}}) \\
& \qquad - I(Q_{S|Y}^*(Q_{\mathbf{y}}); Q_{\mathbf{y}}) - \tilde{\tilde{\delta}}_n]^+) \frac{P_{XY}^n(\mathbf{x},\mathbf{y})}{|T_{Q_{S|Y}^*(Q_{\mathbf{y}})}^n(\mathbf{y})|} \\
& + \sum_{\mathcal{E}_c} \exp(-(n+1)^2).
\end{aligned}
$$

Observing that the summation over $\mathcal{E}_c$ decays super-exponentially, we may safely omit this term, and use the notation $\dot{\leq}$ to denote inequality to the first order of the exponent. Now summing first over types and then over sequences within the type class, we get

$$
\begin{aligned}
P_e \dot{\leq} \sum_{Q_Y} \Big[ & \sum_{Q_{XYS} \in \mathcal{D}_{r,1} \cap (\mathcal{D}_c)^c} \sum_{(\mathbf{x},\mathbf{y},\mathbf{s}) \in T_{Q_{XYS}}^n} \exp(-n[R_1 - H(Q_{\mathbf{x}|\mathbf{s}}|Q_{\mathbf{s}}) - \delta_n]^+) \frac{P_{XY}^n(\mathbf{x},\mathbf{y})}{|T_{Q_{S|Y}^*(Q_{\mathbf{y}})}^n(\mathbf{y})|} \\
& + \sum_{Q_{XYS} \in \mathcal{D}_{r,2} \cap (\mathcal{D}_c)^c} \sum_{(\mathbf{x},\mathbf{y},\mathbf{s}) \in T_{Q_{XYS}}^n} \exp(-n[R_1 + R_2 - H(Q_{\mathbf{x}|\mathbf{s}}|Q_{\mathbf{s}}) \tag{36} \\
& \qquad - I(Q_{S|Y}^*(Q_{\mathbf{y}}); Q_{\mathbf{y}}) - \tilde{\tilde{\delta}}_n]^+) \frac{P_{XY}^n(\mathbf{x},\mathbf{y})}{|T_{Q_{S|Y}^*(Q_{\mathbf{y}})}^n(\mathbf{y})|} \Big], \tag{37}
\end{aligned}
$$

where in the summation over joint types $Q_{XYS}$, the marginal type of $Y$ is fixed to be that set by the earlier summation. Using the following facts

$$
\begin{aligned}
P_{XY}^n(\mathbf{x},\mathbf{y}) &= \exp(-n(D(Q_{\mathbf{xy}}||P_{XY}) + H(Q_{\mathbf{xy}}))) \\
|T_{Q_{XYS}}^n| &\leq \exp(n(H(Q_{XYS}))) \leq \exp(n \log(|\mathcal{X}||\mathcal{Y}||\mathcal{S}|)) \tag{38} \\
|T_{Q_{S|Y}}^n| &\geq (n+1)^{-|\mathcal{Y}||\mathcal{S}|} \exp(n(H(Q_{S|Y}|Q_Y))) \tag{39}
\end{aligned}
$$

and continuing from (36), we can further bound $P_e$ as follows

$$
\begin{aligned}
P_e \dot{\leq} \sum_{Q_Y} \Big[ & \sum_{Q_{XYS} \in \mathcal{D}_{r,1} \cap (\mathcal{D}_c)^c} \exp\Big( -n\Big( [R_1 - H(Q_{X|S}|Q_S) - \delta_n]^+ \\
& + D(Q_{XY}||P_{XY}) + H(Q_{XY}) + H(Q_{S|Y}|Q_Y) - H(Q_{XYS}) \Big) \Big) \\
& + \sum_{Q_{XYS} \in \mathcal{D}_{r,2} \cap (\mathcal{D}_c)^c} \exp\Big( -n\Big( [R_1 + R_2 - H(Q_{X|S}|Q_S) - I(Q_Y; Q_{S|Y}) - \tilde{\tilde{\delta}}_n]^+ \\
& + D(Q_{XY}||P_{XY}) + H(Q_{XY}) + H(Q_{S|Y}|Q_Y) - H(Q_{XYS}) \Big) \Big) \Big]. \tag{40}
\end{aligned}
$$

Next we note that

$$
\begin{aligned}
& D(Q_{XY}||P_{XY}) + H(Q_{XY}) + H(Q_{S|Y}|Q_Y) - H(Q_{XYS}) \\
=& D(Q_{XY}||P_{XY}) + H(Q_{S|Y}|Q_Y) - H(Q_{S|XY}|Q_{XY}) \\
=& D(Q_{XYS}|P_{XY}Q_{S|Y}),
\end{aligned}
$$

and substituting this identity into (40) gives

$$P_e \overset{\cdot}{\le} \sum_{Q_Y} \Bigg[ \sum_{Q_{XYS} \in \mathcal{D}_{r,1} \cap (\mathcal{D}_c)^c} \exp\Big(-n\big([R_1 - H(Q_{X|S}|Q_S) - \delta_n]^+ + D(Q_{XYS}||P_{XY}Q_{S|Y})\big)\Big)$$

$$+ \sum_{Q_{XYS} \in \mathcal{D}_{r,2} \cap (\mathcal{D}_c)^c} \exp\Big(-n\big([R_1 + R_2 - H(Q_{X|S}|Q_S) - I(Q_Y; Q_{S|Y}) - \tilde{\tilde{\delta}}_n]^+ + D(Q_{XYS}||P_{XY}Q_{S|Y})\big)\Big)\Bigg].$$

We may now upper bound the summations by maximizing over the types and optimizing over the choice of test channel $Q_{S|Y}$. We now let $F_1^n$ be defined as in Lemma 7 and apply (22) to yield

$$P_e \overset{\cdot}{\le} |\mathcal{P}^n(\mathcal{X} \times \mathcal{Y} \times \mathcal{S})||\mathcal{P}^n(\mathcal{Y})| \max_{Q_Y} \min_{Q_{S|Y} \in \mathcal{C}^n(Q_Y, \mathcal{S})} \max_{Q_{XYS} \cap (\mathcal{D}_c)^c}$$

$$\exp\Big(-n\big(D(Q_{XYS}||P_{XY}Q_{S|Y}) + F_1^n(Q_{XYS}, R_1, R_2)\big)\Big). \tag{41}$$

Let $F^n$ be as defined in (32). We may move the optimizations appearing in (41) into the exponent and this yields

$$P_e \overset{\cdot}{\le} \exp(-n(F^n(P_{XY}, R_1, R_2))).$$

Then we have

$$\liminf_{n \to \infty} -\frac{1}{n} \log P_e \overset{\cdot}{\ge} \liminf_{n \to \infty} -\frac{1}{n} \log(\exp(-n(F^n(P_{XY}, R_1, R_2))))$$

$$= \liminf_{n \to \infty} F^n(P_{XY}, R_1, R_2)$$

$$\ge F^\infty(P_{XY}, R_1, R_2)$$

where the final line followed by an application of Lemma 7. ∎

# APPENDIX B
# PROOF OF THEOREM 2

Before proving Theorem 2, we prove two technical lemmas. We first prove the cardinality bound on $S$ given in (5). This argument differs from conventional cardinality-bound proofs in that it uses the KKT conditions in addition to Carathéodory's theorem. We then prove a continuity lemma that is similar to Lemma 7. For the purposes of these lemmas define two new quantities

$$\tilde{\eta}_U(P_{XY}, R_1, R_2) \triangleq \inf_{Q_Y} \sup_{\substack{Q_{S|Y}: |\mathcal{S}| \le |\mathcal{X}| \cdot |\mathcal{Y}| + |\mathcal{Y}| + 2 \\ I(Q_Y; Q_{S|Y}) \le R_2}} \inf_{\substack{Q_{X|Y}: \\ H(Q_{X|S}|Q_S) \ge R_1}} D(Q_{XY}||P_{XY})$$

$$\text{and } \overline{\eta}_U(P_{XY}, R_1, R_2) \triangleq \inf_{Q_Y} \sup_{\substack{Q_{S|Y}: \\ I(Q_Y; Q_{S|Y}) \le R_2}} \inf_{\substack{Q_{X|Y}: \\ H(Q_{X|S}|Q_S) \ge R_1}} D(Q_{XY}||P_{XY}).$$

Note that $\tilde{\eta}_U$ differs from $\eta_U$ only in that the inequality in the inner-most infimum is no longer strict, and $\overline{\eta}_U$ differs from $\tilde{\eta}_U$ only in the omission of the cardinality bound on $S$. Since for $R_1 \ge \log|\mathcal{X}_1|$, $\eta_U(P_{XY}, R_1, R_2) = \infty$ and Theorem 2 is trivial, we assume throughout this appendix that $R_1 < \log|\mathcal{X}_1|$.

**Lemma 8.** *If $R_1 < \log|\mathcal{X}_1|$ and $P_{XY}(x, y) > 0$ for all $x$ and $y$, then $\tilde{\eta}_U = \overline{\eta}_U$.*

*Proof:* Clearly $\overline{\eta}_U \ge \tilde{\eta}_U$. To show the reverse inequality, it suffices to show that for all $Q_Y$ and all $Q_{S|Y}$ such that $I(Q_Y; Q_{S|Y}) \le R_2$, there exists $\tilde{Q}_{S|Y}$ such that

1) $I(Q_Y; \tilde{Q}_{S|Y}) \le R_2$
2) $|\mathcal{S}| \le |\mathcal{X}| \cdot |\mathcal{Y}| + |\mathcal{Y}| + 2$

3) $\gamma(Q_Y, Q_{S|Y}) \leq \gamma(Q_Y, \tilde{Q}_{S|Y})$,

where

$$\gamma(Q_Y, Q_{S|Y}) = \inf_{\substack{Q_{X|Y}: \\ H(Q_{X|S}|Q_S) \geq R_1}} D(Q_{XY}||P_{XY}).$$

We will show that $Q^*_{X|Y}$ achieving the infimum in $\gamma$ for a given $Q_{S|Y}$ must satisfy certain KKT conditions. Carathéodory's theorem will then be used to show that $Q_{S|Y}$ can be replaced by a cardinality-limited distribution for which the same $Q^*_{X|Y}$ again satisfies the KKT conditions, and therefore $Q^*_{X|Y}$ attains the infimum in this case also.

Fix $Q_Y$ and $Q_{S|Y}$. For the $P_{XY}$ of the hypothesis, $\gamma(Q_Y, \cdot)$ has a continuous objective and a compact feasible set, so there exists $Q^*_{X|Y}$ such that

$$\gamma(Q_Y, Q_{S|Y}) = D(Q_Y Q^*_{X|Y}||P_{XY})$$

and $H(Q^*_{X|S}|Q_S) \geq R_1$. Since the objective in this optimization problem is convex and the constraint is convex and strictly feasible, the optimizer $Q^*_{X|Y}$ must satisfy the KKT conditions for optimality [37, p.g. 243]: there exists[9]

$$\mu_{x,y} \geq 0 \quad \text{for all } x, y$$
$$\lambda \geq 0$$
$$\nu_y \geq 0 \quad \text{for all } y$$

such that

$$Q(y)\Big(\log \frac{Q^*(x|y)Q(y)}{P(x,y)} + 1 + \lambda\Big) - \mu_{x,y} + \nu_y$$
$$+ \lambda\Big(\sum_s Q(s)\Big(Q(y|s)\log\Big(\sum_{y'} Q^*(x|y')Q(y'|s)\Big)\Big)\Big) = 0 \quad \text{for all } x, y$$

$$\mu_{x,y}Q(x|y) = 0 \quad \text{for all } x, y$$
$$\lambda(H(Q^*_{X|S}|Q_S) - R_1) = 0$$
$$\nu_y(\sum_x Q^*(x|y) - 1) = 0 \quad \text{for all } y.$$

By Carathéodory's theorem [10, Ch. 3, Lemma 3.4], there exists $\tilde{Q}(s)$ such that

$$|s : \tilde{Q}(s) > 0| \leq |\mathcal{X}| \cdot |\mathcal{Y}| + |\mathcal{Y}| + 2$$

and

$$\sum_s \tilde{Q}(s)Q(y|s) = Q(y) \quad \text{for all } y$$

$$Q(y)\Big(\log \frac{Q^*(x|y)Q(y)}{P(x,y)} + 1 + \lambda\Big) - \mu_{x,y} + \nu_y$$
$$+ \lambda\Big(\sum_s \tilde{Q}(s)\Big(Q(y|s)\log\Big(\sum_{y'} Q^*(x|y')Q(y'|s)\Big)\Big)\Big) = 0 \quad \text{for all } x, y$$

$$I(Q_S; Q_{Y|S}) = I(\tilde{Q}_S; Q_{Y|S})$$
$$H(Q^*_{X|S}|Q_S) = H(Q^*_{X|S}|\tilde{Q}_S).$$

---

[9]The assumption that $P_{XY}(x,y) > 0$ for all $x$ and $y$ guarantees that $D(Q_Y Q^*_{X|Y}||P_{XY})$ is finite. If this quantity is infinite, then the KKT conditions may not hold at $Q^*_{X|Y}$.

Define $\tilde{Q}_{S|Y}$ via $Q_{Y|S}\tilde{Q}_S/Q_Y$. Then $\tilde{Q}_{S|Y}$ is feasible because

$$I(Q_S; Q_{Y|S}) = I(\tilde{Q}_S; Q_{Y|S}) \leq R_2$$

Given that the code designer selects the test channel $\tilde{Q}_{S|Y}$ instead of $Q_{S|Y}$, $Q_{X|Y}^*$ is still a feasible choice for nature because

$$H(Q_{X|S}^*|\tilde{Q}_S) \geq R_1$$

Moreover, $Q_{X|Y}^*$ still satisfies the KKT conditions and

$$Q(y)\Big( \log \frac{Q^*(x|y)Q(y)}{P(x,y)} + 1 + \lambda \Big) - \mu_{x,y} + \nu_y$$
$$+ \lambda\Big( \sum_s \tilde{Q}(s)\Big( Q(y|s) \log \Big( \sum_{y'} Q^*(x|y')Q(y'|s) \Big) \Big) \Big) = 0 \quad \text{for all } x, y$$

$$\mu_{x,y}Q(x|y) = 0$$
$$\lambda(H(Q_{X|S}^*|\tilde{Q}_S) - R_1) = 0$$
$$\nu_y(\sum_x Q^*(x|y) - 1) = 0 \quad \text{for all } y.$$

Since $\gamma(Q_Y, \cdot)$ is convex, the KKT conditions are also sufficient for optimality, and we have

$$\gamma(Q_Y, \tilde{Q}_{S|Y}) = D(Q_{XY}||P_{XY}) = \gamma(Q_Y, Q_{S|Y}).$$

∎

**Lemma 9.** *For $R_1 < \log |\mathcal{X}_1|$, we have*

$$\lim_{\epsilon \to 0} \tilde{\eta}_U(P_{XY}, R_1 + \epsilon, R_2 + \epsilon) = \eta_U(P_{XY}, R_1, R_2).$$

*Proof:* Clearly $\tilde{\eta}_U(P_{XY}, R_1 + \epsilon, R_2 + \epsilon) \geq \eta_U(P_{XY}, R_1, R_2)$ for all $\epsilon > 0$. To show the reverse inequality, fix a sequence $\epsilon_n \downarrow 0$. Note that there exists $Q_Y^*$ such that[10]

$$\sup_{\substack{Q_{S|Y}: \\ I(Q_Y^*; Q_{S|Y}) \leq R_2}} \inf_{\substack{Q_{X|Y}: \\ H(Q_{X|S}|Q_S) > R_1}} D(Q_Y^*Q_{X|Y}||P_{XY}) \leq \inf_{Q_Y} \sup_{\substack{Q_{S|Y}: \\ I(Q_Y; Q_{S|Y}) \leq R_2}} \inf_{\substack{Q_{X|Y}: \\ H(Q_{X|S}|Q_S) > R_1}} D(Q_{XY}||P_{XY}) + \delta.$$

For each $n$, there exists $Q_{S|Y}^{(n)}$ such that

$$\inf_{\substack{Q_{X|Y}: \\ H(Q_{X|S}|Q_S^{(n)}) \geq R_1 + \epsilon_n}} D(Q_{X|Y}Q_Y^*||P_{XY}) \geq \sup_{\substack{Q_{S|Y}: \\ I(Q_Y^*; Q_{S|Y}) \leq R_2 + \epsilon_n}} \inf_{\substack{Q_{X|Y}: \\ H(Q_{X|S}|Q_S) \geq R_1 + \epsilon_n}} D(Q_{X|Y}Q_Y^*||P_{XY}) - \delta.$$

By considering subsequences, we may assume that

$$Q_{S|Y}^{(n)} \to Q_{S|Y}^\infty.$$

Then there exists $Q_{X|Y}^\infty$ such that

$$H(Q_{X|S}^\infty|Q_S^\infty) > R_1,$$

and

$$D(Q_{X|Y}^\infty Q_Y^*||P_{XY}) \leq \inf_{\substack{Q_{X|Y}: \\ H(Q_{X|S}|Q_S^\infty) > R_1}} D(Q_{X|Y}Q_Y^*||P_{XY}) + \delta.$$

---

[10]Throughout this proof, $Q_{S|Y}$ is assumed to satisfy the cardinality bound (5).

Note that for all sufficiently large $n$, we have

$$H(Q_{X|S}^\infty|Q_S^{(n)}) \geq R_1 + \epsilon_n.$$

Then for all sufficiently large $n$,

$$
\begin{aligned}
\tilde{\eta}_U(P_{XY}, R_1 + \epsilon_n, R_2 + \epsilon_n) &\leq \sup_{\substack{Q_{S|Y}: \\ I(Q_Y^*; Q_{S|Y}) \leq R_2 + \epsilon_n}} \inf_{\substack{Q_{X|Y}: \\ H(Q_{X|S}|Q_S) \geq R_1 + \epsilon_n}} D(Q_{X|Y}Q_Y^*||P_{XY}) \\
&\leq \inf_{\substack{Q_{X|Y}: \\ H(Q_{X|S}|Q_S^{(n)}) \geq R_1 + \epsilon_n}} D(Q_{X|Y}Q_Y^*||P_{XY}) + \delta \\
&\leq D(Q_{X|Y}^\infty Q_Y^*||P_{XY}) + \delta.
\end{aligned}
$$

Thus

$$\limsup_{n \to \infty} \tilde{\eta}_U(P_{XY}, R_1 + \epsilon_n, R_2 + \epsilon_n) \leq D(Q_{X|Y}^\infty Q_Y^*||P_{XY}) + \delta. \tag{42}$$

On the other hand, we have

$$
\begin{aligned}
D(Q_{X|Y}^\infty Q_Y^*||P_{XY}) &\leq \inf_{\substack{Q_{X|Y}: \\ H(Q_{X|S}|Q_S^\infty) > R_1}} D(Q_{X|Y}Q_Y^*||P_{XY}) + \delta \\
&\leq \sup_{\substack{Q_{S|Y}: \\ I(Q_Y^*; Q_{S|Y}) \leq R_2}} \inf_{\substack{Q_{X|Y}: \\ H(Q_{X|S}|Q_S) > R_1}} D(Q_{X|Y}Q_Y^*||P_{XY}) + \delta \\
&\leq \eta_U(P_{XY}, R_1, R_2) + 2\delta.
\end{aligned}
$$

Combining this with (42) yields

$$\limsup_{n \to \infty} \tilde{\eta}_U(P_{XY}, R_1 + \epsilon_n, R_2 + \epsilon_n) \leq \eta_U(P_{XY}, R_1, R_2) + 3\delta,$$

but $\delta > 0$ and $\epsilon_n \to 0$ were arbitrary. ∎

*Proof of Theorem 2:* Recall that we may assume $R_1 < \log|\mathcal{X}_1|$. As we are eventually considering small $\epsilon$, we may assume that $R_1 + 2\epsilon \leq \log|\mathcal{X}_1|$. Take $n$ sufficiently large so that $\frac{1}{n} \leq \frac{\epsilon}{2}$.

Let $(f_1^n, f_2^n, g^n)$ be any code satisfying (2) and let

$$\mathcal{E}^n(f_1^n, f_2^n, g^n) = \{(\mathbf{x}, \mathbf{y}): g^n(f_1^n(\mathbf{x}), f_2^n(\mathbf{y})) \neq \mathbf{x}\}.$$

denote its erroneous sequences. Take any $Q_{XY}$ such that

$$H_{Q_{XY}}(X^n|f_2^n(Y^n)) \geq n(R_1 + 2\epsilon). \tag{43}$$

We first show that for this choice of $Q_{XY}$ the following inequality holds

$$Q_{XY}^n(\mathcal{E}^n(f_1^n, f_2^n, g^n)) \geq \frac{\epsilon}{2\log|\mathcal{X}|} > 0, \tag{44}$$

we will then apply a change of measure argument. Fano's inequality gives

$$Q_{XY}^n(\mathcal{E}^n(f_1^n, f_2^n, g^n)) \geq \frac{H(X^n|f_1^n(X^n), f_2^n(Y^n)) - 1}{\log|\mathcal{X}^n|}. \tag{45}$$

But

$$
\begin{aligned}
H(X^n, f_1^n(X^n)|f_2^n(Y^n)) &= H(X^n|f_2^n(Y^n)) + H(f_1^n(X^n)|X^n, f_2^n(Y^n)) = H(X^n|f_2^n(Y^n)) \\
&= H(f_1^n(X^n)|f_2^n(Y^n)) + H(X^n|f_1^n(X^n), f_2^n(Y^n)).
\end{aligned}
$$

Therefore

$$
\begin{aligned}
H(X^n|f_1^n(X^n), f_2^n(Y^n)) &= H(X^n|f_2^n(Y^n)) - H(f_1^n(X^n)|f_2^n(Y^n)) \\
&\geq H(X^n|f_2^n(Y^n)) - H(f_1^n(X^n)) \\
&\geq H(X^n|f_2^n(Y^n)) - n(R_1 + \epsilon) \\
&\geq n\epsilon.
\end{aligned}
\tag{46}
$$

The fact that $\frac{1}{n} \leq \frac{\epsilon}{2}$ along with equations (45) and (46) gives (44). For $\delta > 0$ define the set

$$
\mathcal{D}^n = \left\{ (\mathbf{x}, \mathbf{y}) : \left| \frac{1}{n} \log \frac{Q_{XY}^n(\mathbf{x}, \mathbf{y})}{P_{XY}^n(\mathbf{x}, \mathbf{y})} - D(Q_{XY}||P_{XY}) \right| \leq \delta \right\}.
$$

Fix $0 < \alpha < \infty$ such that for all distributions $Q_{XY}$,

$$
\mathbb{E}_Q \left[ \log^2 \frac{Q(X,Y)}{P(X,Y)} \right] \leq \alpha.
$$

Such an $\alpha$ exists because the alphabet is finite and $P(x,y) > 0$ by assumption. By Chebyshev's inequality we have

$$
\begin{aligned}
Q_{XY}^n(\mathcal{D}^n) &= 1 - Q_{XY}^n((\mathcal{D}^n)^c) \\
&\geq 1 - (\delta^{-2}) \mathbb{E}_Q \left[ \left( \frac{1}{n} \sum_i \log \frac{Q(X_i, Y_i)}{P(X_i, Y_i)} - D(Q_{XY}||P_{XY}) \right)^2 \right] \\
&\geq 1 - \frac{\mathbb{E}_Q \left[ \log^2 \frac{Q(X,Y)}{P(X,Y)} \right]}{n\delta^2} \\
&\geq 1 - \frac{\alpha}{\delta^2 n}
\end{aligned}
$$

We may bound the error probability as follows

$$
\begin{aligned}
P_{XY}^n(\mathcal{E}^n(f_1^n, f_2^n, g^n)) &\geq P_{XY}^n(\mathcal{E}^n(f_1^n, f_2^n, g^n) \cap \mathcal{D}^n) \\
&= \sum_{\mathcal{E}^n(f_1^n, f_2^n, g^n) \cap \mathcal{D}^n} Q_{XY}^n(\mathbf{x}, \mathbf{y}) \exp \left( -\log \frac{Q_{XY}^n(\mathbf{x}, \mathbf{y})}{P_{XY}^n(\mathbf{x}, \mathbf{y})} \right) \\
&\geq Q_{XY}^n(\mathcal{E}^n(f_1^n, f_2^n, g^n) \cap \mathcal{D}^n) \exp \left( -n(D(Q_{XY}||P_{XY}) + \delta) \right) \\
&\geq \left( \frac{\epsilon}{2 \log |\mathcal{X}|} - \frac{\alpha}{\delta^2 n} \right) \exp \left( -n(D(Q_{XY}||P_{XY}) + \delta) \right).
\end{aligned}
\tag{47}
$$

However, for $n$ large enough

$$
\frac{\epsilon}{2 \log |\mathcal{X}|} - \frac{\alpha}{\delta^2 n} \geq \frac{\epsilon}{4 \log |\mathcal{X}|} \triangleq \beta > 0,
$$

thus, observing that the argument above holds for every $Q_{XY}$ satisfying (43) we see that

$$
P_{XY}^n(\mathcal{E}^n(f_1^n, f_2^n, g^n)) \geq \sup_{Q_{XY} : H_Q(X^n|f_2^n(Y^n)) \geq n(R_1 + 2\epsilon)} \beta \exp \left( -n(D(Q_{XY}||P_{XY}) + \delta) \right).
$$

Now we note that the above holds for every code satisfying (2), thus, observing that the right hand side does not depend on $f_1^n, g^n$, we conclude that

$$
\min_{f_1^n, f_2^n, g^n} P_{XY}^n(\mathcal{E}^n(f_1^n, f_2^n, g^n)) \geq \min_{f_2^n} \sup_{Q_{XY} : H_Q(X^n|f_2^n(Y^n)) \geq n(R_1 + 2\epsilon)} \beta \exp \left( -n(D(Q_{XY}||P_{XY}) + \delta) \right).
$$

We now move the optimizations into the exponent and focus our attention there.

$$\max_{\substack{f_2^n: \\ \log|f_2^n| \leq n(R_2+\epsilon)}} \quad \inf_{Q_{XY}: H_Q(X^n|f_2^n(Y^n)) \geq n(R_1+2\epsilon)} D(Q_{XY}||P_{XY})$$

$$= \max_{\substack{f_2^n: \\ \log|f_2^n| \leq n(R_2+\epsilon)}} \inf_{Q_Y} \quad \inf_{\substack{Q_{X|Y}: \\ H_Q(X^n|f_2^n(Y^n)) \geq n(R_1+2\epsilon)}} D(Q_{XY}||P_{XY})$$

$$\leq \inf_{Q_Y} \max_{\substack{f_2^n: \\ \log|f_2^n| \leq n(R_2+\epsilon)}} \quad \inf_{\substack{Q_{X|Y}: \\ H_Q(X^n|f_2^n(Y^n)) \geq n(R_1+2\epsilon)}} D(Q_{XY}||P_{XY})$$

$$\leq \inf_{Q_Y} \max_{\substack{f_2^n: \\ I(Y^n; f_2^n(Y^n)) \leq n(R_2+\epsilon)}} \quad \inf_{\substack{Q_{X|Y}: \\ H_Q(X^n|f_2^n(Y^n)) \geq n(R_1+2\epsilon)}} D(Q_{XY}||P_{XY})$$

$$\leq \inf_{Q_Y} \sup_{\substack{Q_{U|Y^n}: \\ I(Y^n;U) \leq n(R_2+\epsilon)}} \quad \inf_{\substack{Q_{X|Y}: \\ H_Q(X^n|U) \geq n(R_1+2\epsilon)}} D(Q_{XY}||P_{XY}) \tag{48}$$

In the previous line, we note that the deterministic functions are still feasible and on deterministic functions the previous two bounds agree. Henceforth the joint distribution of $X, Y, U$ is $Q_Y Q_{U|Y} Q_{X|Y}$, so that $X, Y$ and $U$ form a Markov chain. To continue we use the following, obtained via the chain rule

$$H(X^n|U) = \sum_{i=1}^n H(X_i|U, X_1^{i-1})$$

$$\geq \sum_{i=1}^n H(X_i|U, X_1^{i-1}, Y_1^{i-1})$$

$$= \sum_{i=1}^n H(X_i|U, Y_1^{i-1}) \tag{49}$$

where on the final line we used the fact that $X_i - (U, Y_1^{i-1}) - X_1^{i-1}$. The following identity also holds

$$I(Y^n; U) = \sum_{i=1}^n I(Y_i; U|Y_1^{i-1})$$

$$= \sum_{i=1}^n H(Y_i|Y_1^{i-1}) - H(Y_i|Y_1^{i-1}, U)$$

$$= \sum_{i=1}^n I(Y_i; Y_1^{i-1}, U). \tag{50}$$

Substituting (49) into (48) makes the feasible set smaller because of the inequality. After substituting (50), we can continue to bound the exponent by

$$\leq \inf_{Q_Y} \sup_{\substack{Q_{U|Y^n}: \\ \frac{1}{n}\sum_{i=1}^n I(Y_i;Y_1^{i-1},U) \leq R_2+\epsilon}} \quad \inf_{\substack{Q_{X|Y}: \\ \frac{1}{n}\sum_{i=1}^n H(X_i|U,Y_1^{i-1}) \geq R_1+2\epsilon}} D(Q_{XY}||P_{XY})$$

$$= \inf_{Q_Y} \sup_{\substack{Q_{U|Y^n}: \\ \frac{1}{n}\sum_{i=1}^n I(Y_i;V_i) \leq R_2+\epsilon}} \quad \inf_{\substack{Q_{X|Y}: \\ \frac{1}{n}\sum_{i=1}^n H(X_i|V_i) \geq R_1+2\epsilon}} D(Q_{XY}||P_{XY})$$

where on the previous line, we let $V_i = (Y_1^{i-1}, U)$. Let $T$ denote a time sharing random variable, uniformly distributed on $\{1, \ldots, n\}$ and independent of everything else. Then the quantity above can be written

$$\inf_{Q_Y} \sup_{\substack{Q_{U|Y^n}: \\ I(Y_T; V_T, T) \leq R_2 + \epsilon}} \inf_{\substack{Q_{X|Y}: \\ H(X_T|V_T, T) \geq R_1 + 2\epsilon}} D(Q_{XY}||P_{XY})$$

$$= \inf_{Q_Y} \sup_{\substack{Q_{U|Y^n}: \\ I(Y_T; W) \leq R_2 + \epsilon}} \inf_{\substack{Q_{X|Y}: \\ H(X_T|W) \geq R_1 + 2\epsilon}} D(Q_{XY}||P_{XY}). \tag{*}$$

where we set $W = (V_T, T) = (Y_1^{T-1}, U, T)$. Since $(X_T, Y_T) \overset{d}{=} (X, Y)$, the above quantity is upper bounded by

$$\inf_{Q_Y} \sup_{\substack{Q_{S|Y}: \\ I(Y; S) \leq R_2 + 2\epsilon}} \inf_{\substack{Q_{X|Y}: \\ H(X|S) \geq R_1 + 2\epsilon}} D(Q_{XY}||P_{XY}) = \overline{\eta}_U(P_{XY}, R_1 + 2\epsilon, R_2 + 2\epsilon).$$

To see this, we note that every choice in $(*)$ is a feasible choice in $F$. In particular for a given $Q_Y$, let $U^*$ denote a choice for $Q_{U|Y^n}$ in $(*)$. Then choosing $S$ so that $(Y, S) \overset{d}{=} (Y, Y_1^{T-1}, U^*, T)$, is feasible. By Lemma 8, this quantity equals $\tilde{\eta}_U(P_{XY}, R_1 + 2\epsilon, R_2 + 2\epsilon)$. Thus we have shown that

$$\min_{f_1^n, f_2^n, g^n} P_{XY}^n(\mathcal{E}^n(f_1^n, f_2^n, g^n)) \geq \beta \exp\left(-n(\tilde{\eta}_U(P_{XY}, R_1 + 2\epsilon, R_2 + 2\epsilon) + \delta)\right).$$

Taking logs and the $\limsup$ as $n \to \infty$, and letting $\delta \downarrow 0$ and $\epsilon \downarrow 0$ (and invoking Lemma 9) gives the result. ∎

## Appendix C
## Proof of Theorem 3

### A. Scheme

For a given blocklength $n$, we operate on a type-by-type basis and define the encoder and decoder functions as follows. For each type $Q_X$, fix a conditional type $Q_{Z|X}^*(Q_X) \in \mathcal{C}^n(Q_X, \mathcal{Y})$, a decoding function $f(Q_X, Q_Y) \in \mathcal{F}$, and randomly choose a set of codewords $B^n(Q_X)$ in the following way. The size of $B^n(Q_X)$ is an integer satisfying

$$\exp(nI(Q_X; Q_{Z|X}^*(Q_X)) + (|\mathcal{X}||\mathcal{Z}| + 2)\log(n+1))$$
$$\leq |B^n(Q_X)| \tag{51}$$
$$\leq \exp(nI(Q_X; Q_{Z|X}^*(Q_X)) + (|\mathcal{X}||\mathcal{Z}| + 4)\log(n+1))$$

and the codewords are drawn uniformly, with replacement, from the marginal type class $T_{Q_Z}^n$ induced by $Q_X$ and $Q_{Z|X}^*(Q_X)$.

Define $Z : T_{Q_\mathbf{x}}^n \to B^n(Q_\mathbf{x})$ as follows. Let $\mathcal{G}(\mathbf{x}) \triangleq B^n(Q_\mathbf{x}) \cap T_{Q_{Z|X}^*(Q_\mathbf{x})}^n(\mathbf{x})$. If $\mathcal{G}(\mathbf{x})$ is non-empty, then the output of $Z(\mathbf{x})$ is drawn uniformly at random from $\mathcal{G}(\mathbf{x})$[11]. If $\mathcal{G}(\mathbf{x})$ is empty the output of $Z(\mathbf{x})$ is drawn uniformly at random from $B^n(Q_\mathbf{x})$. The function $Z(\cdot)$ determines the codeword sent by the encoder to the decoder. We define $Z^n = Z(X^n)$ and define the encoder's message set as follows

$$\mathcal{M} = \mathcal{M}_1 \times \mathcal{M}_2,$$
$$\mathcal{M}_1 = \{1, 2, \ldots, M_1 \triangleq \exp(nR)\},$$
$$\mathcal{M}_2 = \{1, 2, \ldots, (n+1)^{|\mathcal{X}|}\}.$$

*Operation of the encoder:* To encode a sequence $\mathbf{x} \in T_{Q_X}^n$, the encoder sends the type of $\mathbf{x}$ and an index, $U(Z(\mathbf{x}))$, of the codeword $Z(\mathbf{x})$. There are two cases to consider:

---

[11]Codewords that appear multiple times are proportionally more likely to be selected

1) $\log |B^n(Q_X)| < nR$, in which case we can map each member of $B^n(Q_X)$ to an element of $\mathcal{M}_1$ in a one-to-one manner.

2) $\log |B^n(Q_X)| \geq nR$, in which case we assign each distinct member of $B^n(Q_X)$ to $\mathcal{M}_1$ uniformly at random.

Let $U(Z(\mathbf{x}))$ denote the element to which $Z(\mathbf{x})$ is mapped. The encoder can be expressed mathematically as

$$\psi(\mathbf{x}) = (U(Z(\mathbf{x})), k(Q_X)) \text{ for } \mathbf{x} \in T_{Q_X}^n \tag{52}$$

*Operation of the Decoder:* The decoder operates in a two-step manner. First it attempts to recover the codeword $Z^n$:

1) If $|B^n(Q_X)| < nR$ then $Z^n$ can be decoded without error,

2) If $|B^n(Q_X)| \geq nR$ the decoder receives a bin index and uses the side information to pick the "best" $\mathbf{z}$ from the bin in the minimum conditional entropy sense: it searches for a $\hat{\mathbf{z}}$ in the received bin so that among all $\tilde{\mathbf{z}}$ in the bin, $H(\tilde{\mathbf{z}}|\mathbf{y}) > H(\hat{\mathbf{z}}|\mathbf{y})$. If there is no such $\hat{\mathbf{z}}$ it picks uniformly at random from the bin.

Let

$$\varphi_1(i, k(Q_X), \mathbf{y}) = \begin{cases} \hat{\mathbf{z}} & \hat{\mathbf{z}} \in \text{Bin}(i) \text{ and } \forall \tilde{\mathbf{z}} \in \text{Bin}(i), \\ & \tilde{\mathbf{z}} \neq \hat{\mathbf{z}} : H(\tilde{\mathbf{z}}|\mathbf{y}) > H(\hat{\mathbf{z}}|\mathbf{y}) \\ \text{any } \tilde{\mathbf{z}} \in \text{Bin}(i) & \text{if no such } \hat{\mathbf{z}} \in \text{Bin}(i) \end{cases} \tag{53}$$

where $\text{Bin}(i) = \{\mathbf{z} : \mathbf{z} \in B^n(Q_X) \text{ and } U(\mathbf{z}) = i\}$ denotes the set of codewords that are assigned to index $i$. Second, the decoder uses the estimation function, $f$, to combine the side information $\mathbf{y}$ with codeword $\mathbf{z}$ to give the reproduction $\hat{\mathbf{x}}$. This is expressed mathematically as

$$\varphi(i, k(Q_X), \mathbf{y}) = \hat{\mathbf{x}} \text{ s.t. } \hat{\mathbf{x}}_j = f(\varphi_1(i, k(Q_X), \mathbf{y})_j, \mathbf{y}_j). \tag{54}$$

### B. Error probability calculation

It will be convenient to consider the following subsets of the sequence space

$$\begin{aligned} \mathcal{E}_b = &\Big\{ (\mathbf{x}, \mathbf{y}, \mathbf{z}) : \mathbf{z} \in T_{Q_{Z|X}^*(Q_\mathbf{x})}^n(\mathbf{x}), d(\mathbf{x}, f(\mathbf{y}, \mathbf{z})) < \Delta, \\ &\quad \log |B^n(Q_\mathbf{x})| \geq nR \Big\} \\ \mathcal{E}_c = &\Big\{ (\mathbf{x}, \mathbf{y}, \mathbf{z}) : \mathbf{z} \notin T_{Q_{Z|X}^*(Q_\mathbf{x})}^n(\mathbf{x}) \Big\} \\ \mathcal{E}_d = &\Big\{ (\mathbf{x}, \mathbf{y}, \mathbf{z}) : \mathbf{z} \in T_{Q_{Z|X}^*(Q_\mathbf{x})}^n(\mathbf{x}), d(\mathbf{x}, f(\mathbf{y}, \mathbf{z})) \geq \Delta \Big\} \end{aligned}$$

$\mathcal{E}_b$ corresponds to a potential binning error, $\mathcal{E}_c$ to a covering error and $\mathcal{E}_d$ to a distortion error. We will consider the errors on these sets separately. Equivalently we can view these error events as properties of the joint type, so we define

$$\begin{aligned} \mathcal{D}_b = &\{Q_{XYZ} : \mathbb{E}[d(X, f(Y, Z))] < \Delta, Q_{Z|X} = Q_{Z|X}^*(Q_X) \\ &\quad \log |B^n(Q_X)| \geq nR \} \\ \mathcal{D}_c = &\{Q_{XYZ} : Q_{Z|X} \neq Q_{Z|X}^*(Q_X) \} \\ \mathcal{D}_d = &\{Q_{XYZ} : \mathbb{E}[d(X, f(Y, Z))] \geq \Delta, \\ &\quad Q_{Z|X} = Q_{Z|X}^*(Q_X) \}. \end{aligned}$$

Before we proceed with the proof of Theorem 1, we establish the following useful facts.

**Lemma 10.** *Let $X^n, Y^n, Z^n = \hat{Z}(X^n)$ be generated according to our scheme and suppose that $(\mathbf{x}, \mathbf{y}, \mathbf{z})$ is in $(\mathcal{E}_c)^c$, i.e. that $\mathbf{z} \in T^n_{Q^*_{Z|X}(Q_{\mathbf{x}})}(\mathbf{x})$. Then*

$$\Pr(X^n = \mathbf{x}, \, Y^n = \mathbf{y}, Z^n = \mathbf{z}) \tag{55}$$

$$\leq P^n_{XY}(\mathbf{x}, \mathbf{y}) \frac{1}{|T^n_{Q^*_{Z|X}(Q_{\mathbf{x}})}(\mathbf{x})|}. \tag{56}$$

*Proof:* The proof mirrors that of Lemma 1 and is omitted. ∎

**Lemma 11.** *Let $X^n, Y^n, Z^n = Z(X^n)$ be generated according to our scheme and suppose that $(\mathbf{x}, \mathbf{y}, \mathbf{z}) \in \mathcal{E}_c$. Then*

$$\Pr(X^n = \mathbf{x}, \, Y^n = \mathbf{y}, Z^n = \mathbf{z})$$

$$\leq \exp(-(n+1)^2). \tag{57}$$

*Proof:* The proof mirrors that of Lemma 2 and is omitted. ∎

**Lemma 12.** *For all strings $\mathbf{x}, \mathbf{z}$ such that $\mathbf{z} \in T^n_{Q_{\mathbf{z}}}$,*

$$\Pr(\mathbf{z} \in B^n(Q_{\mathbf{x}})) \leq$$

$$(n+1)^{|\mathcal{Z}|(1+|\mathcal{X}|)+4}$$

$$\times \exp(n(I(Q_{\mathbf{x}}; Q^*_{Z|X}(Q_{\mathbf{x}})) - H(Q_{\mathbf{z}}))).$$

*Proof:* By the construction of $B^n(Q_{\mathbf{x}})$, each of the codewords is chosen with replacement from the set $T^n_{Q_{\mathbf{z}}}$. Thus each string has probability $|T^n_{Q_{\mathbf{z}}}|^{-1}$ and we make $|B^n(Q_{\mathbf{x}})|$ such choices (bounded by (51)). From [10, lemma 2.3] we have

$$|T_{Q_{\mathbf{z}}}| \geq (n+1)^{-|\mathcal{Z}|} \exp(nH(Q_{\mathbf{z}})).$$

Invoking the union bound gives the result. ∎

**Lemma 13.** *Let $(\mathbf{x}, \mathbf{y}, \mathbf{z}) \in (\mathcal{E}_c \cup \mathcal{E}_d)^c$. Then*

$$\Pr(d(X^n, \hat{X}^n) > \Delta | X^n = \mathbf{x}, Y^n = \mathbf{y}, Z^n = \mathbf{z})$$

$$\leq \exp\left(-n\left((R - J(Q_{\mathbf{xyz}}) - \delta_b^n)^+\right)\right) \tag{58}$$

*where*

$$J(Q_{\mathbf{xyz}}) = I(Q_{\mathbf{x}}; Q^*_{Z|X}(Q_{\mathbf{x}})) - I(Q_{\mathbf{y}}; Q_{\mathbf{z}|\mathbf{y}})$$

$$and \quad \delta_b^n = \frac{1}{n} \log(n+1)^{|\mathcal{Z}|(|\mathcal{Y}|+1+|\mathcal{X}|)+4}.$$

*Moreover, if $\log |B^n(Q_{\mathbf{x}})| < nR$ then*

$$\Pr(d(X^n, \hat{X}^n) > \Delta | X^n = \mathbf{x}, Y^n = \mathbf{y}, Z^n = \mathbf{z}) = 0.$$

*Proof:* For the given sequence $(\mathbf{x}, \mathbf{y}, \mathbf{z})$ let $L$ be the event that $\mathbf{z} \neq \varphi_1(\psi(\mathbf{x}), \mathbf{y})$. (Observe that $L$ occurs when the decoder decodes the wrong codeword and that $\Pr(d(X^n, \hat{X}^n) > \Delta | X^n = \mathbf{x}, Y^n = \mathbf{y}, Z^n = \mathbf{z})$ is upper bounded by $\Pr(L | X^n = \mathbf{x}, Y^n = \mathbf{y}, Z^n = \mathbf{z})$.)

If $Q_{\mathbf{x}}$ is such that $\log |B^n(Q_{\mathbf{x}})| < nR$, then

$$\Pr(L | X^n = \mathbf{x}, Y^n = \mathbf{y}, Z^n = \mathbf{z}) = 0.$$

For the case where $\log |B^n(Q_{\mathbf{x}})| \geq nR$ (i.e. $(\mathbf{x}, \mathbf{y}, \mathbf{z}) \in \mathcal{E}_b$), we note that the set $S(\mathbf{z}|\mathbf{y})$ contains all strings $\tilde{\mathbf{z}}$ having the property that $\tilde{\mathbf{z}}$ has the same type as $\mathbf{z}$ and conditional empirical entropy with $\mathbf{y}$ that does not exceed $H(\mathbf{z}|\mathbf{y})$.

$$\Pr(L|X^n = \mathbf{x}, Y^n = \mathbf{y}, Z^n = \mathbf{z})$$
$$\leq \sum_{\substack{\tilde{\mathbf{z}} \in S(\mathbf{z}|\mathbf{y}) \\ \tilde{\mathbf{z}} \neq \mathbf{z}}} \Pr(\tilde{\mathbf{z}} \in B^n(Q_{\mathbf{x}}), U(\tilde{\mathbf{z}}) = U(\mathbf{z})|X^n = \mathbf{x}, Y^n = \mathbf{y}, Z^n = \mathbf{z})$$
$$\overset{1}{\leq} \sum_{\substack{\tilde{\mathbf{z}} \in S(\mathbf{z}|\mathbf{y}) \\ \tilde{\mathbf{z}} \neq \mathbf{z}}} \Pr(\tilde{\mathbf{z}} \in B^n(Q_{\mathbf{x}})|X^n = \mathbf{x}, Y^n = \mathbf{y}) \Pr(U(\tilde{\mathbf{z}}) = U(\mathbf{z})|\tilde{\mathbf{z}} \in B^n(Q_{\mathbf{x}}))$$
$$\overset{2}{\leq} \sum_{\substack{\tilde{\mathbf{z}} \in S(\mathbf{z}|\mathbf{y}) \\ \tilde{\mathbf{z}} \neq \mathbf{z}}} (n+1)^{|\mathcal{Z}|(1+|\mathcal{X}|)+4} \frac{1}{M_1} \exp(n(I(Q_{\mathbf{x}}; Q^*_{Z|X}(Q_{\mathbf{x}})) - H(Q_{\mathbf{z}}))) \tag{59}$$

where inequality [1] follows from a reasoning similar to that used in the proof of Lemma 6 and [2] follows from Lemma 12. Next,

$$\Pr(L|X^n = \mathbf{x}, Y^n = \mathbf{y}, Z^n = \mathbf{z})$$
$$\leq (n+1)^{|\mathcal{Z}|(|\mathcal{Y}|+1+|\mathcal{X}|)+4} \exp(nH(Q_{\mathbf{z}|\mathbf{y}}|Q_{\mathbf{y}}))$$
$$\times \exp(n(I(Q_{\mathbf{x}}; Q^*_{Z|X}(Q_{\mathbf{x}})) - H(Q_{\mathbf{z})))\frac{1}{M_1}$$
$$= (n+1)^{|\mathcal{Z}|(|\mathcal{Y}|+1+|\mathcal{X}|)+4} \exp(-n(R - J(Q_{\mathbf{xyz}})))$$

where the first line follows from Lemma 4. Also, since $\Pr(L|X^n = \mathbf{x}, Y^n = \mathbf{y}, Z^n = \mathbf{z}) \leq 1$ we get

$$\Pr(L|X^n = \mathbf{x}, Y^n = \mathbf{y}, Z^n = \mathbf{z})$$
$$= \exp(-n(R - J(Q_{\mathbf{xyz}}) - \delta_b^n)^+).$$

∎

**Lemma 14.** *Let* $\delta_b^n \to 0$, $\delta_r^n \to 0$ *as* $n \to \infty$,

$$G^n[Q_{XYZ}, P_{XY}, f, d, \Delta, R] =$$
$$\begin{cases} D(Q_{XYZ}||P_{XY}Q_{Z|X}) & \mathbb{E}_Q[d(X, f(Y, Z))] \geq \Delta \\ D(Q_{XYZ}||P_{XY}Q_{Z|X}) \\ \quad + (R - I(Q_X; Q^*_{Z|X}(Q_X)) \\ \quad + I(Q_Y; Q_{Z|Y}) - \delta_b^n)^+ & \mathbb{E}_Q[d(X, f(Y, Z))] < \Delta \\ & I(Q_X; Q_{Z|X}) \geq R - \delta_r^n \\ \infty & otherwise \end{cases}$$

*and*

$$\theta^n(P_{XY}, d, \Delta, R) = \min_{Q_X} \max_{Q_{Z|X} \in \mathcal{C}^n(\mathcal{X} \to \mathcal{Z})} \min_{Q_Y} \max_{f \in \mathcal{F}} \min_{Q_{XYZ}} G^n(Q_{XYZ}, P_{XY}, f, d, \Delta, R),$$
$$\theta^\infty(P_{XY}, d, \Delta, R) = \inf_{Q_X} \sup_{Q_{Z|X}} \inf_{Q_Y} \sup_{f \in \mathcal{F}} \inf_{Q_{XYZ}} G(Q_{XYZ}, P_{XY}, f, d, \Delta, R).$$

*In* $\theta^n$ *the minimizations and maximizations on* $Q_X, Q_{Z|X}, Q_Y$ *and* $Q_{XYZ}$ *are over types/conditional types, and in* $\theta^\infty$ *they are over distributions. And, in the optimization of* $Q_{XYZ}$ *the marginal type/distribution* *of $X$ and $Y$ and conditional type/distribution of $Z$ given $X$ are taken to be those specified earlier in the optimization.* Then

$$\liminf_{n \to \infty} \theta^n(P_{XY}, d, \Delta, R) \geq \theta^\infty(P_{XY}, d, \Delta, R) \tag{60}$$

*Proof:* Let $Q_X^{(n)}, Q_{Z|X}^{(n)}, Q_Y^{(n)}, Q_{XYZ}^{(n)}$ and $f^{(n)}$ be such that

$$\theta^n(P_{XY}, d, \Delta, R) = G^n(Q_{XYZ}^{(n)}, P_{XY}, f^{(n)}, d, \Delta, R).$$

For convenience, henceforth we omit writing the arguments $P_{XY}, d, \Delta$ and $R$ in $G(\cdot)$ and $G^n(\cdot)$. Also, when necessary for clarity, we expand $Q_{XYZ} = Q_X, Q_{Z|X}, Q_Y, Q_{Y|XZ}$ in the argument to $G$ and $G^n(\cdot)$.

By boundedness there exists a subsequence of $(Q_X^{(n)}, Q_{Z|X}^{(n)}, Q_Y^{(n)}, Q_{XYZ}^{(n)})$ with index $n'$ such that the sequence $(Q_X^{(n')}, Q_{Z|X}^{(n')}, Q_Y^{(n')}, Q_{XYZ}^{(n')}, f^{(n')})$ converges to a limit $(Q_X^\infty, Q_{Z|X}^\infty, Q_Y^\infty, Q_{XYZ}^\infty, f^\infty)$. Let $\delta > 0$, then there exists $\tilde{Q}_{Z|X}^\infty$ so that

$$\inf_{Q_Y} \sup_f \inf_{Q_{Y|XZ}} G(Q_X^\infty, \tilde{Q}_{Z|X}^\infty, Q_Y, Q_{XYZ}, f) \geq \sup_{Q_{Z|X}} \inf_{Q_Y} \sup_f \inf_{Q_{Y|XZ}} G(Q_X^\infty, Q_{Z|X}, Q_Y, Q_{XYZ}, f) - \delta$$

and there is a sequence $\tilde{Q}_{Z|X}^{(n')}$ converging to $\tilde{Q}_{Z|X}^\infty$. Let

$$\tilde{Q}_Y^{(n')} = \arg \min_{\bar{Q}_Y} \max_f \min_{\substack{Q_{XYZ}: \\ Q_X = Q_X^{(n')} \\ Q_{Z|X} = \tilde{Q}_{Z|X}^{(n')} \\ Q_Y = \bar{Q}_Y}} G^{n'}(Q_{XYZ}, f)$$

and by considering a further subsequence we may assume that $\tilde{Q}_Y^{(n')} \to \tilde{Q}_Y^\infty$. Then there exists $\tilde{f}^\infty$ so that

$$\inf_{Q_{Y|XZ}} G(Q_X^\infty, \tilde{Q}_{Z|X}^\infty, \tilde{Q}_Y^\infty, Q_{Y|XZ}, \tilde{f}^\infty) \geq \max_f \inf_{Q_{Y|XZ}} G(Q_X^\infty, \tilde{Q}_{Z|X}^\infty, \tilde{Q}_Y^\infty, Q_{Y|XZ}, f)$$

and we set $\tilde{f}^{(n')} = \tilde{f}^\infty$. Let

$$Q_{XYZ}^{(n')} = \arg \min_{\substack{Q_{XYZ}: \\ Q_X = Q_X^{(n')} \\ Q_{Z|X} = \tilde{Q}_{Z|X}^{(n')} \\ Q_Y = Q_Y^{(n')}}} G^{n'}(Q_{XYZ}, \tilde{f}^{(n')})$$

and by considering a further subsequence we may assume that $\tilde{Q}_{XYZ}^{(n')} \to \tilde{Q}_{XYZ}^\infty$. Observe that

$$\begin{aligned}
\theta^{n'}(P_{XY}, d, \Delta, R) &= \max_{Q_{Z|X} \in \mathcal{C}^{n'}(\mathcal{X} \to \mathcal{Z})} \min_{Q_Y} \max_{f \in \mathcal{F}} \min_{Q_{Y|XZ}} G^{n'}(Q_X^{(n')}, Q_{Z|X}, Q_Y, Q_{Y|XZ}, f) \\
&\geq \min_{Q_Y} \max_{f \in \mathcal{F}} \min_{Q_{Y|XZ}} G^{n'}(Q_X^{(n')}, \tilde{Q}_{Z|X}^{(n')}, Q_Y, Q_{Y|XZ}, f) \\
&= \max_{f \in \mathcal{F}} \min_{Q_{Y|XZ}} G^{n'}(Q_X^{(n')}, \tilde{Q}_{Z|X}^{(n')}, \tilde{Q}_Y^{(n')}, Q_{Y|XZ}, f) \\
&\geq \min_{Q_{Y|XZ}} G(Q_X^{(n')}, Q_{Z|X}^{(n')}, Q_Y^{(n')}, Q_{Y|XZ}, \tilde{f}^{(n')}) \\
&= G^{n'}(\tilde{Q}_{XYZ}^{(n')}, \tilde{f}^{(n')}).
\end{aligned}$$

We now verify that

$$\liminf_{n \to \infty} G^{n'}(\tilde{Q}_{XYZ}^{(n')}, \tilde{f}^{(n')}) \geq G(\tilde{Q}_{XYZ}^\infty, \tilde{f}^\infty).$$

If $(\tilde{Q}^{\infty}_{XYZ}, \tilde{f}^{\infty})$ are such that $\mathbb{E}_{\tilde{Q}^{\infty}_{XYZ}}[d(X, \tilde{f}^{\infty}(Y, Z))] \geq \Delta$, then the result holds using the semicontinuity of the information measures. If $(\tilde{Q}^{\infty}_{XYZ}, \tilde{f}^{\infty})$ are such that $\mathbb{E}_{\tilde{Q}^{\infty}_{XYZ}}[d(X, \tilde{f}^{\infty}(Y, Z))] < \Delta$ and $I(\tilde{Q}^{\infty}_{X}; \tilde{Q}^{\infty}_{Z|X}) \geq R$, then the sequence $G^{n'}(\tilde{Q}^{(n')}_{XYZ}, \tilde{f}^{(n')})$ is either $\infty$ or equal to $G(\tilde{Q}^{\infty}_{XYZ}, \tilde{f}^{\infty})$ on account of the continuity of $\mathbb{E}_{Q}[\cdot]$. In the final case that $(\tilde{Q}^{\infty}_{XYZ}, \tilde{f}^{\infty})$ are such that $\mathbb{E}_{\tilde{Q}^{\infty}_{XYZ}}[d(X, \tilde{f}^{\infty}(Y, Z))] < \Delta$ and $I(\tilde{Q}^{\infty}_{X}; \tilde{Q}^{\infty}_{Z|X}) < R$, then we must have $\limsup_{n \to \infty} I(\tilde{Q}^{\infty}_{X}; \tilde{Q}^{\infty}_{Z|X}) + \delta^{n}_{r} < R$. Therefore

$$
\begin{aligned}
\liminf_{n' \to \infty} G^{n'}(Q^{(n')}_{X}, \tilde{Q}^{(n')}_{Z|X}, \tilde{Q}^{(n')}_{Y}, \tilde{Q}^{\infty}_{XYZ}, \tilde{f}^{\infty}) &\geq G(Q^{\infty}_{X}, \tilde{Q}^{\infty}_{Z|X}, \tilde{Q}^{\infty}_{Y}, \tilde{Q}^{\infty}_{XYZ}, \tilde{f}^{\infty}) \\
&\geq \inf_{Q_{XYZ}} G(Q^{\infty}_{X}, \tilde{Q}^{\infty}_{Z|X}, \tilde{Q}^{\infty}_{Y}, Q_{XYZ}, \tilde{f}^{\infty}) \\
&= \sup_{f} \inf_{Q_{XYZ}} G(Q^{\infty}_{X}, \tilde{Q}^{\infty}_{Z|X}, Q^{\infty}_{Y}, Q_{XYZ}, f) \\
&\geq \inf_{Q_Y} \sup_{f} \inf_{Q_{XYZ}} G(Q^{\infty}_{X}, \tilde{Q}^{\infty}_{Z|X}, Q_{Y}, Q_{XYZ}, f) \\
&\geq \sup_{Q_{Z|X}} \inf_{Q_Y} \sup_{f} \inf_{Q_{XYZ}} G(Q^{\infty}_{X}, Q_{Z|X}, Q_{Y}, Q_{XYZ}, f) - \delta \\
&\geq \inf_{Q_X} \sup_{Q_{Z|X}} \inf_{Q_Y} \sup_{f} \inf_{Q_{XYZ}} G(Q_{X}, Q_{Z|X}, Q_{Y}, Q_{XYZ}, f) - \delta \\
&= \theta^{\infty}(P_{XY}, d, \Delta, R) - \delta
\end{aligned}
$$

Hence $\liminf_{n' \to \infty} \theta^{n'}(P_{XY}, d, \Delta, R) \geq \liminf_{n' \to \infty} G(\tilde{Q}^{(n')}_{XYZ}, \tilde{f}^{(n')}) \geq \theta(P_{XY}, d, \Delta, R) - \delta$. Letting $\delta \downarrow 0$ gives the result. ∎

We are now in a position to prove Theorem 3. We will accomplish this by giving an upper bound on the probability of error by considering the error events separately.

*Proof of Theorem 3:* We start by noting that for $n$ sufficiently large the constraint of equation (9) is satisfied. Summing over sequences gives

$$
\begin{aligned}
&\Pr(d(X^n, \hat{X}^n) > \Delta) \\
&= \sum_{\mathbf{x}, \mathbf{y}, \mathbf{z}} \Pr(d(X^n, \hat{X}^n) > \Delta | X^n = \mathbf{x}, Y^n = \mathbf{y}, Z^n = \mathbf{z}) \times \Pr(X^n = \mathbf{x}, Y^n = \mathbf{y}, Z^n = \mathbf{z}) \\
&\leq \sum_{\mathcal{E}_b} \Big[ \Pr(d(X^n, \hat{X}^n) > \Delta | X^n = \mathbf{x}, Y^n = \mathbf{y}, Z^n = \mathbf{z}) \times \Pr(X^n = \mathbf{x}, Y^n = \mathbf{y}, Z^n = \mathbf{z}) \Big] \\
&\quad + \sum_{\mathcal{E}_c} \Pr(X^n = \mathbf{x}, Y^n = \mathbf{y}, Z^n = \mathbf{z}) \\
&\quad + \sum_{\mathcal{E}_d} \Pr(X^n = \mathbf{x}, Y^n = \mathbf{y}, Z^n = \mathbf{z})
\end{aligned}
$$

where the last inequality followed from upper bounding the conditional error probability by 1 in the summations over $\mathcal{E}_c$ and $\mathcal{E}_d$, and by zero (Lemma 13) on $(\mathcal{E}_b \cup \mathcal{E}_c \cup \mathcal{E}_d)^c$ (the sequences omitted from the sum). Next, we bound the sequence probabilities using Lemma 10 on $\mathcal{E}_b$ and $\mathcal{E}_d$ and Lemma 11 on $\mathcal{E}_c$. We bound the conditional error probability on $\mathcal{E}_b$ using Lemma 13.

$$
\begin{aligned}
&\Pr(d(X^n, \hat{X}^n) > \Delta) \\
&\leq \sum_{\mathcal{E}_b} \left[ \exp\left(-n\left(R - J(Q_{\mathbf{xyz}}) - \delta^n_b\right)^+\right) \times P^n_{XY}(\mathbf{x}, \mathbf{y}) \frac{1}{|T^n_{Q^*_{Z|X}(Q_{\mathbf{x}})}(\mathbf{x})|} \right] \\
&\quad + \sum_{\mathcal{E}_c} \exp(-(n+1)^2) \\
&\quad + \sum_{\mathcal{E}_d} P^n_{XY}(\mathbf{x}, \mathbf{y}) \frac{1}{|T^n_{Q^*_{Z|X}(Q_{\mathbf{x}})}(\mathbf{x})|}
\end{aligned}
$$

Observing that the summation over $\mathcal{E}_c$ decays super-exponentially, we may safely omit this term, and use the notation $\dot{\leq}$ to denote inequality to the first order of the exponent. We can rewrite the above by first summing over types and then over sequences within each type class. This gives us

$$
\Pr(d(X^n, \hat{X}^n) > \Delta)
$$

$$
\dot{\leq} \sum_{Q_X} \sum_{Q_Y} \Bigg[ \Bigg( \sum_{Q_{XYZ} \in \mathcal{D}_b} \sum_{(\mathbf{x}, \mathbf{y}, \mathbf{z}) \in T^n_{Q_{XYZ}}} P^n_{XY}(\mathbf{x}, \mathbf{y}) \frac{1}{|T^n_{Q_{Z|X}(Q_\mathbf{x})}(\mathbf{x})|}
$$

$$
\times \exp\left( -n \left( R - J(Q_{\mathbf{xyz}}) - \delta^n_b \right)^+ \right) \Bigg)
$$

$$
+ \Bigg( \sum_{Q_{XYZ} \in \mathcal{D}_d} \sum_{(\mathbf{x}, \mathbf{y}, \mathbf{z}) \in T^n_{Q_{XYZ}}} P^n_{XY}(\mathbf{x}, \mathbf{y}) \frac{1}{|T^n_{Q^*_{Z|X}(Q_\mathbf{x})}(\mathbf{x})|} \Bigg).
$$

Note that in the summation over joint types $Q_{XYZ}$, the marginal types of $X$ and $Y$ are fixed to be those set by the earlier summations. Proceeding in a similar manner as was taken in going from (36) to (40) in the SCCSI proof (with $Z$ taking the role of $S$) we obtain

$$
\Pr(d(X^n, \hat{X}^n) > \Delta)
$$

$$
\leq \sum_{Q_X} \sum_{Q_Y} \Bigg[ \sum_{Q_{XYZ} \in \mathcal{D}_b} \exp\Big( -n \big( D(Q_{XYZ} \| P_{XY} Q_{Z|X})
$$

$$
+ R - J(Q_{XYZ}) - \delta^n_b \big)^+ \Big)
$$

$$
+ \sum_{Q_{XYZ} \in \mathcal{D}_d} \exp\Big( -n D(Q_{XYZ} \| P_{XY} Q_{Z|X}) \Big)
$$

$$
+ \exp(-(n+1)^2 + n \log(|\mathcal{X}||\mathcal{Y}||\mathcal{Z}|)) \Bigg]
$$

Next, we use $a + b \leq 2 \max(a, b)$ to combine the first two terms. We can then upper bound the summations by maximizing over the types, and since the choice of test channel $Q^*_{Z|X}$ and estimation function $f$ were arbitrary, we can optimize to give

$$
\Pr(d(X^n, \hat{X}^n) > \Delta)
$$

$$
\dot{\leq} \Big[ |\mathcal{P}^n(\mathcal{X})| \max_{Q_X} \min_{Q^*_{Z|X}} |\mathcal{P}^n(\mathcal{Y})| \max_{Q_Y} 2 |\mathcal{P}^n(\mathcal{X} \times \mathcal{Y} \times \mathcal{Z})|
$$

$$
\min_{f \in \mathcal{F}} \max_{\substack{Q_{XYZ}: \\ Q_{Z|X} = Q^*_{Z|X}}} \tilde{G}^n[Q_{XYZ}, P_{XY}, f, \Delta, R, n] \Big],
$$

where we used the definition of $G^n$ from Lemma 14, taking $\delta^n_r = n^{-1}(|\mathcal{X}||\mathcal{Z}| + 4) \log(n+1)$. Moving the optimizations into the exponent we get

$$
\Pr(d(X^n, \hat{X}^n) > \Delta)
$$

$$
\dot{\leq} 2 |\mathcal{P}^n(\mathcal{X})||\mathcal{P}^n(\mathcal{Y})||\mathcal{P}^n(\mathcal{X} \times \mathcal{Y} \times \mathcal{Z})| \exp\Big( -n \Big[ \min_{Q_X} \max_{Q^*_{Z|X}} \min_{Q_Y}
$$

$$
\max_{f \in \mathcal{F}} \min_{\substack{Q_{XYZ}: \\ Q_{Z|X} = Q^*_{Z|X}}} G^n[Q_{XYZ}, P_{XY}, f, d, \Delta, R] \Big] \Big).
$$

We can absorb the set cardinalities $\delta_2 = \frac{1}{n}[1 + \log(n+1)^{|\mathcal{X}| + |\mathcal{Y}| + |\mathcal{X}||\mathcal{Y}| + |\mathcal{Y}||\mathcal{Z}|}]$ and observe that in the limit as $n \to \infty$, $\delta_2$ vanishes. Hence we have

$$\liminf_{n \to \infty} -\frac{1}{n} \log \Pr(d(X^n, \hat{X}^n) > \Delta)$$

$$\geq \liminf_{n \to \infty} \min_{Q_X} \max_{\substack{Q_{Z|X} \in \\ \mathcal{C}^n(Q_X, \mathcal{Z})}} \min_{Q_Y} \max_{f \in \mathcal{F}} \min_{Q_{XYZ}}$$

$$G^n\left[Q_{XYZ}, P_{XY}, f, d, \Delta, R\right]$$

$$\geq \inf_{Q_X} \sup_{Q_{Z|X}} \inf_{Q_Y} \sup_{f \in \mathcal{F}} \inf_{Q_{XYZ}} G\left[Q_{XYZ}, P_{XY}, f, \Delta, R\right],$$

where the final line followed from application of Lemma 14. ∎

## Appendix D
### Gaussian Type-classes

For the Gaussian case ($\mathcal{X} = \mathcal{Y} = \mathbb{R}$), we need the following definitions[12]. These are a modification of the Gaussian types used by Arikan and Merhav [34]. The difference is that here the type-classes are disjoint and the conditions specifying joint types are independent. This significantly simplifies the subsequent analysis and might prove useful in other applications.

**Definition 1.** *For a given $0 < \epsilon < 1$ and $\sigma_X^2 > 0$, a Gaussian type-class $T_{\sigma_X^2}^\epsilon$ is defined as the set of $n$-sequences*

$$T_{\sigma_X^2}^\epsilon = \left\{ \mathbf{x} \in \mathbb{R}^n : |\mathbf{x}^t \mathbf{x} - n\sigma_X^2| \leq n\epsilon \right\}.$$

*For such a type-class, it can be shown that (see the calculation at the end of this section)*

$$\left(1 - \frac{2\sigma_X^4}{n\epsilon^2}\right) \exp\left(n\left(h(\sigma_X^2) - \frac{\epsilon}{2\sigma_X^2}\right)\right)$$

$$\leq \mathrm{Vol}(T_{\sigma_X^2}^\epsilon) \leq \exp\left(n\left(h(\sigma_X^2) + \frac{\epsilon}{2\sigma_X^2}\right)\right). \tag{61}$$

*Similarly, for a given $0 < \epsilon < 1$ and covariance matrix*

$$K = \begin{bmatrix} \sigma_X^2 & \rho\sigma_X\sigma_Y \\ \rho\sigma_X\sigma_X & \sigma_Y^2 \end{bmatrix},$$

*with non-zero variances, a joint Gaussian type-class $T_K^\epsilon$ is defined as the set of pairs of $n$-sequences*

$$T_K^\epsilon = \left\{ (\mathbf{x}, \mathbf{y}) \in \mathbb{R}^n \times \mathbb{R}^n : |\mathbf{x}^t \mathbf{x} - n\sigma_X^2| \leq n\epsilon \right.$$

$$|\mathbf{y}^t \mathbf{y} - n\sigma_Y^2| \leq n\epsilon$$

$$\left. |\mathbf{x}^t \mathbf{y} - \rho\sqrt{\mathbf{x}^t \mathbf{x} \mathbf{y}^t \mathbf{y}}| \leq \epsilon\sqrt{\mathbf{x}^t \mathbf{x} \mathbf{y}^t \mathbf{y}} \right\}.$$

*This set has the corresponding volume bound*

$$\mathrm{Vol}(T_K^\epsilon) \leq \exp\left(n\left(h(K) + o_\epsilon(1)\right)\right), \tag{62}$$

*where we use $o_\epsilon(1)$ to denote a quantity $g(\epsilon) > 0$ having the property that $\lim_{\epsilon \to 0} g(\epsilon) = 0$.*

*Furthermore, for a given $\mathbf{x} \in T_{\sigma_X^2}^\epsilon$, we define the conditional Gaussian type-class $T_K^\epsilon(\mathbf{x})$ as the $\mathbf{x}$-set of $n$-sequences*

$$T_K^\epsilon(\mathbf{x}) = \{ \mathbf{y} \in \mathbb{R}^n : (\mathbf{x}, \mathbf{y}) \in T_K^\epsilon \}.$$

---

[12]For more than two jointly Gaussian random variables, these definitions can be extended in the obvious way.

*For this set one can show that (see the calculation at the end of this section)*

$$
\begin{aligned}
&\text{Vol}(T_K^\epsilon(\mathbf{x})) \\
&\geq \left(1 - \frac{1}{no_\epsilon(1)} + o_\epsilon(1)\right) \exp(n(h(K_{Y|X}) - \tilde{f}_\epsilon)).
\end{aligned}
\tag{63}
$$

*where $\tilde{f}_\epsilon$ is an $o_\epsilon(1)$ term whose value is determined in the proof. In Appendix D-3 we show for a Gaussian distribution $f_K(\cdot, \cdot)$, if $(\mathbf{x}, \mathbf{y}) \in T_{\tilde{K}}^\epsilon$, where $\tilde{K}$ is any positive definite covariance matrix, then*

$$
f_{XY}^n(\mathbf{x}, \mathbf{y}) \leq \exp\left(-n\left(D(\tilde{K}\|K) + h(Q_{\tilde{K}}) - o_\epsilon(1)\right)\right).
\tag{64}
$$

The analysis for the Gaussian case requires that we "quantize" the space of $3 \times 3$ covariance matrices. Unlike discrete memoryless sources, Gaussian sources require use of a "bounding box" to limit the number of types. To this end, fix $0 < M_L < 1$ and $M_U > M_L$; both will be chosen later. For a fixed $0 < \epsilon < M_L$ define $\sigma^2(i) = M_L + 2i\epsilon$ and for $i, j, \epsilon$, given define $\eta_{ij}(r) = \sqrt{\sigma^2(i)\sigma^2(j)}(-1 + 2\epsilon(r-1))$. We will consider type-classes indexed by matrices of the form

$$
K_\epsilon(i, j, k, r, s, t) = \begin{bmatrix} \sigma^2(i) & \eta_{ij}(r) & \eta_{ik}(s) \\ \eta_{ij}(r) & \sigma^2(j) & \eta_{jk}(t) \\ \eta_{ik}(s) & \eta_{jk}(t) & \sigma^2(k) \end{bmatrix}
$$

and $i, j, k, r, s, t \geq 1$; note that not all of these matrices are positive semidefinite.

We let $\mathcal{P}_X^\epsilon = \{i : \exists \mathbf{x} \in T_{\sigma^2(i)}^\epsilon \text{ with } \mathbf{x}^t\mathbf{x} \leq M_U\}$ and similarly $\mathcal{P}_{XYZ}^\epsilon = \{(i, j, k, r, s, t) : \exists (\mathbf{x}, \mathbf{y}, \mathbf{z}) \in T_{K(i,j,k,r,s,t)}^\epsilon \text{ with } \mathbf{x}^t\mathbf{x} \leq nM_U \text{ and } \mathbf{y}^t\mathbf{y} \leq nM_U \text{ and } \mathbf{z}^t\mathbf{z} \leq nM_U\}$, where $M_U \gg M_L$. With $\mathcal{S}_L = \{(\mathbf{x}, \mathbf{y}, \mathbf{z}) : \mathbf{x}^t\mathbf{x} \leq n(M_L + \epsilon) \text{ or } \mathbf{y}^t\mathbf{y} \leq n(M_L + \epsilon) \text{ or } \mathbf{z}^t\mathbf{z} \leq n(M_L + \epsilon)\}$, $\mathcal{S}_U = \{(\mathbf{x}, \mathbf{y}, \mathbf{z}) : \mathbf{x}^t\mathbf{x} > nM_U \text{ or } \mathbf{y}^t\mathbf{y} > nM_U \text{ or } \mathbf{z}^t\mathbf{z} > nM_U\}$, the union of the shells $T_{K(i,j,k,r,s,t)}^\epsilon$, and the set $\mathcal{S}_L$ cover $\mathbb{R}^{3n}$ entirely and we define $\mathcal{R}^{3n} = \mathbb{R}^{3n} \backslash (\mathcal{S}_L \cup \mathcal{S}_U)$. We denote by $\nu(\mathbf{x})$ the *index* of the shell containing the string $\mathbf{x}$, i.e. $\mathbf{x} \in T_{\sigma^2(\nu(\mathbf{x}))}^\epsilon$, which is uniquely defined almost everywhere in $\mathcal{R}^{3n}$.

*1) Proof of* (61)*:* Let $X \sim \mathcal{N}(0, \sigma_X^2)$. Then

$$
\begin{aligned}
1 &\geq \int_{T_{\sigma_X^2}^\epsilon} (2\pi\sigma_X^2)^{-\frac{n}{2}} \exp\left(-\frac{\mathbf{x}^t\mathbf{x}}{2\sigma_X^2}\right) d\mathbf{x} \\
&\geq \int_{T_{\sigma_X^2}^\epsilon} (2\pi\sigma_X^2)^{-\frac{n}{2}} \exp\left(-\frac{n(\sigma_X^2 + \epsilon)}{2\sigma_X^2}\right) d\mathbf{x} \\
&= \exp\left(-n\left(\frac{1}{2}\log(2\pi\sigma_X^2) + \frac{1}{2}\right) - \frac{n\epsilon}{2\sigma_X^2}\right) \text{Vol}(T_{\sigma_X^2}^\epsilon),
\end{aligned}
$$

which gives the upper bound. For the lower bound,

$$
\begin{aligned}
\Pr(T_{\sigma_X^2}^\epsilon) &= \int_{T_{\sigma_X^2}^\epsilon} (2\pi\sigma_X^2)^{-\frac{n}{2}} \exp\left(-\frac{\mathbf{x}^t\mathbf{x}}{2\sigma_X^2}\right) d\mathbf{x} \\
&\leq \int_{T_{\sigma_X^2}^\epsilon} (2\pi\sigma_X^2)^{-\frac{n}{2}} \exp\left(-\frac{n(\sigma_X^2 - \epsilon)}{2\sigma_X^2}\right) d\mathbf{x} \\
&= \text{Vol}(T_{\sigma_X^2}^\epsilon) \exp\left(-n\left(\frac{1}{2}\log(2\pi e\sigma_X^2)\right) + \frac{n\epsilon}{2\sigma_X^2}\right).
\end{aligned}
$$

Conversely, by Chebyshev's inequality

$$
\begin{aligned}
1 - \Pr(T^\epsilon_{\sigma^2_X}) &= \Pr\left(|\mathbf{x}^t\mathbf{x} - n\sigma^2_X| > n\epsilon\right) \\
&\leq \mathbb{E}\left[\frac{(\mathbf{x}^t\mathbf{x} - n\sigma^2_X)^2}{n^2\epsilon^2}\right] \\
&= \frac{2\sigma^4_X}{n\epsilon^2}
\end{aligned}
$$

Combining these two calculations gives the lower bound. ∎

*2) Proof of* (63): Let $\mathbf{x} \in T^\epsilon_{\sigma^2_X}$, then

$$
T^\epsilon_K(\mathbf{x}) = \Big\{ \mathbf{y} \in \mathbb{R}^n : |\mathbf{y}^t\mathbf{y} - n\sigma^2_Y| \leq n\epsilon
$$
$$
\left|\frac{\mathbf{y}^t\mathbf{x}}{n} - \rho\sqrt{\frac{\mathbf{x}^t\mathbf{x}}{n}\frac{\mathbf{y}^t\mathbf{y}}{n}}\right| \leq \epsilon\sqrt{\frac{\mathbf{x}^t\mathbf{x}}{n}\frac{\mathbf{y}^t\mathbf{y}}{n}} \Big\}.
$$

By the triangle inequality

$$
\left|\frac{\mathbf{y}^t\mathbf{x}}{n} - \rho\sqrt{\frac{\mathbf{x}^t\mathbf{x}}{n}\frac{\mathbf{y}^t\mathbf{y}}{n}}\right| \leq \left|\frac{\mathbf{y}^t\mathbf{x}}{n} - \rho\sqrt{\frac{\mathbf{x}^t\mathbf{x}}{n}}\sigma_Y\right| + \left|\rho\sqrt{\frac{\mathbf{x}^t\mathbf{x}}{n}}\sigma_Y - \rho\sqrt{\frac{\mathbf{x}^t\mathbf{x}}{n}\frac{\mathbf{y}^t\mathbf{y}}{n}}\right|,
$$

whence

$$
T^\epsilon_K(\mathbf{x}) \supset A(\mathbf{x}) \triangleq \Big\{ \mathbf{y} \in \mathbb{R}^n : |\mathbf{y}^t\mathbf{y} - n\sigma^2_Y| \leq n\epsilon
$$
$$
\left|\frac{\mathbf{y}^t\mathbf{x}}{n} - \rho\sqrt{\frac{\mathbf{x}^t\mathbf{x}}{n}}\sigma_Y\right| \leq \frac{\epsilon}{2}\sqrt{\sigma^2_Y - \epsilon}\sqrt{\frac{\mathbf{x}^t\mathbf{x}}{n}}
$$
$$
\left|\rho\sqrt{\frac{\mathbf{x}^t\mathbf{x}}{n}}\sigma_Y - \rho\sqrt{\frac{\mathbf{x}^t\mathbf{x}}{n}\frac{\mathbf{y}^t\mathbf{y}}{n}}\right| \leq \frac{\epsilon|\rho|}{2}\sqrt{\sigma^2_Y - \epsilon}\sqrt{\frac{\mathbf{x}^t\mathbf{x}}{n}} \Big\}.
$$

Let $\mathbf{V}$ be a Gaussian random vector whose law is $\mathcal{N}(0, I\sigma^2_Y(1-\rho^2))$, and let $\mathbf{Y} = \frac{\rho\sigma_Y}{\sqrt{\frac{\mathbf{x}^t\mathbf{x}}{n}}}\mathbf{x} + \mathbf{V}$. Applying the union bound gives

$$
\Pr(A(\mathbf{x})^c) \leq \Pr\left(|\mathbf{Y}^t\mathbf{Y} - n\sigma^2_Y| > n\epsilon\right) + \Pr\left(\left|\frac{\mathbf{Y}^t\mathbf{x}}{n} - \rho\sqrt{\frac{\mathbf{x}^t\mathbf{x}}{n}}\sigma_Y\right| > \frac{\epsilon}{2}\sqrt{\sigma^2_Y - \epsilon}\sqrt{\frac{\mathbf{x}^t\mathbf{x}}{n}}\right)
$$
$$
+ \Pr\left(\left|\rho\sqrt{\frac{\mathbf{x}^t\mathbf{x}}{n}}\sigma_Y - \rho\sqrt{\frac{\mathbf{x}^t\mathbf{x}}{n}\frac{\mathbf{Y}^t\mathbf{Y}}{n}}\right| > \frac{\epsilon|\rho|}{2}\sqrt{\sigma^2_Y - \epsilon}\sqrt{\frac{\mathbf{x}^t\mathbf{x}}{n}}\right).
$$

The event in the third probability on the right is equivalent to

$$
\Big\{ \left|\frac{\mathbf{Y}^t\mathbf{Y}}{n} - \sigma^2_Y - \frac{\epsilon^2(\sigma^2_Y - \epsilon)}{4}\right| > \epsilon\sigma_Y\sqrt{\sigma^2_Y - \epsilon} \Big\}.
$$

Using this fact and bounding each of the probabilities using Chebyshev's inequality yields

$$
\Pr(A(\mathbf{x})^c) \le \mathbb{E}\left[\frac{(\mathbf{Y}^t\mathbf{Y} - n\sigma_Y^2)^2}{n^2\epsilon^2}\right] + \mathbb{E}\left[\frac{(\frac{\mathbf{Y}^t\mathbf{x}}{n}\sqrt{\frac{n}{\mathbf{x}^t\mathbf{x}}} - \rho\sigma_Y)^2}{\epsilon^2(\sigma_Y^2 - \epsilon)/4}\right]
$$
$$
+ \mathbb{E}\left[\frac{(\frac{\mathbf{Y}^t\mathbf{y}}{n} - \sigma_Y^2 - \epsilon^2(\sigma_Y^2 - \epsilon)/4)^2}{\epsilon^2\sigma_Y^2(\sigma_Y^2 - \epsilon)}\right]
$$
$$
= \frac{2\sigma_Y^4}{n\epsilon^2} + \frac{\sigma_Y^2(1-\rho^2)}{n(\epsilon^2(\sigma_Y^2 - \epsilon))/4} + \frac{2\sigma_Y^4}{n\epsilon^2\sigma_Y^2(\sigma_Y^2 - \epsilon)} + \frac{\epsilon^2(\sigma_Y^2 - \epsilon)}{16\sigma_Y^2}
$$
$$
= \frac{1}{no_\epsilon(1)} + o_\epsilon(1) \tag{*}
$$

To bound the volume we note that under the law above

$$
\Pr(A(\mathbf{x})) = \int_{A(\mathbf{x})} f_{\mathbf{V}}\left(\mathbf{y} - \frac{\sqrt{n}\rho\sigma_Y\mathbf{x}}{\sqrt{\mathbf{x}^t\mathbf{x}}}\right) d\mathbf{y}
$$
$$
= \int_{A(\mathbf{x})} (2\pi\sigma_Y^2(1-\rho^2))^{-n/2}\exp\left(\frac{-\sum_i(y_i - \frac{\sqrt{n}\rho\sigma_Y}{\sqrt{\mathbf{x}^t\mathbf{x}}}x_i)^2}{2(\sigma_Y^2(1-\rho^2))}\right) d\mathbf{y}.
$$

We can get an upper bound on the density by lower bounding the summand in the exponent

$$
\sum_i\left(y_i - \frac{\sqrt{n}\rho\sigma_Y}{\sqrt{\mathbf{x}^t\mathbf{x}}}x_i\right)^2 = \mathbf{y}^t\mathbf{y} - 2\rho\sigma_Y\frac{\sqrt{n}}{\sqrt{\mathbf{x}^t\mathbf{x}}}\mathbf{y}^t\mathbf{x} + n\rho^2\sigma_Y^2
$$
$$
\ge n(\sigma_Y^2 - \epsilon) + n\rho^2\sigma_Y^2 - 2\rho\sigma_Y n\left(\rho\sigma_Y + \mathrm{sgn}(\rho)\epsilon/2\sqrt{\sigma_Y^2 - \epsilon}\right)
$$
$$
= n(\sigma_Y^2(1-\rho^2) - f_\epsilon(\rho, \sigma_Y))
$$

where $f_\epsilon(\rho, \sigma_Y) = \epsilon(1 + \rho\,\mathrm{sgn}(\rho)\sigma_Y\sqrt{\sigma_Y^2 - \epsilon})$ goes to zero with $\epsilon$. Thus

$$
\Pr(A(\mathbf{x})) \le \mathrm{Vol}(A(\mathbf{x}))\exp\left(-n\left(\frac{1}{2}\log(2\pi\sigma_Y^2(1-\rho^2)) - \frac{1}{2} - \tilde{f}_\epsilon(\rho, \sigma_Y)\right)\right)
$$
$$
= \mathrm{Vol}(A(\mathbf{x}))\exp\left(-n\left(\frac{1}{2}\log(2\pi e\sigma_Y^2(1-\rho^2)) - \tilde{f}_\epsilon(\rho, \sigma_Y)\right)\right),
$$

where $\tilde{f}_\epsilon = f_\epsilon/(2(\sigma_Y^2(1-\rho^2)))$. Combining this with $(*)$ and using the fact that $\mathrm{Vol}(T_K^\epsilon(\mathbf{x})) \ge \mathrm{Vol}(A(\mathbf{x}))$ gives the result. ∎

*3) Proof of* (64): Let $(X, Y) \sim \mathcal{N}(0, K)$ and $(\mathbf{x}, \mathbf{y}) \in T_{\tilde{K}}^\epsilon$. Then

$$
f(\mathbf{x}, \mathbf{y}) = [(2\pi)^2|K|]^{-\frac{n}{2}}
$$
$$
\times \exp\left(-\frac{1}{2(1-\rho^2)}\left(\frac{\mathbf{x}^t\mathbf{x}}{\sigma_X^2} + \frac{\mathbf{y}^t\mathbf{y}}{\sigma_Y^2} - \frac{2\rho\mathbf{x}^t\mathbf{y}}{\sigma_X\sigma_Y}\right)\right).
$$

Applying the bounds from the definition of $T_{\tilde{K}}^\epsilon$ allows us to continue the inequality with

$$
\le \exp\left(-\frac{n}{2}\log((2\pi)^2|K|) - \frac{1}{2(1-\rho^2)}\left(\frac{n(\tilde{\sigma}_X^2 - \epsilon)}{\sigma_X^2}\right.\right.
$$
$$
\left.\left. + \frac{n(\tilde{\sigma}_Y^2 - \epsilon)}{\sigma_Y^2} - \frac{2\rho n\sqrt{(\tilde{\sigma}_X^2 + \epsilon)(\tilde{\sigma}_Y^2 + \epsilon)}(\tilde{\rho} + \mathrm{sgn}(\rho)\epsilon)}{\sigma_X\sigma_Y}\right)\right).
$$

For $f_\epsilon(\sigma_X, \sigma_Y, \tilde{\sigma}_X, \tilde{\sigma}_Y, \rho, \tilde{\rho})$ (which goes to zero with $\epsilon$), we can write

$$\leq \exp -n \left( \frac{1}{2} \Big( \log((2\pi)^2 |K|) + \frac{\tilde{\sigma}_X^2}{\sigma_X^2(1-\rho^2)} + \frac{\tilde{\sigma}_Y^2}{\sigma_Y^2(1-\rho^2)} \right.$$
$$\left. - \frac{2\rho \tilde{\sigma}_X \tilde{\sigma}_Y \tilde{\rho}}{\sigma_X \sigma_Y (1-\rho^2)} - f_\epsilon(\sigma_X, \sigma_Y, \tilde{\sigma}_X, \tilde{\sigma}_Y, \rho, \tilde{\rho}) \Big) \right).$$

Finally, using the identity

$$D(\tilde{K}||K) = \frac{1}{2} \left( \log \frac{|K|}{|\tilde{K}|} + \mathrm{Tr}(K^{-1}\tilde{K}) - 2 \right)$$

gives

$$f(\mathbf{x}, \mathbf{y}) \leq \exp -n \Big( D(\tilde{K}||K) + \frac{1}{2} \log(2\pi e)^2 |\tilde{K}|$$
$$- f_\epsilon(\sigma_X, \sigma_Y, \tilde{\sigma}_X, \tilde{\sigma}_Y, \rho, \tilde{\rho}) \Big) \blacksquare$$

## Appendix E
## Proof of Theorem 5

### A. Scheme

Let $\epsilon > 0$ and $M_L, M_U$ as defined in Appendix D. For each blocklength $n$, and for each shell of $n$-length $\mathbf{x}$ sequences, $T^\epsilon_{\sigma^2(i)}$ we choose a Gaussian test channel. The test channel is specified by selecting integers $k(i)$ and $s(i)$ (such that $\sigma^2(k(i)) < M_U$) so that if $X \sim \mathcal{N}(0, \sigma^2(i))$ is the input to the channel then $(X, Z) \sim \mathcal{N}(0, \overline{\sigma^2(i)})$; where the bar applied to a scalar results in

$$\overline{\sigma^2(i)} = \begin{bmatrix} \sigma^2(i) & \eta_{i,k(i)}(s(i)) \\ \eta_{i,k(i)}(s(i)) & \sigma^2(k(i)) \end{bmatrix}. \tag{65}$$

The codebook for the $i$th shell of $\mathbf{x}$ sequences is a randomly chosen set of codewords, $B^n(i)$, selected in the following way. The size of $B^n(i)$ is an integer satisfying

$$\exp(n(I_{\overline{\sigma^2(i)}}(X; Z) + 2g_\epsilon)) \leq |B^n(i)| \leq \exp(n(I_{\overline{\sigma^2(i)}}(X; Z) + 3g_\epsilon)) \tag{66}$$

where $g_\epsilon = \tilde{f}_\epsilon + \epsilon/2\sigma^2(k(i))$ (cf. (63)) and the codewords are chosen uniformly from the shell $T_{\sigma^2(k(i))}$.

For $\mathbf{x} \in T^\epsilon_{\sigma^2(i)}$, define $Z(\mathbf{x}) : T^\epsilon_{\sigma^2(i)} \to B^n(i)$ as follows. We can cover the shell $T^\epsilon_{\sigma^2(i)}$ with conditional type-classes $T^\epsilon_{\overline{\sigma^2(i)}}(B^n(i)[j])$, where $B^n(i)[j]$ is the $j$th codeword. This covering induces a partition of sequences in $T^\epsilon_{\sigma^2(i)}$, with the partition being based on the set of possible codewords in $B^n(i)$ that have the correct joint type with the sequences. For each set generated by this partition, we choose the codeword for that set uniformly among the covering conditional type-classes. For the sets not covered by any class, the codeword is selected at random from $B^n(i)$. We define $Z^n = Z(X^n)$. Finally, let the encoder's message set be defined as $\mathcal{M} = \mathcal{M}_1 \times \mathcal{M}_2$, where

$$\mathcal{M}_1 = \{1, \ldots, M_1 \triangleq \exp(nR)\}, \mathcal{M}_2 = \{1, 2, \ldots, |\mathcal{P}^\epsilon_X|\}.$$

*Operation of the Encoder:* To encode a sequence $\mathbf{x} \in T^\epsilon_{\sigma^2(i)}$, the encoder sends $i$, the "type" of $\mathbf{x}$ and an index, $U(Z(\mathbf{x}))$, of the codeword $Z(\mathbf{x})$. If $\log |B^n(i)| \geq nR$ we use random binning of the codewords, and $U(Z(\mathbf{x}))$ denotes the element of $\mathcal{M}_1$ to which $Z(\mathbf{x})$ is mapped. For sequences with $\mathbf{x}^t \mathbf{x} \notin (n(M_L + \epsilon), nM_U]$ the encoder declares an error. The encoder can be expressed mathematically as

$$\psi(\mathbf{x}) = (U(Z(\mathbf{x})), i) \text{ for } \mathbf{x} \in T^\epsilon_{\sigma^2(i)} \tag{67}$$

*Operation of the Decoder:* The decoder operates in a two-step manner. First it attempts to recover the codeword $Z^n$:

1) If $\log |B^n(i)| < nR$ then $Z^n$ can be decoded without error,
2) If $\log |B^n(i)| \geq nR$ the decoder receives a bin index and uses the side information to pick the $\mathbf{z}$ from the bin by searching for a $\hat{\mathbf{z}}$ in the received bin so that among all $\tilde{\mathbf{z}}$ in the bin, $\rho_{\hat{\mathbf{z}},\mathbf{y}}^2 < \rho_{\tilde{\mathbf{z}},\mathbf{y}}^2$. If there is no such $\hat{\mathbf{z}}$, the encoder picks uniformly at random from the bin.

Let

$$\varphi_1(l, i, \mathbf{y}) = \begin{cases} \hat{\mathbf{z}} & \hat{\mathbf{z}} \in \mathrm{Bin}(l) \text{ and } \forall \tilde{\mathbf{z}} \neq \hat{\mathbf{z}} \in \mathrm{Bin}(l), \\ & \qquad \rho_{\hat{\mathbf{z}},\mathbf{y}}^2 < \rho_{\tilde{\mathbf{z}},\mathbf{y}}^2 \\ \text{any } \tilde{\mathbf{z}} & \text{if no such } \hat{\mathbf{z}} \in \mathrm{Bin}(l) \end{cases} \tag{68}$$

where $\mathrm{Bin}(l) = \{\mathbf{z} : \mathbf{z} \in B^n(i) \text{ and } U(\mathbf{z}) = l\}$ denotes the set of codewords that are assigned to bin $l$. The *marginal* types $i, j$ of $\mathbf{x}$ and $\mathbf{y}$ are known, and for each pair $i, j$ we choose an estimation function. We restrict our attention to estimation functions that are linear in the side information and the codeword, i.e. $\lambda_{i,j}(y, z) = \alpha(i, j)y + \beta(i, j)z$, where $\alpha(i, j) = \nu\epsilon, \beta(i, j) = \kappa\epsilon$ for integers $\nu, \kappa$ so that $\alpha(i, j), \beta(i, j) \in [-M_\lambda, M_\lambda]$. $\alpha$ and $\gamma$ will be optimized later and $M_\lambda > 0$ is an arbitrary positive constant. For the second step the decoder uses the estimation function, $\lambda$, to combine the side information $\mathbf{y}$ with codeword $\mathbf{z}$ to give the reproduction $\hat{\mathbf{x}}$. This is expressed mathematically as

$$\varphi(l, i, \mathbf{y}) = \hat{\mathbf{x}} \tag{69}$$
$$\text{s.t. } \hat{\mathbf{x}}_m = \alpha(i, \nu(\mathbf{y}))\mathbf{y}_m + \beta(i, \nu(\mathbf{y}))\varphi_1(l, i, \mathbf{y})_m.$$

## B. Key events

The following subsets of $\mathbb{R}^{3n}$ will be of interest.

$$\mathcal{E}_b = \left\{(\mathbf{x}, \mathbf{y}, \mathbf{z}) \in \mathcal{R}^{3n} : \mathbf{z} \in T_{\sigma^2(\nu(\mathbf{x}))}^\epsilon(\mathbf{x}), \right.$$
$$\left. \frac{1}{n}\|\mathbf{x} - \lambda_{\nu(\mathbf{x}),\nu(\mathbf{y})}(\mathbf{y}, \mathbf{z})\|_2^2 < \Delta, \log |B^n(\nu(\mathbf{x}))| \geq nR \right\}$$
$$\mathcal{E}_c = \left\{(\mathbf{x}, \mathbf{y}, \mathbf{z}) \in \mathcal{R}^{3n} : \mathbf{z} \notin T_{\sigma^2(\nu(\mathbf{x}))}^\epsilon(\mathbf{x}) \right\}$$
$$\mathcal{E}_d = \left\{(\mathbf{x}, \mathbf{y}, \mathbf{z}) \in \mathcal{R}^{3n} : \mathbf{z} \in T_{\sigma^2(\nu(\mathbf{x}))}^\epsilon(\mathbf{x}), \right.$$
$$\left. \frac{1}{n}\|\mathbf{x} - \lambda_{\nu(\mathbf{x}),\nu(\mathbf{y})}(\mathbf{y}, \mathbf{z})\|_2^2 \geq \Delta \right\}.$$

On $\mathcal{E}_b$, the distortion constraint is violated only if there is a decoding error. On $\mathcal{E}_c$ we say there is a "covering" error: the encoder cannot find a codeword with the desired joint type with the source sequence. On $\mathcal{E}_d$, the distortion constraint will be violated even if the codeword is decoded correctly by the decoder.

For $\mathbf{x} \in T_{\sigma^2(i)}^\epsilon$, $F$ is defined to be the event that there exists $\tilde{\mathbf{z}} \in B^n(i)$ such that $\tilde{\mathbf{z}} \in T_{\sigma^2(i)}^n(\mathbf{x})$.

## C. Error Probability Calculation

We will first state several useful lemmas, which are "Gaussian versions" of the discrete memoryless Wyner-Ziv lemmas.

**Lemma 15.** *Let $X^n, Y^n, Z^n = Z(X^n)$ be generated according to our scheme and suppose that $A \subset (\mathcal{E}_c)^c \cap \mathcal{R}^{3n}$. Then*

$$\Pr((X^n, Y^n, Z^n) \in A)$$
$$\leq \int_A f_{XY}^n(\mathbf{x}, \mathbf{y}) \frac{1}{\mathrm{Vol}(T_{\sigma^2(\nu(\mathbf{x}))}^\epsilon(\mathbf{x}))} d\mathbf{x}\mathbf{y}\mathbf{z}. \tag{70}$$

*Proof:* For the $\mathbf{x}, \mathbf{y}, \mathbf{z} \in A$ in this lemma, $\{X^n = \mathbf{x}, Y^n = \mathbf{y}, Z^n = \mathbf{z}\}$ implies that the event $F$ has occurred. Let $A_{XY}$ be the projection of $A$ onto $XY$ space, i.e. $A_{XY} = \{(\mathbf{x}, \mathbf{y}) : (\mathbf{x}, \mathbf{y}, \mathbf{z}) \in A \text{ for some } \mathbf{z}\}$ and $A^{\mathbf{x}, \mathbf{y}} = \{\mathbf{z} : (\mathbf{x}, \mathbf{y}, \mathbf{z}) \in A\}$. Then

$$
\begin{aligned}
\Pr((X^n, & Y^n, Z^n) \in A) \\
&= \Pr((X^n, Y^n, Z^n) \in A, F) \\
&= \int_{A_{XY}} f_{XY}^n(\mathbf{x}, \mathbf{y}) \Pr(F | X^n = \mathbf{x}, Y^n = \mathbf{y}) \\
&\quad \times \Pr(Z^n \in A^{\mathbf{x}, \mathbf{y}} | X^n = \mathbf{x}, Y^n = \mathbf{y}, F) d\mathbf{xy} \\
&\leq \int_{A_{XY}} f_{XY}^n(\mathbf{x}, \mathbf{y}) \\
&\quad \times \Pr(Z^n \in A^{\mathbf{x}, \mathbf{y}} | X^n = \mathbf{x}, Y^n = \mathbf{y}, F) d\mathbf{xy} \\
&= \int_{A_{XY}} f_{XY}^n(\mathbf{x}, \mathbf{y}) \int_{A^{\mathbf{x}, \mathbf{y}}} f_{Z|X,Y,F}(\mathbf{z} | \mathbf{x}, \mathbf{y}) d\mathbf{z} d\mathbf{xy} \\
&= \int_A f_{XY}^n(\mathbf{x}, \mathbf{y}) \frac{1}{\mathrm{Vol}(T^\epsilon_{\sigma^2(\nu(\mathbf{x}))}(\mathbf{x}))} d\mathbf{xyz}
\end{aligned}
$$

where in the final line we used that conditional on $F$ and $X^n = \mathbf{x}$, $Z^n$ is uniformly distributed over $T^\epsilon_{\sigma^2(\nu(\mathbf{x}))}(\mathbf{x})$ and independent of $Y$. ∎

**Lemma 16.** *Let $X^n, Y^n, Z^n = Z(X^n)$ be generated according to our scheme. Then for $n$ sufficiently large*

$$
\Pr((X^n, Y^n, Z^n) \in \mathcal{E}_c) \leq |\mathcal{P}_X^\epsilon| \exp(-\exp(no_\epsilon(1))) \tag{71}
$$

*Proof:* For $(\mathbf{x}, \mathbf{y}, \mathbf{z}) \in \mathcal{E}_c$, $\{X^n = \mathbf{x}, Y^n = \mathbf{y}, Z^n = \mathbf{z}\}$ implies that the event $F^c$ has occurred. Thus

$$
\begin{aligned}
\Pr((X^n, & Y^n, Z^n) \in \mathcal{E}_c) \\
&= \Pr((X^n, Y^n, Z^n) \in \mathcal{E}_c, F^c) \\
&\leq \sum_{T^\epsilon_{\sigma^2(i)} \in \mathcal{P}_X^\epsilon} \Pr(X^n \in T^\epsilon_{\sigma^2(i)}) \Pr(F^c | X^n \in T^\epsilon_{\sigma^2(i)}) \\
&\quad \times \Pr((X^n, Y^n, Z^n) \in \mathcal{E}_c | X^n \in T^\epsilon_{\sigma^2(i)}, F^c) \\
&\leq \sum_{T^\epsilon_{\sigma^2(i)} \in \mathcal{P}_X^\epsilon} \Pr(F^c | X^n \in T^\epsilon_{\sigma^2(i)}) \Pr(X^n \in T^\epsilon_{\sigma^2(i)}).
\end{aligned}
$$

$\Pr(F^c | X^n \in T^\epsilon_{\sigma^2(i)})$ is the probability that there is no $\tilde{\mathbf{z}} \in B^n(i)$ so that $\tilde{\mathbf{z}} \in T^\epsilon_{\sigma(i)}(X^n)$. We will now give an upper bound on this probability using the properties of the codeword set. Let $m = |B^n(i)|$ and $B^n(i)[j]$ be the $j$th codeword in the set $B^n(i)$. Then

$$
\begin{aligned}
\Pr(F^c | X^n \in T^\epsilon_{\sigma^2(i)}) &= \prod_{j=1}^m \Pr(B^n(i)[j] \notin T^\epsilon_{\sigma^2(i)}(X^n)) \\
&= \prod_{j=1}^m [1 - \Pr(B^n(i)[j] \in T^\epsilon_{\sigma^2(i)}(X^n))] \\
&= \left(1 - \frac{\mathrm{Vol}(T^\epsilon_{\sigma^2(i)}(X^n))}{\mathrm{Vol}(T^\epsilon_{\sigma^2(k(i))})}\right)^m \\
&\leq \exp\left(-\frac{\mathrm{Vol}(T^\epsilon_{\sigma^2(i)}(X^n))}{\mathrm{Vol}(T^\epsilon_{\sigma^2(k(i))})} m\right)
\end{aligned}
$$

where the last line followed by applying the inequality $(1-t)^m \leq \exp(-tm)$. Next, using (61) and (63) to bound the volume of the shells,

$$\Pr(F^c | X^n \in T^\epsilon_{\sigma^2(i)})$$
$$\leq \exp\left(-\left(1 - \frac{1}{no_\epsilon(1)} - o_\epsilon(1)\right) m \exp\left(-n\left(I_{\overline{\sigma^2(i)}}(X;Z) + g_\epsilon\right)\right)\right)$$
$$\leq \exp(-\exp(no_\epsilon(1)))$$

where the final line followed by substitution our choice of $m$ from (66). ∎

**Lemma 17.** *For any positive definite covariance matrix $K$,*

$$\Pr(T^\epsilon_K \cap (\mathcal{E}_d \cup \mathcal{E}_b))$$
$$\leq \exp\left(-n\left(D(K||\bar{K}) - o_\epsilon(1) - \delta_p\right)\right) \tag{72}$$

*where $\bar{K}$ is defined in* (14) *and*

$$and \ \delta_p = \frac{1}{n} \log\left(1 - \frac{1}{no_\epsilon(1)} - o_\epsilon(1)\right)^{-1}$$

*Proof:* Lemma 15 gives an upper bound for the probability density on $\mathcal{E}_b$ and $\mathcal{E}_d$. Applying this lemma with (61) and (64), we get

$$\Pr(T^\epsilon_K \cap (\mathcal{E}_d \cup \mathcal{E}_b)) \leq \int_{T^\epsilon_K} f^n_\Sigma(\mathbf{x}, \mathbf{y}) \frac{1}{\mathrm{Vol}(T^\epsilon_{\sigma^2(\nu(\mathbf{x}))}(\mathbf{x}))} d\mathbf{xyz}$$
$$\leq \int_{T^\epsilon_K} \exp\left(-n\left(D(K_{XY}||\Sigma) + h(K_{XY}) - o_\epsilon(1)\right)\right)$$
$$\times \left(1 - \frac{1}{no_\epsilon(1)} - o_\epsilon(1)\right)^{-1} \exp(-n(h(K_{Z|X}) - o_\epsilon(1))) d\mathbf{xyz}$$
$$= \mathrm{Vol}(T^\epsilon_K) \exp\left(-n\left(D(K_{XY}||\Sigma) + h(K_{XY}) - o_\epsilon(1)\right)\right)$$
$$\times \left(1 - \frac{1}{no_\epsilon(1)} - o_\epsilon(1)\right)^{-1} \exp(-n(h(K_{Z|X}) - o_\epsilon(1))) d\mathbf{xyz}.$$

Bounding the volume term using (62) and applying the identity

$$D(K||\bar{K}) = D(K_{XY}||\Sigma) + h(K_{Z|X}) - h(K_{Z|XY})$$

gives the result. ∎

**Lemma 18.** *Let $\mathbf{y}, \mathbf{z}$ be two strings with empirical correlation $\rho_{\mathbf{z},\mathbf{y}}$ and let*

$$A(\mathbf{z}, \mathbf{y}) = \{\tilde{\mathbf{z}} \in T^\epsilon_{\sigma^2} : \rho^2_{\tilde{\mathbf{z}},\mathbf{y}} \geq \rho^2_{\mathbf{z},\mathbf{y}}\}.$$

*Then*

$$\mathrm{Vol}(A(\mathbf{z}, \mathbf{y})) \leq 2\exp\left(\frac{n}{2} \log\left(2\pi e\sigma^2(1 - \rho^2_{\mathbf{z},\mathbf{y}})\right) + no_\epsilon(1)\right).$$

*Proof:* The empirical correlation does not change if we scale the vectors, so we may assume that $\mathbf{z}^t\mathbf{z} = \mathbf{y}^t\mathbf{y} = n(\sigma^2 + \epsilon)$. Suppose $\mathbf{z}^t\mathbf{y} \geq 0$, in which case

$$A(\mathbf{z}, \mathbf{y}) = \{\tilde{\mathbf{z}} \in T^\epsilon_{\sigma^2} : \rho_{\tilde{\mathbf{z}},\mathbf{y}} \geq \rho_{\mathbf{z},\mathbf{y}}\} \cup \{\tilde{\mathbf{z}} \in T^\epsilon_{\sigma^2} : \rho_{\tilde{\mathbf{z}},\mathbf{y}} \leq -\rho_{\mathbf{z},\mathbf{y}}\}$$

and by symmetry the two sets on the right hand side have the same volume and it suffices to consider one of them.

$$\{\tilde{\mathbf{z}} \in T_{\sigma^2}^\epsilon : \rho_{\tilde{\mathbf{z}},\mathbf{y}} \geq \rho_{\mathbf{z},\mathbf{y}}\} = \Big\{ \tilde{\mathbf{z}} \in T_{\sigma^2}^\epsilon : \frac{\tilde{\mathbf{z}}^t \mathbf{y}}{\sqrt{\tilde{\mathbf{z}}^t \tilde{\mathbf{z}} \mathbf{y}^t \mathbf{y}}} \geq \frac{\mathbf{z}^t \mathbf{y}}{\sqrt{\mathbf{z}^t \mathbf{z} \mathbf{y}^t \mathbf{y}}} \Big\}$$

$$= \Big\{ \tilde{\mathbf{z}} \in T_{\sigma^2}^\epsilon : -2 \ \rho_{\mathbf{z},\mathbf{y}} \tilde{\mathbf{z}}^t \mathbf{y} \leq -2 \ \rho_{\mathbf{z},\mathbf{y}} \mathbf{z}^t \mathbf{y} \frac{\sqrt{\tilde{\mathbf{z}}^t \tilde{\mathbf{z}}}}{\sqrt{\mathbf{z}^t \mathbf{z}}} \Big\}$$

$$\triangleq B(\mathbf{z}, \mathbf{y})$$

We now bound the volume of $B(\mathbf{z}, \mathbf{y})$. Let $\mathbf{X} \sim \mathcal{N}(\rho_{\mathbf{x},\mathbf{y}} \mathbf{y}, \sigma^2(1 - \rho_{\mathbf{z},\mathbf{y}}^2) I)$. Then

$$
\begin{aligned}
1 &\geq \int_{B(\mathbf{z},\mathbf{y})} f_{\mathbf{X}}(\mathbf{x}) \, d\mathbf{x} \\
&= \int_{B(\mathbf{z},\mathbf{y})} (2\pi\sigma^2(1 - \rho_{\mathbf{z},\mathbf{y}}^2))^{-n/2} \exp\Big( -\frac{\sum(x_i - \rho_{\mathbf{z},\mathbf{y}} y_i)^2}{2(\sigma^2(1 - \rho_{\mathbf{z},\mathbf{y}}^2))} \Big) d\mathbf{x}.
\end{aligned}
\tag{*}
$$

To continue we upper bound the summand in the exponent as follows

$$
\begin{aligned}
\sum (x_i - \rho_{\mathbf{x},\mathbf{y}} y_i)^2 &= \mathbf{x}^t \mathbf{x} - 2\rho_{\mathbf{z},\mathbf{y}} \mathbf{x}^t \mathbf{y} + \rho_{\mathbf{z},\mathbf{y}}^2 \mathbf{y}^t \mathbf{y} \\
&\leq n(\sigma^2 + \epsilon) - 2\rho_{\mathbf{z},\mathbf{y}} \mathbf{x}^t \mathbf{y} + \rho_{\mathbf{z},\mathbf{y}}^2 n(\sigma^2 + \epsilon) \\
&\leq n(\sigma^2 + \epsilon) - 2\rho_{\mathbf{z},\mathbf{y}}^2 n(\sigma^2 + \epsilon)(1 - o_\epsilon(1)) + \rho_{\mathbf{z},\mathbf{y}}^2 n(\sigma^2 + \epsilon) \\
&\leq n(\sigma^2(1 - \rho_{\mathbf{z},\mathbf{y}}^2) + o_\epsilon(1)).
\end{aligned}
$$

Substituting the above into (*) gives

$$
\begin{aligned}
\mathrm{Vol}(B(\mathbf{z}, \mathbf{y})) &\leq \exp\Big( n\big(\tfrac{1}{2} \log(2\pi\sigma^2(1 - \rho_{\mathbf{z},\mathbf{y}}^2)) + \tfrac{1}{2} + o_\epsilon(1)\big) \Big) \\
&= \exp\Big( n\big(\tfrac{1}{2} \log(2\pi e \sigma^2(1 - \rho_{\mathbf{z},\mathbf{y}}^2)) + o_\epsilon(1)\big) \Big).
\end{aligned}
$$

Observing that an identical argument holds for $\mathbf{z}^t \mathbf{y} \leq 0$ we are done. ∎

**Lemma 19.** *Let* $(\mathbf{x}, \mathbf{y}, \mathbf{z}) \in (\mathcal{E}_c \cup \mathcal{E}_d)^c \cap \mathcal{R}^{3n}$. *Then*

$$
\begin{aligned}
\Pr\Big( \frac{1}{n} \|X^n &- \hat{X}^n\|_2^2 > \Delta | X^n = \mathbf{x}, Y^n = \mathbf{y}, Z^n = \mathbf{z} \Big) \\
&\leq \exp\big( -n \left( R - J(K) - o_\epsilon(1) - \delta_b \right)^+ \big)
\end{aligned}
\tag{73}
$$

*where* $K = K(i, j, k(i), r, s(i), t)$ *is the type containing* $(\mathbf{x}, \mathbf{y}, \mathbf{z})$ *and*

$$J(K) = I_K(X; Z) - I_K(Y; Z),$$

$$\delta_b = \frac{1}{n} \log \left( 2 \left( 1 - \frac{2\sigma^4(k(i))}{n\epsilon^2} \right)^{-1} \right)$$

*Moreover, if* $\log |B^n(\nu(\mathbf{x}))| < nR$ *then*

$$\Pr\Big( \frac{1}{n} \|X^n - \hat{X}^n\|_2^2 > \Delta | X^n = \mathbf{x}, Y^n = \mathbf{y}, Z^n = \mathbf{z} \Big) = 0.$$

*Proof:* Let $L$ be the event that $Z^n \neq \varphi_1(\psi(X^n), Y^n)$. Observe that $L$ occurs when the decoder decodes the wrong codeword and that $\Pr\Big( \frac{1}{n} \|X^n - \hat{X}^n\|_2^2 > \Delta | X^n = \mathbf{x}, Y^n = \mathbf{y}, Z^n = \mathbf{z} \Big)$ is upper bounded by $\Pr(L | X^n = \mathbf{x}, Y^n = \mathbf{y}, Z^n = \mathbf{z})$.

If $i$ is such that $\log |B^n(i)| < nR$, then

$$\Pr(L | X^n = \mathbf{x}, Y^n = \mathbf{y}, Z^n = \mathbf{z}) = 0.$$

For the opposite we case we argue as follows. Quantize the set $T_{\sigma^2(k(i))}$ in cells of diameter at-most $\delta$ so that each cell is either entirely typical with respect to $\mathbf{x}$ or not (except possibly the boundaries). We will now study

$$\Pr(L|X^n = \mathbf{x}, Y^n = \mathbf{y}, Z^n \in [\mathbf{z}]^\delta),$$

where $[\mathbf{z}]^\delta$ is the cell containing the fixed $\mathbf{z}$. In order that $L$ occurs there must be a cell containing a codeword $\tilde{\mathbf{z}}$ such that $\rho^2_{\tilde{\mathbf{z}}, \mathbf{y}} \geq \rho^2_{Z^n, \mathbf{y}}$ and $U(\tilde{\mathbf{z}}) = U(Z^n)$. Let $B(\mathbf{y}, [\mathbf{z}]^\delta)$ be the set of cells containing at least one $\tilde{\mathbf{z}}$ such that $\rho^2_{\tilde{\mathbf{z}}, \mathbf{y}} \geq \rho^2_{\mathbf{z}', \mathbf{y}}$ for some $\mathbf{z}' \in [\mathbf{z}]^\delta$. By summing over each cell (written as $[\tilde{\mathbf{z}}]$) we have

$$\Pr(L|X^n = \mathbf{x}, Y^n = \mathbf{y}, Z^n \in [\mathbf{z}]^\delta)$$

$$\leq \sum_{\substack{[\tilde{\mathbf{z}}]^\delta \in B(\mathbf{y}, [\mathbf{z}]^\delta) \\ [\tilde{\mathbf{z}}]^\delta \neq [\mathbf{z}]^\delta}} \Pr(\exists j : B^n(i)[j] \in [\tilde{\mathbf{z}}]^\delta, U(B^n(i)[j]) = U(Z^n)|X^n = \mathbf{x}, Y^n = \mathbf{y}, Z^n \in [\mathbf{z}]^\delta) \qquad (74)$$

$$+ \Pr(\exists j : B^n(i)[j] \neq Z^n, B^n(i)[j] \in [\mathbf{z}]^\delta, U(B^n(i)[j]) = U(Z^n)|X^n = \mathbf{x}, Y^n = \mathbf{y}, Z^n \in [\mathbf{z}]^\delta). \quad (75)$$

The inequality follows by the union bound and also because the right-hand side assumes that a codeword in any of the cells in $B(\mathbf{y}, [\mathbf{z}]^\delta)$ will lead to an error.

We observe that

$$\lim_{\delta \to 0} \Pr(\exists j : B^n(i)[j] \neq Z^n, B^n(i)[j] \in [\mathbf{z}]^\delta, U(B^n(i)[j]) = U(Z^n)|X^n = \mathbf{x}, Y^n = \mathbf{y}, Z^n \in [\mathbf{z}]^\delta) = 0,$$

because the probability of a cell containing two codewords is negligible as $\delta$ tends to zero.

Turning now to the first summand, applying the union bound gives

$$\Pr(\exists j : B^n(i)[j] \in [\tilde{\mathbf{z}}]^\delta, U(B^n(i)[j]) = U(Z^n)|X^n = \mathbf{x}, Y^n = \mathbf{y}, Z^n \in [\mathbf{z}]^\delta)$$

$$\leq \sum_{j=1}^{|B^n(i)|} \Pr(B^n(i)[j] \in [\tilde{\mathbf{z}}]^\delta|X^n = \mathbf{x}, Y^n = \mathbf{y}, Z^n \in [\mathbf{z}]^\delta)$$

$$\times \Pr(U(B^n(i)[j]) = U(Z^n)|Z^n \in [\mathbf{z}]^\delta, B^n(i)[j] \in [\tilde{\mathbf{z}}]^\delta).$$

Conditioned on $\{Z^n \in [\mathbf{z}]^\delta, B^n(i)[j] \in [\tilde{\mathbf{z}}]^\delta\}$, the chance that the two (necessarily different) codewords share the same bin is $\exp(-nR)$ by the code construction. We will now show that

$$\Pr(B^n(i)[j] \in [\tilde{\mathbf{z}}]^\delta|X^n = \mathbf{x}, Y^n = \mathbf{y}, Z^n \in [\mathbf{z}]^\delta) \leq \Pr(B^n(i)[j] \in [\tilde{\mathbf{z}}]^\delta|X^n = \mathbf{x}, Y^n = \mathbf{y}), \qquad (76)$$

to establish this we will show

$$\Pr(Z^n \in [\mathbf{z}]^\delta|X^n = \mathbf{x}, Y^n = \mathbf{y}, B^n(i)[j] \in [\tilde{\mathbf{z}}]^\delta) \leq \Pr(Z^n \in [\mathbf{z}]^\delta|X^n = \mathbf{x}, Y^n = \mathbf{y})$$

which implies the inequality by reversing the conditioning. Suppose first that $[\tilde{\mathbf{z}}]^\delta$ was a cell not typical with respect to $\mathbf{x}$. Observe that

$$\Pr(Z^n \in [\mathbf{z}]^\delta|X^n = \mathbf{x}, Y^n = \mathbf{y}, B^n(i)[j] \in [\tilde{\mathbf{z}}]^\delta)$$

$$= \Pr(Z^n \in [\mathbf{z}]^\delta, F|X^n = \mathbf{x}, Y^n = \mathbf{y}, B^n(i)[j] \in [\tilde{\mathbf{z}}]^\delta)$$

$$= \Pr(F|X^n = \mathbf{x}, Y^n = \mathbf{y}, B^n(i)[j] \in [\tilde{\mathbf{z}}]^\delta)$$

$$\quad \times \Pr(Z^n \in [\mathbf{z}]^\delta|X^n = \mathbf{x}, Y^n = \mathbf{y}, B^n(i)[j] \in [\tilde{\mathbf{z}}]^\delta, F)$$

$$\leq \Pr(F|X^n = \mathbf{x}, Y^n = \mathbf{y})$$

$$\quad \times \Pr(Z^n \in [\mathbf{z}]^\delta|X^n = \mathbf{x}, Y^n = \mathbf{y}, B^n(i)[j] \in [\tilde{\mathbf{z}}]^\delta, F)$$

$$= \Pr(F|X^n = \mathbf{x}, Y^n = \mathbf{y})$$

$$\quad \times \Pr(Z^n \in [\mathbf{z}]^\delta|X^n = \mathbf{x}, Y^n = \mathbf{y}, F).$$

The inequality follows because $F$ has a greater chance to occur if one of the slots is not occupied by a non-typical codeword. The final equality follows because $Z^n$ is independent of $\{B^n(i)[j] \in [\tilde{\mathbf{z}}]^\delta\}$ conditional on $F$. Therefore

$$\Pr(Z^n \in [\mathbf{z}]^\delta | X^n = \mathbf{x}, Y^n = \mathbf{y}, B^n(i)[j] \in [\tilde{\mathbf{z}}]^\delta)$$
$$\leq \Pr(Z^n \in [\mathbf{z}]^\delta, F | X^n = \mathbf{x}, Y^n = \mathbf{y})$$
$$= \Pr(Z^n \in [\mathbf{z}]^\delta | X^n = \mathbf{x}, Y^n = \mathbf{y}).$$

To argue the case when $[\tilde{\mathbf{z}}]^\delta$ is a cell that is typical one may use a straightforward coupling argument. Together these two cases establish (76). Now

$$\Pr(B^n(i)[j] \in [\tilde{\mathbf{z}}]^\delta | X^n = \mathbf{x}, Y^n = \mathbf{y}) = \int_{[\tilde{\mathbf{z}}]^\delta} \operatorname{Vol}(T^\epsilon_{\sigma^2(k(i))})^{-1} d\mathbf{z}',$$

as $\delta \to 0$ it follows that

$$\Pr(B^n(i)[j] \in [\tilde{\mathbf{z}}]^\delta | X^n = \mathbf{x}, Y^n = \mathbf{y}) \to \Pr(B^n(i)[j] \in [\tilde{\mathbf{z}}]^{d\mathbf{z}} | X^n = \mathbf{x}, Y^n = \mathbf{y})$$
$$= \operatorname{Vol}(T^\epsilon_{\sigma^2(k(i))})^{-1} d\mathbf{z}'.$$

Also as $\delta \to 0$ we have that

$$\sum_{\substack{[\tilde{\mathbf{z}}]^\delta \in B(\mathbf{y}, [\mathbf{z}]^\delta) \\ [\tilde{\mathbf{z}}]^\delta \neq [\mathbf{z}]^\delta}} \to \int_{\tilde{\mathbf{z}}: \rho^2_{\tilde{\mathbf{z}}, \mathbf{y}} \geq \rho^2_{\mathbf{z}, \mathbf{y}}} .$$

Therefore in the limit as $\delta \to 0$,

$$\Pr(L | X^n = \mathbf{x}, Y^n = \mathbf{y}, Z^n = \mathbf{z}) \leq \int_{\tilde{\mathbf{z}}: \rho^2_{\tilde{\mathbf{z}}, \mathbf{y}} \geq \rho^2_{\mathbf{z}, \mathbf{y}}} d\tilde{\mathbf{z}} |B^n(i)| \exp(-nR) \operatorname{Vol}(T^\epsilon_{\sigma^2(k(i))})^{-1}$$
$$= \operatorname{Vol}(A(\mathbf{z}, \mathbf{y})) |B^n(i)| \exp(-nR) \operatorname{Vol}(T^\epsilon_{\sigma^2(k(i))})^{-1},$$

where we used the set $A(\mathbf{z}, \mathbf{y})$ from Lemma 18. Now using the result of Lemma 18 and (61) we obtain

$$\leq 2 \left( 1 - \frac{2\sigma^4(k(i))}{n\epsilon^2} \right)^{-1} \exp\left( -n\left( R + I(\mathbf{y}; \mathbf{z}) \right. \right.$$
$$\left. \left. - I_{\overline{\sigma^2(k(i))}}(X; Z) - o_\epsilon(1) \right) \right)$$
$$\leq 2 \left( 1 - \frac{2\sigma^4(k(i))}{n\epsilon^2} \right)^{-1} \exp\left( -n\left( R - J(\mathbf{x}, \mathbf{y}, \mathbf{z}) - o_\epsilon(1) \right) \right).$$

Also, since $\Pr(L | X^n = \mathbf{x}, Y^n = \mathbf{y}, Z^n = \mathbf{z}) \leq 1$ we get

$$\Pr(L | X^n = \mathbf{x}, Y^n = \mathbf{y}, Z^n = \mathbf{z})$$
$$\exp\left( -n\left( (R - J(\mathbf{x}, \mathbf{y}, \mathbf{z}) - o_\epsilon(1) - \delta_b)^+ \right) \right).$$

$\blacksquare$

**Lemma 20.** *Let $\delta_p, \delta_b$ be sequences going to zero as $n \to \infty$,*

$$G^n_\epsilon(K, \Sigma, \lambda, \Delta, R) =$$
$$\begin{cases} D(K || \bar{K}) - o_\epsilon(1) - \delta_p & \mathbb{E}_K[(X - \lambda(Y, Z))^2] \geq \Delta - o_\epsilon(1) \\ D(K || \bar{K}) - o_\epsilon(1) - \delta_p & \\ \quad + \left( R - I_K(X; Z) \right) & \mathbb{E}_K[(X - \lambda(Y, Z))^2] < \Delta - o_\epsilon(1) \\ \quad + I_K(Y; Z) - o_\epsilon(1) - \delta_b \big)^+ & \text{and } I_K(X; Z) \geq R - o_\epsilon(1) \\ \infty & \text{otherwise,} \end{cases}$$

$$\pi_\epsilon^n(R, \Delta, \Sigma) = \min_i \max_{k,s} \min_j \max_\lambda \min_{r,t} G_\epsilon^n(K_\epsilon, \Sigma, \lambda, \Delta, R),$$

*and*

$$\pi(R, \Delta, \Sigma) = \inf_{\sigma_X} \sup_{\sigma_Z, \rho_{xz}} \inf_{\sigma_Y} \sup_\lambda \inf_{\rho_{xy}, \rho_{yz}} G_G(K, \Sigma, \lambda, \Delta, R),$$

*where $K_\epsilon$ is shorthand for $K_\epsilon(i, j, k, r, s, t)$ and $K$ is a covariance matrix with entries $(\sigma_X, \sigma_Y, \sigma_Z, \rho_{xy}, \rho_{xz}, \rho_{yz})$. Then*

$$\liminf_{\epsilon \to 0} \liminf_{n \to \infty} \pi_\epsilon^n(R, \Delta, \Sigma) \geq \pi(R, \Delta, \Sigma).$$

*Proof:* Let $\delta > 0$. For $\epsilon > 0$ define

$$G_\epsilon(K, \Sigma, \lambda, \Delta, R) =$$
$$\begin{cases} D(K||\bar{K}) - o_\epsilon(1) & \mathbb{E}_K[(X - \lambda(Y, Z))^2] \geq \Delta - o_\epsilon(1) \\ D(K||\bar{K}) - o_\epsilon(1) \\ \quad + (R - I_K(X; Z)) & \mathbb{E}_K[(X - \lambda(Y, Z))^2] < \Delta - o_\epsilon(1) \\ \quad + I_K(Y; Z) - o_\epsilon(1))^+ & \text{and } I_K(X; Z) \geq R - o_\epsilon(1) \\ \infty & \text{otherwise,} \end{cases}$$

and

$$\pi_\epsilon(R, \Delta, \Sigma) \triangleq \min_i \max_{k,s} \min_j \max_\lambda \min_{r,t} G_\epsilon(K_\epsilon, \Sigma, \lambda, \Delta, R).$$

Then for any choice of arguments and $n$ sufficiently large $G_\epsilon - G_\epsilon^n \leq \frac{\delta}{3}$. Hence

$$\liminf_{n \to \infty} \pi_\epsilon^n(R, \Delta, \Sigma) \geq \pi_\epsilon(R, \Delta, \Sigma) - \frac{\delta}{3}.$$

Via the use of the functions $\sigma^2(\cdot)$ and $\eta(\cdot, \cdot, \cdot)$ we write the optimization above as follows

$$\pi_\epsilon(R, \Delta, \Sigma) = \min_{\sigma_X} \max_{\sigma_Z, \rho_{xz}} \min_{\sigma_Y} \max_\lambda \min_{\rho_{xy}, \rho_{yz}} G_\epsilon(K_\epsilon, \Sigma, \lambda, \Delta, R),$$

where the use of $\max, \min$ are justified since we optimizing over finite sets.

Take any sequence $\epsilon_m \to 0$. Let $K^{(m)} = K^{(m)}(\sigma_X^{(m)}, \sigma_Z^{(m)}, \rho_{xz}^{(m)}, \sigma_Y^{(m)}, \rho_{xy}^{(m)}, \rho_{yz}^{(m)})$ and $\lambda^{(m)}$ be such that

$$\pi_{\epsilon_m}(R, \Delta, \Sigma) = G_{\epsilon_m}(K^{(m)}, \Sigma, \lambda^{(m)}, \Delta, R).$$

By considering subsequences, we may assume that $K^{(m)} \to K^\infty$ and $\lambda^{(m)} \to \lambda^\infty$. Then there exists $\tilde{\sigma}_Z^\infty, \tilde{\rho}_{xz}^\infty$ so that

$$\inf_{\sigma_Y} \sup_\lambda \inf_{\rho_{xy}, \rho_{yz}} G_G(K(\sigma_X^\infty, \sigma_Y, \tilde{\sigma}_Z^\infty, \tilde{\rho}_{xz}^\infty, \rho_{xy}, \rho_{yz}), \Sigma, \lambda, \Delta, R)$$

$$\geq \sup_{\sigma_Z, \rho_{xz}} \inf_{\sigma_Y} \sup_\lambda \inf_{\rho_{xy}, \rho_{yz}} G_G(K(\sigma_X^\infty, \sigma_Y, \sigma_Z, \rho_{xz}, \rho_{xy}, \rho_{yz}), \Sigma, \lambda, \Delta, R) - \frac{\delta}{3}$$

and there are sequences $\tilde{\rho}_{xz}^{(m)}, \tilde{\sigma}_Z^{(m)}$ converging to $\tilde{\rho}_{xz}^\infty$ and $\tilde{\sigma}_Z^\infty$ respectively. Let

$$\tilde{\sigma}_Y^{(m)} \in \arg\min_{\sigma_Y} \max_\lambda \min_{\rho_{xy}, \rho_{yz}} G_{\epsilon_m}(K(\sigma_X^\infty, \sigma_Y, \tilde{\sigma}_Z^{(m)}, \tilde{\rho}_{xz}^{(m)}, \rho_{xy}, \rho_{yz}), \Sigma, \lambda, \Delta, R)$$

and by taking a further subsequence we can assume $\tilde{\sigma}_Y^{(m)} \to \tilde{\sigma}_Y^\infty$. Then there exists $\tilde{\lambda}^\infty$ such that

$$\inf_{\rho_{xy}, \rho_{yz}} G_G(K(\sigma_X^\infty, \tilde{\sigma}_Y^\infty, \tilde{\sigma}_Z^\infty, \tilde{\rho}_{xz}^\infty, \rho_{xy}, \rho_{yz}), \Sigma, \tilde{\lambda}^\infty, \Delta, R)$$

$$\geq \sup_\lambda \inf_{\rho_{xy}, \rho_{yz}} G_G(K(\sigma_X^\infty, \tilde{\sigma}_Y^\infty, \tilde{\sigma}_Z^\infty, \tilde{\rho}_{xz}^\infty, \rho_{xy}, \rho_{yz}), \Sigma, \lambda, \Delta, R) - \frac{\delta}{3}$$

and we let $\tilde{\lambda}^{(m)}$ be a sequence converging to $\tilde{\lambda}^\infty$. Let

$$(\tilde{\rho}_{xz}^{(m)}, \tilde{\rho}_{yz}^{(m)}) \in \arg\min_{\rho_{xy}, \rho_{yz}} G_{\epsilon_m}(K(\sigma_X^{(m)}, \tilde{\sigma}_Y^{(m)}, \tilde{\sigma}_Z^{(m)}, \tilde{\rho}_{xz}^{(m)}, \rho_{xy}, \rho_{yz}), \Sigma, \lambda^{(m)}, \Delta, R).$$

Define $\tilde{K}^{(m)} \triangleq \tilde{K}^{(m)}(\sigma_X^{(m)}, \tilde{\sigma}_Y^{(m)}, \tilde{\sigma}_Z^{(m)}, \tilde{\rho}_{xz}^{(m)}, \rho_{xy}^{(m)}, \rho_{yz}^{(m)})$, then observe that

$$\pi_{\epsilon_m}(R, \Delta, \Sigma)$$
$$= \max_{\sigma_Z, \rho_{xz}} \min_{\sigma_Y} \max_{\lambda} \min_{\rho_{xy}, \rho_{yz}} G_{\epsilon_m}(K^{(m)}(\sigma_X^{(m)}, \sigma_Y, \sigma_Z, \rho_{xz}, \rho_{xy}, \rho_{yz}), \Sigma, \lambda, \Delta, R)$$
$$\geq \min_{\sigma_Y} \max_{\lambda} \min_{\rho_{xy}, \rho_{yz}} G_{\epsilon_m}(K^{(m)}(\sigma_X^{(m)}, \sigma_Y, \tilde{\sigma}_Z^{(m)}, \tilde{\rho}_{xz}^{(m)}, \rho_{xy}, \rho_{yz}), \Sigma, \lambda, \Delta, R)$$
$$= \max_{\lambda} \min_{\rho_{xy}, \rho_{yz}} G_{\epsilon_m}(K^{(m)}(\sigma_X^{(m)}, \tilde{\sigma}_Y^{(m)}, \tilde{\sigma}_Z^{(m)}, \tilde{\rho}_{xz}^{(m)}, \rho_{xy}, \rho_{yz}), \Sigma, \lambda, \Delta, R)$$
$$\geq \min_{\rho_{xy}, \rho_{yz}} G_{\epsilon_m}(K^{(m)}(\sigma_X^{(m)}, \tilde{\sigma}_Y^{(m)}, \tilde{\sigma}_Z^{(m)}, \tilde{\rho}_{xz}^{(m)}, \rho_{xy}, \rho_{yz}), \Sigma, \tilde{\lambda}^{(m)}, \Delta, R)$$
$$= G_{\epsilon_m}(\tilde{K}^{(m)}, \Sigma, \tilde{\lambda}^{(m)}, \Delta, R)$$

By examining the various cases and using the continuity of expectation and the information measures, one can show that

$$\liminf_{m \to \infty} G_{\epsilon_m}(\tilde{K}^{(m)}, \Sigma, \tilde{\lambda}^{(m)}, R, \Delta) \geq G_G(\tilde{K}^\infty, \Sigma, \tilde{\lambda}^\infty, R, \Delta).$$

Furthermore,

$$G_G(\tilde{K}^\infty, \Sigma, \tilde{\lambda}^\infty, R, \Delta)$$
$$\geq \inf_{\rho_{xz}, \rho_{yz}} G_G(K(\sigma_X^\infty, \tilde{\sigma}_Y^\infty, \tilde{\sigma}_Z^\infty, \tilde{\rho}_{xz}^\infty, \rho_{xy}, \rho_{yz}), \Sigma, \tilde{\lambda}^\infty, R, \Delta)$$
$$\geq \sup_{\lambda} \inf_{\rho_{xy}, \rho_{yz}} G_G(K(\sigma_X^\infty, \tilde{\sigma}_Y^\infty, \tilde{\sigma}_Z^\infty, \tilde{\rho}_{xz}^\infty, \rho_{xy}, \rho_{yz}), \Sigma, \lambda, \Delta, R) - \frac{\delta}{3}$$
$$\geq \inf_{\sigma_Y} \sup_{\lambda} \inf_{\rho_{xy}, \rho_{yz}} G_G(K(\sigma_X^\infty, \sigma_Y, \tilde{\sigma}_Z^\infty, \tilde{\rho}_{xz}^\infty, \rho_{xy}, \rho_{yz}), \Sigma, \lambda, \Delta, R) - \frac{\delta}{3}$$
$$\geq \sup_{\sigma_Z, \rho_{xz}} \inf_{\sigma_Y} \sup_{\lambda} \inf_{\rho_{xy}, \rho_{yz}} G_G(K(\sigma_X^\infty, \sigma_Y, \sigma_Z, \rho_{xz}, \rho_{xy}, \rho_{yz}), \Sigma, \lambda, \Delta, R) - \frac{2\delta}{3}$$
$$\geq \pi(R, \Delta, \Sigma) - \frac{2\delta}{3}$$

Hence

$$\liminf_{m \to \infty} \liminf_{n \to \infty} \pi_{\epsilon_m}^n(R, \Delta, \Sigma) \geq \pi(R, \Delta, \Sigma) - \delta.$$

But $\epsilon \to 0$ and $\delta > 0$ were arbitrary. ∎

*Proof of Theorem 5:*

$$\Pr\left(\frac{1}{n}\|X^n - \hat{X}^n\|_2^2 > \Delta\right) \tag{77}$$
$$= \Pr\left(\frac{1}{n}\|X^n - \hat{X}^n\|_2^2 > \Delta \big| (X^n, Y^n, Z^n) \in (\mathcal{R}^{3n})^c\right) \Pr((\mathcal{R}^{3n})^c)$$
$$+ \Pr\left(\frac{1}{n}\|X^n - \hat{X}^n\|_2^2 > \Delta \big| (X^n, Y^n, Z^n) \in \mathcal{R}^{3n}\right) \Pr(\mathcal{R}^{3n})$$
$$\leq \int_{\mathcal{R}^{3n}} \Pr\left(\frac{1}{n}\|X^n - \hat{X}^n\|_2^2 > \Delta | \mathbf{x}, \mathbf{y}, \mathbf{z}\right) dF(\mathbf{xyz}) + \Pr((\mathcal{R}^{3n})^c)$$

For now we focus on the integral and will deal with $\Pr((\mathcal{R}^{3n})^c)$ separately.

Observe first that the error probability on $(\mathcal{E}_b \cup \mathcal{E}_c \cup \mathcal{E}_d)^c$ is zero, thus we can we can split the integral as follows, allowing us to deal with the various key events defined in Section E-B.

$$\int_{\mathcal{R}^{3n} \cap \mathcal{E}_c} \Pr\left(\frac{1}{n}\|X^n - \hat{X}^n\|_2^2 > \Delta | \mathbf{x}, \mathbf{y}, \mathbf{z}\right) dF(\mathbf{xyz})$$
$$+ \int_{\mathcal{R}^{3n} \cap \mathcal{E}_d} \Pr\left(\frac{1}{n}\|X^n - \hat{X}^n\|_2^2 > \Delta | \mathbf{x}, \mathbf{y}, \mathbf{z}\right) dF(\mathbf{xyz})$$
$$+ \int_{\mathcal{R}^{3n} \cap \mathcal{E}_b} \Pr\left(\frac{1}{n}\|X^n - \hat{X}^n\|_2^2 > \Delta | \mathbf{x}, \mathbf{y}, \mathbf{z}\right) dF(\mathbf{xyz}).$$

Bounding the error probability on $\mathcal{E}_c$ and $\mathcal{E}_d$ by 1 gives

$$\Pr(\mathcal{E}_c \cap \mathcal{R}^{3n}) + \Pr(\mathcal{E}_d \cap \mathcal{R}^{3n}) \tag{78}$$
$$+ \int_{\mathcal{R}^{3n} \cap \mathcal{E}_b} \Pr\left(\frac{1}{n}\|X^n - \hat{X}^n\|_2^2 > \Delta | \mathbf{x}, \mathbf{y}, \mathbf{z}\right) dF(\mathbf{xyz}).$$

By Lemma 16, $\Pr(\mathcal{E}_c \cap \mathcal{R}^{3n})$ tends to zero double exponentially with the block length and can therefore also be neglected. Let

$$\mathcal{D}_d = \{K : T_K^\epsilon \cap \mathcal{E}_d \neq \emptyset\}.$$

Then applying Lemma 17 gives

$$\Pr(\mathcal{E}_d \cap \mathcal{R}^{3n}) \leq \sum_i \sum_j \sum_{r,t:K(i,j,k(i),r,s(i),t) \in \mathcal{D}_d}$$
$$\exp(-n(D(K||\bar{K}) - o_\epsilon(1) - \delta_p))$$
$$\leq \sum_i \sum_j |\mathcal{P}_{XYZ}^\epsilon| \max_{r,t:K(i,j,k(i),r,s(i),t) \in \mathcal{D}_d}$$
$$\exp(-n(D(K||\bar{K}) - o_\epsilon(1) - \delta_p)).$$

where we have written $K$ for $K_\epsilon(i,j,k,r,s(i),t)$ and likewise $\bar{K}$ for $\overline{K_\epsilon(i,j,k(i),r,s(i),t)}$. Next let

$$\mathcal{D}_b = \{K : T_K^\epsilon \cap \mathcal{E}_b \neq \emptyset\}.$$

Addressing the integral in (78),

$$\int_{\mathcal{R}^{3n} \cap \mathcal{E}_b} \Pr\left(\frac{1}{n}\|X^n - \hat{X}^n\|_2^2 > \Delta | \mathbf{x}, \mathbf{y}, \mathbf{z}\right) dF(\mathbf{xyz})$$
$$\overset{(a)}{\leq} \sum_{K \in \mathcal{D}_b} \int_{T_K \cap \mathcal{E}_b} \exp(-n(R - J(K) - o_\epsilon(1) - \delta_b)^+)$$
$$dF(\mathbf{xyz})$$
$$\leq \sum_{K \in \mathcal{D}_b} \exp(-n(D(K||\overline{K}) - o_\epsilon(1) + (R - J(K) - o_\epsilon(1) - \delta_b)^+ - \delta_p))$$
$$= \sum_i \sum_j \sum_{(r,t):K(i,j,k(i),r,s(i),t) \in \mathcal{D}_b}$$
$$\exp(-n(D(K||\overline{K}) - o_\epsilon(1) + (R - J(K) - o_\epsilon(1) - \delta_b)^+ - \delta_p))$$
$$\leq \sum_i \sum_j |\mathcal{P}_{XYZ}^\epsilon| \max_{(r,t):K(i,j,k(i),r,s(i),t) \in \mathcal{D}_b}$$
$$\exp(-n(D(K||\overline{K}) - o_\epsilon(1) + (R - J(K) - o_\epsilon(1) - \delta_b)^+ - \delta_p)),$$

where (a) follows from Lemma 19 and (b) follows from Lemma 17.

Turning to $(\mathcal{R}^{3n})^c$, using well-known large-deviations results for the Gaussian distribution, we obtain

$$\Pr((\mathcal{R}^{3n})^c) \leq 2\Pr(\mathbf{x}^t\mathbf{x} < n(M_L + \epsilon)) + 2\Pr(\mathbf{x}^t\mathbf{x} > nM_U)$$
$$\leq 2\exp\left(-\frac{n}{2}\left((M_L + \epsilon) - \log(M_L + \epsilon) - 1 - o_\epsilon(1)\right)\right)$$
$$+ 2\exp\left(-\frac{n}{2}\left(M_U - \log M_U - 1 - o_\epsilon(1)\right)\right).$$

Now $M_U$ and $M_L$ can be chosen so that this term does not dominate the exponent and can therefore be neglected. Combining the various bounds (and neglecting the terms in the previous equation) gives

$$\Pr\left(\frac{1}{n}\|X^n - \hat{X}^n\|_2^2 > \Delta \cap \mathcal{R}^{3n}\right)$$
$$\leq \sum_{i,j} |\mathcal{P}_{XYZ}^\epsilon| \Big[ \max_{(r,t):K\in\mathcal{D}_d} \exp(-n(D(K\|\overline{K}) - \delta_p - o_\epsilon(1)))$$
$$+ \max_{(r,t):K\in\mathcal{D}_b} \exp(-n(D(K\|\overline{K}) - o_\epsilon(1) + (R - J(K) - o_\epsilon(1)$$
$$- \delta_b)^+) - \delta_p)\Big].$$

Using the formula $a + b \leq 2\max(a, b)$, we can upper bound the quantity in square brackets by

$$2\max\Big(\max_{(r,t):K\in\mathcal{D}_d} \exp(-n(D(K\|\overline{K}) - \delta_p - o_\epsilon(1))),$$
$$\max_{(r,t):K\in\mathcal{D}_b} \exp(-n(D(K\|\overline{K}) - o_\epsilon(1) + (R - J(K) - o_\epsilon(1) - \delta_b)^+) - \delta_p)\Big).$$

Note that the sets $\mathcal{D}_b$ and $\mathcal{D}_d$ may overlap. However, without loss of generality, we may assume that the $o_\epsilon(1)$ terms are such that the objective in the $\mathcal{D}_d$ max is no smaller than the objective in the $\mathcal{D}_b$ max. This quantity can then be further upper bounded by replacing the maximum over $(r, t)$ such that $K \in \mathcal{D}_b$ with a maximum over $(r, t)$ such that $K \in \mathcal{D}_b \backslash \mathcal{D}_d$. This yields

$$2|\mathcal{P}_{XYZ}^\epsilon| \max_{(r,t)} H(K),$$

with $H(K) = \exp(-nG_\epsilon^n(K))$, where $G_\epsilon^n(K)$ is as in Lemma 20.

Thus

$$P\left(\frac{1}{n}\|X^n - \hat{X}^n\|_2^2 > \Delta\right) \leq \sum_i \sum_j 2|\mathcal{P}_{XYZ}^\epsilon| \max_{r,t} H(K).$$

Since $\lambda$ and the choice of the test channel were arbitrary, the right-hand side is upper bounded by

$$2|\mathcal{P}_{XYZ}^\epsilon|^3 \max_i \min_{k,s} \max_j \min_\lambda \max_{r,t} H(K).$$

We then let take logs, divide by $n$, and let $n$ tend to infinity and $\epsilon$ tend to zero, invoking Lemma 20 to obtain the desired result. ∎

## Appendix F
### Proof of Theorem 6

*Proof:*

Let $f^n, g^n$ be a code for the two-sided Gaussian rate distortion problem with conditional rate distortion function $R_{X|Y}$ and define

$$\mathcal{E}_\Delta^n \triangleq \{(\mathbf{x}, \mathbf{y}) : \|\mathbf{x} - g^n(f^n(\mathbf{x}, \mathbf{y}), \mathbf{y})\|_2^2 > n\Delta\}$$

and

$$\mathcal{E}_K^n \triangleq \{(\mathbf{x}, \mathbf{y}) : nK \geq \|\mathbf{x} - g^n(f^n(\mathbf{x}, \mathbf{y}), \mathbf{y})\|_2^2\},$$

where $K \in \mathbb{R}^+$ is to be specified later. For $R$ fixed, choose a covariance matrix $\Pi$ so that

$$R_{X|Y}(f_\Pi, \Delta) > R. \tag{79}$$

Let $\Delta'$ be the solution to $R_{X|Y}(f_\Pi, \Delta') = R$ and define $\bar{\Delta}(f^n, g^n) \triangleq \mathbb{E}_\Pi[\frac{1}{n}\|X^n - g^n(f^n(X^n, Y^n), Y^n)\|_2^2]$. Then according to [36, section 4]

$$R_{X|Y}(f_\Pi, \Delta) > R = R_{X|Y}(f_\Pi, \Delta') \geq R_{X|Y}(f_\Pi, \bar{\Delta}) \tag{80}$$

for every $n$ and code $(f^n, g^n)$ with rate at most $R$. Monotonicity of the rate distortion function implies that $\bar{\Delta}(f^n, g^n) \geq \Delta' > \Delta$.

To continue we modify our original code to give $(\tilde{f}^n, \tilde{g}^n)$. The modification comprises adding a new codeword such that the decoder emits the string $\mathbf{0}$ on receipt of this codeword. Encoder $\tilde{f}^n$, knowing the side information can choose to send this codeword if the choice by $f_n$ results in a higher distortion than $\frac{1}{n}\|X^n\|_2^2$. If we let $\hat{X}^n = g^{(n)}(f^{(n)}(X^n, Y^n), Y^n)$ and $\hat{\tilde{X}}^n = \tilde{g}^{(n)}(\tilde{f}^{(n)}(X^n, Y^n), Y^n)$ then we see that $n^{-1}(X^n - \hat{\tilde{X}}^n)^2 \leq \frac{1}{n}\|X^n\|_2^2$ a.s. Modifying the code in this way only reduces the squared error, hence defining

$$\tilde{\mathcal{E}}_\Delta^n = \{(\mathbf{x}, \mathbf{y}) : \|\mathbf{x} - \tilde{g}^n(\tilde{f}^n(\mathbf{x}, \mathbf{y}), \mathbf{y})\|_2^2 > n\Delta\}$$

(and correspondingly $\tilde{\mathcal{E}}_K^n$) we see that $\mathcal{E}_\Delta \supset \tilde{\mathcal{E}}_\Delta$. In the following all expectations and probabilities are with respect to the law $f_\Pi$ unless stated otherwise.

$$\mathbb{E}[\|X^n - \hat{\tilde{X}}^n\|_2^2 \mathbf{1}_{\tilde{\mathcal{E}}_\Delta^n \cap (\tilde{\mathcal{E}}_K^n)^c}] \leq \mathbb{E}[\|X^n\|_2^2 \mathbf{1}_{\tilde{\mathcal{E}}_\Delta^n \cap (\tilde{\mathcal{E}}_K^n)^c}]$$
$$\leq \mathbb{E}[\|X^n\|_2^2 \mathbf{1}_{\{\|X^n\|_2^2 > nK\}}].$$

Next, applying the Cauchy-Schwarz inequality gives

$$\leq \sqrt{\mathbb{E}[(\|X^n\|_2^2)^2] \Pr(\|X^n\|_2^2 > nK)}$$
$$= \sqrt{\mathbb{E}\left[\sum_{i=1}^n \sum_{j=1}^n X_i^2 X_j^2\right] \Pr(\|X^n\|_2^2 > nK)}$$
$$= \sqrt{(n\mathbb{E}[X_1^4] + (n^2 - n)\mathbb{E}[X_1^2]\mathbb{E}[X_1^2]) \Pr(\|X^n\|_2^2 > nK)}.$$

Choosing $K = \mathbb{E}[X_1^2] + \epsilon$ and applying Chebyshev's inequality to the probability allows us to further bound this quantity by

$$\leq \sqrt{(n\mathbb{E}[X_1^4] + (n^2 - n)\mathbb{E}[X_1^2]\mathbb{E}[X_1^2])}$$
$$\times \sqrt{\frac{\mathbb{E}[X_1^4] - \mathbb{E}[X_1^2]^2}{n\epsilon^2}}.$$

Hence

$$\mathbb{E}[n^{-1}\|X^n - \hat{\tilde{X}}^n\|^2 \mathbf{1}_{\tilde{\mathcal{E}}_\Delta^n \cap (\tilde{\mathcal{E}}_K^n)^c}]$$
$$\leq \sqrt{(n^{-1}\mathbb{E}[X_1^4] + (1 - n^{-1})\mathbb{E}[X_1^2]\mathbb{E}[X_1^2])}$$
$$\times \sqrt{\frac{\mathbb{E}[X_1^4] - \mathbb{E}[X_1^2]^2}{n^3\epsilon^2}}$$

which goes to zero with $n$. We note that this new code has rate $\tilde{R} = R + n^{-1}\log(1 + \exp(-nR)) = R + o_n(1)$. Let $\Delta' - \Delta > \delta_1 > 0$, and $\tilde{\Delta}$ be the solution to $\tilde{R} = R(f_\Pi, \tilde{\Delta})$. Then for $n$ sufficiently large

$$\Delta' - \tilde{\Delta} < \delta_1.$$

We also note that $\bar{\Delta}(\tilde{f}^n, \tilde{g}^n) \geq \tilde{\Delta}$. One may decompose the space into different events to see that

$$
\begin{aligned}
\bar{\Delta}(\tilde{f}^n, \tilde{g}^n) &= \mathbb{E}[n^{-1}\|X^n - \tilde{\tilde{X}}^n\|_2^2] \\
&= \mathbb{E}[n^{-1}\|X^n - \tilde{\tilde{X}}^n\|_2^2 \mathbf{1}_{(\tilde{\mathcal{E}}_\Delta^n)^c}] \\
&\quad + \mathbb{E}[n^{-1}\|X^n - \tilde{\tilde{X}}^n\|_2^2 \mathbf{1}_{\tilde{\mathcal{E}}_\Delta^n \cap \tilde{\mathcal{E}}_K^n}] \\
&\quad + \mathbb{E}[n^{-1}\|X^n - \tilde{\tilde{X}}^n\|_2^2 \mathbf{1}_{\tilde{\mathcal{E}}_\Delta^n \cap (\tilde{\mathcal{E}}_K^n)^c}] \\
&\leq \Delta \Pr((\tilde{\mathcal{E}}_\Delta^n)^c) + K \Pr(\tilde{\mathcal{E}}_\Delta^n \cap \tilde{\mathcal{E}}_K^n) \\
&\quad + \mathbb{E}[n^{-1}\|X^n - \tilde{\tilde{X}}^n\|_2^2 \mathbf{1}_{\tilde{\mathcal{E}}_\Delta^n \cap (\tilde{\mathcal{E}}_K^n)^c}] \\
&\leq \Delta(1 - \Pr(\tilde{\mathcal{E}}_\Delta^n)) + K \Pr(\tilde{\mathcal{E}}_\Delta^n) \\
&\quad + \mathbb{E}[n^{-1}\|X^n - \tilde{\tilde{X}}^n\|_2^2 \mathbf{1}_{\tilde{\mathcal{E}}_\Delta^n \cap (\tilde{\mathcal{E}}_K^n)^c}]
\end{aligned}
$$

i.e.

$$\Pr(\tilde{\mathcal{E}}_\Delta^n) \geq \frac{\bar{\Delta}(\tilde{f}^n, \tilde{g}^n) - \Delta - \mathbb{E}[n^{-1}\|X^n - \tilde{\tilde{X}}^n\|_2^2 \mathbf{1}_{\tilde{\mathcal{E}}_\Delta^n \cap (\tilde{\mathcal{E}}_K^n)^c}]}{K - \Delta}. \tag{81}$$

Thus

$$
\begin{aligned}
\Pr(\mathcal{E}_\Delta^n) &\geq \Pr(\tilde{\mathcal{E}}_\Delta^n) \\
&\geq \frac{\tilde{\Delta} - \Delta - \mathbb{E}[n^{-1}\|X^n - \hat{X}^n\|_2^2 \mathbf{1}_{\tilde{\mathcal{E}}_\Delta^n \cap (\tilde{\mathcal{E}}_K^n)^c}]}{K - \Delta} \\
&\geq \frac{\Delta' - \Delta - \delta_2}{K - \Delta} \triangleq \alpha > 0
\end{aligned}
$$

for all $n > n_1$ (where $\delta_2 \triangleq \delta_1 + \mathbb{E}[n^{-1}(X^n - \hat{X}^n)^2 \mathbf{1}_{\mathcal{E}_\Delta^n \cap (\mathcal{E}_K^n)^c}]$). Next, we set

$$G^n = \left\{ (\mathbf{x}, \mathbf{y}) : \left| \frac{1}{n}\log\frac{f_\Pi(\mathbf{x}, \mathbf{y})}{f_\Sigma(\mathbf{x}, \mathbf{y})} - D(\Pi||\Sigma) \right| < \delta_3 \right\}.$$

By the law of large numbers,

$$\int_{G^n} f_\Pi(\mathbf{x}, \mathbf{y}) \, d\mathbf{x}\mathbf{y} > 1 - \frac{1}{2}\alpha$$

for all $n$ sufficiently large. Combining everything, this gives

$$
\begin{aligned}
\Pr_\Sigma(\mathcal{E}_\Delta^n) &= \int_{\mathcal{E}_\Delta^n} f_\Sigma(\mathbf{x}, \mathbf{y}) \, d\mathbf{x}\mathbf{y} \\
&\geq \int_{\mathcal{E}_\Delta^n \cap G^n} f_\Sigma(\mathbf{x}, \mathbf{y}) \, d\mathbf{x}\mathbf{y} \\
&= \int_{\mathcal{E}_\Delta^n \cap G^n} f_\Pi(\mathbf{x}, \mathbf{y}) \exp\left( -\log\frac{f_\Pi(\mathbf{x}, \mathbf{y})}{f_\Sigma(\mathbf{x}, \mathbf{y})} \right) d\mathbf{x}\mathbf{y} \\
&\geq \frac{1}{2}\alpha \exp(-n(D(\Pi||\Sigma) + \delta_3)).
\end{aligned}
$$

We observe that this inequality holds for all codes of rate at most $R$ and $\Pi$ satisfying (79). To complete the proof it suffices to show that

$$\lim_{\epsilon \to 0} \inf_{\Pi : R(\Pi, \Delta) > R + \epsilon} D(\Pi||\Sigma) = \inf_{\Pi : R(\Pi, \Delta) > R} D(\Pi||\Sigma)$$

The first direction ($\geq$) is obvious. For the reverse inequality, choose $\Pi^*$ to achieve within $\delta$ of the infimum on the right-hand side. Let $\Pi^{(\epsilon)}$ be a collection of covariance matrices converging to $\Pi^*$ such that $R(\Pi^{(\epsilon)}, \Delta) > R + \epsilon$. That such a choice is possible follows by continuity of the rate distortion function. Then

$$\lim_{\epsilon \to 0} \inf_{\Pi : R(\Pi, \Delta) > R + \epsilon} D(\Pi||\Sigma) \leq \lim_{\epsilon \to 0} D(\Pi^{(\epsilon)}||\Sigma) = D(\Pi^*||\Sigma) \leq \inf_{\Pi : R(\Pi, \Delta) > R} D(\Pi||\Sigma) + \delta$$

by continuity of relative entropy. But $\delta$ was arbitrary. ∎

### Acknowledgment

The authors gratefully acknowledge the careful reviews of the anonymous referees which improved the quality of this work. The authors wish to thank Aggelos Vamvatsikos for his assistance during the initial stages of this work. This research was supported by the National Science Foundation under grant CCF 06-42925 (CAREER) and CCF 08-0830496 and by the Air Force Office of Scientific Research (AFOSR) under grant FA9550-08-1-0060.